\documentclass[12pt, oneside, openany]{article} 

\usepackage{preamble}  

\begin{document}

\begin{titlepage}
        \title{\large{\textsc{Evaluating financial tail risk forecasts: \\  
                        Testing Equal Predictive Ability}}}
        \author{{\normalsize Lukas \textsc{Bauer}\({}^{1}\) }}
        \date{\today}
        \maketitle
        \thispagestyle{empty}
        \begin{abstract}                
                \vspace{0.1 cm}
                This paper provides comprehensive simulation results on the finite sample properties of the Diebold-Mariano (DM) test by \textcite{Diebold.1995} and the model confidence set (MCS) testing procedure by \textcite{Hansen.2011} applied to the asymmetric loss functions specific to financial tail risk forecasts, such as Value-at-Risk (VaR) and Expected Shortfall (ES). We focus on statistical loss functions that are strictly consistent in the sense of \textcite{Gneiting.2011}. We find that the tests show little power against models that underestimate the tail risk at the most extreme quantile levels, while the finite sample properties generally improve with the quantile level and the out-of-sample size. For the small quantile levels and out-of-sample sizes of up to two years, we observe heavily skewed test statistics and non-negligible type III errors, which implies that researchers should be cautious about using standard normal or bootstrapped critical values. We demonstrate both empirically and theoretically how these unfavorable finite sample results relate to the asymmetric loss functions and the time varying volatility inherent in financial return data.
                \vspace{1 cm}  
        \end{abstract}
        \thispagestyle{empty}
        \footnotetext[1]{Chair of Statistics and Econometrics, Institute of Economics, University of Freiburg, Rempartstr. 16, 79098, Freiburg, Germany; University of Freiburg, e-mail: lukas.bauer@vwl.uni-freiburg.de, telephone: +49 761 203-2337}
\end{titlepage}

\setcounter{tocdepth}{1}

\begingroup
\let\clearpage\relax
\section{Introduction}\label{ch:introduction}
Since the worldwide financial crisis of 2007 and 2008, both investors and regulators have become increasingly focused on predicting financial risk, in particular on rare but extreme events. Two important downside risk measures are the Value-at-Risk (VaR) and Expected Shortfall (ES). To assess VaR and ES forecasts, econometricians, institutions and regulators may choose among different measures. They may choose to backtest how often financial losses exceed the quantile forecasts, to employ economic objective functions or to use statistical loss functions. 

In this paper, we restrict our attention to statistical losses and examine the finite sample properties of two prominent tests for equal predictive ability, the Diebold-Mariano (DM) test by \textcite{Diebold.1995} and the model confidence set (MCS) testing procedure by \textcite{Hansen.2011} applied to out-of-sample VaR and ES forecasts evaluated using asymmetric statistical loss functions. Furthermore, we empirically and theoretically shed light on the underlying mechanisms that influence the finite sample properties of these two tests.

Not only the DM test, but also the MCS is widely popular among applied econometricians and regularly used to assess the predictive ability of models that make VaR or ES forecasts. For instance, \textcite{Bernardi.2017} successfully use the MCS to determine model weights for combined forecasts. Both \textcite{Taylor.2020} and \textcite{Dimitriadis.2022} employ it to compare models that generate VaR and ES forecasts.  

Unfortunately, little is known about the finite sample properties of the two testing procedures when comparing VaR or ES forecasts, though the losses that result from tail risk forecasts differ strongly from losses that are used in standard comparison problems. We find that the estimators of both mean and variance of the out-of-sample loss have huge variances, which may influence the tests inference about the competing models' predictive ability. Frequently, the estimated loss differentials show a different sign than their expectation, i.e., they do not reflect the correct ranking of two models in terms of their expected losses.

To study the finite sample properties of the tests under `realistic' conditions, i.e., to examine losses that resemble those that we observe empirically, we perform Monte Carlo simulations and simulate forecasts. The forecasting targets of this paper are 1-day-ahead VaR and ES at the 1\%, 2.5\%, 5\% and 10\% level that result from location-scale models with time-varying conditional variances.

To evaluate the VaR and the joint VaR and ES forecasts, we use statistical loss functions that are \textit{strictly consistent} as defined by \textcite{Gneiting.2011}. A strictly consistent loss function ensures that the correct forecast of the target variable is the unique minimizer of the expected loss. Appropriate loss functions to evaluate quantile forecasts live in the generalized piecewise linear (GPL) class. Though there is no strictly consistent loss function to evaluate ES forecasts on their own, \textcite{Fissler.2016} provide a class of strictly consistent loss functions to jointly evaluate VaR and ES.  

Realistically, we may regularly encounter situations when none of the forecasting methods consistently forecasts the true value of the target variable. \textcite{Patton.2020} demonstrates that under model misspecification, estimation error or nonnested information sets of competing forecasting models, different consistent loss functions may induce different rankings of the models. Consequently, we understand the predictive ability (PA) of a forecasting method as the associated expected value of a specific loss function \( L \). 

Regardless the evaluation criterion, researchers must often choose from several or even many models that are available for forecasting. 
Over the last decades, researchers have thus proposed various statistical procedures to rank models according to their predictive ability. An early `one-step' procedure is the Reality Check for data snooping by \textcite{White.2000}, developed to compare a baseline model to a number of other models. A rejection, however, only admits to conclude that there exists at least one superior alternative model. \textcite{Hansen.2005} adjusted his procedure by studentizing the test statistics to improve the power of the testing procedure and to make it more robust against irrelevant alternatives. Yet, both procedures do not deliver a complete subset of models with equal predictive ability (EPA). They only inform if there exists an alternative that produces more accurate forecasts. To address this issue, \textcite{Romano.2005} propose an iterative step-wise procedure to identify those superior alternatives. Their paper is a straightforward extension of \textcite{White.2000} and requires the researcher to choose a certain benchmark. 
Yet, the researcher may lack an obvious benchmark model or face several competing models that are asymptotically statistically equivalent. Thus, \textcite{Hansen.2011} choose a different approach with the Model Confidence Set (MCS) testing procedure. After a series of tests of EPA, the testing procedure returns a subset of models that contains the best model(s) with a given confidence level. For all models in the subset, we cannot reject the null of EPA at a specified level of the test \( \alpha \). The term `best' refers to some suitable criterion, in this paper the smallest expected out-of-sample loss.\footnote{Alternative measures are e.g.\ Sharpe ratios or information criteria (\textcite{Hansen.2011}).}\footnote{As the MCS takes losses as primitives, the notion of the true model is less important. Yet, a strictly consistent loss function ensures that the true model is associated with the smallest expected loss.}

In our simulations, we find that the tests show little power against models that underestimate the tail risk at the extreme quantile levels \( p \in \{0.01, 0.025 \} \), where, for out-of-sample sizes of up to two years, we observe heavily skewed test statistics and non-negligible type III errors. We demonstrate how these unfavorable finite sample results relate to the asymmetric loss functions and the time varying volatility inherent in financial return data. For instance, when comparing a superior, more conservative model to an inferior one that underestimates the risk, an extreme quantile level \( p \) implies that only a tiny fraction of the observations correctly indicate the true ranking of the two models, which leads to huge sampling variability and explains the type III errors. 
The finite sample properties of the tests improve with the quantile level and the out-of-sample size. 
For two and more years of observations, we observe higher power when one performs joint forecast evaluation of ES and VaR as rather than of standalone VaR forecasts. 
Our findings imply that the tests need long evaluation windows of several years to make reliable inference about the predictive ability of competing models for extreme quantiles, which may be problematic due to structural breaks. 

When using different parametrizations of the loss functions for both standalone VaR forecasts and joint VaR and ES forecasts, respectively, we do not find uniform patterns in the finite sample properties of the tests. 

The remainder of this paper is structured as follows. Section \ref{sec:framework} introduces the criterion and procedure to evaluate forecasts. Next, we present the simulation and corresponding results in Section \ref{sec:simulation}. Section \ref{sec:conclusion} concludes.

\section{Framework}\label{sec:framework}

In this paper, we focus on evaluating one-day-ahead \( VaR_{p}\) and \( ES_{p}\) forecasts. 
Formally, \( VaR_{p}\) and \( ES_{p}\) are defined as
\begin{align}
    VaR_{p} &:= F^{-1}(p) := \inf \{ z \in \mathbb{R}: F(z) \geq p \}. \\
    ES_{p} & := \dfrac{1}{p} \int^{p}_{0} VaR_{u} du, \; p \in (0,1], 
\end{align}
where \( F \) is the cumulative distribution function of \( Y \), and \( ES_{0} = ess \inf Y \).

We assess competing forecasting methods through their global out-of-sample loss:
\begin{align}
    \bar{L} = \dfrac{1}{P} \displaystyle \sum^{P}_{t=1} L(x_{t}, y_{t}),
\end{align}
where \( P \in \mathbb{N} \) denotes the number of forecasts, \( x_{1}, \ldots, x_{P} \) are the forecasts and \( y_{1}, \ldots, y_{P} \) the observations. \( L \) is a \textit{strictly consistent} scoring or loss function.

\textcite{Gneiting.2011b} characterizes a class of strictly consistent scoring or loss functions to evaluate quantile forecasts, the so called `generalized piecewise linear' (GPL) class. The loss functions take the form  
\begin{align}\label{eq:gpl_loss}
    L(x,y) = (\mathbbm{1}\{y<x\} - p) \times (g(x) - g(y)),
\end{align}
where \( g\) is a strictly increasing function. This family nests a homogeneous parametric GPL family that \textcite{Patton.2020} defines as:
\begin{align}\label{eq:hom_gpl_loss}
    L(x, y) =  (\mathbbm{1}\{y \leq x\} - p) \times (\sgn (x) |x|^{b} - \sgn(y) |y|^{b})/b , \; b>0.
\end{align}

For \( b = 1 \), the latter equation delivers the popular "tick" loss function.  
\begin{align}\label{eq:tick_loss}
    L(x, y) =  (\mathbbm{1}\{y \leq x\} - p) \times (\sgn (x) |x| - \sgn(y) |y|).
\end{align}

We restrict our attention to the homogeneous parametric GPL family and set \( b = 0.5, 1, 2 \).  

While no strictly consistent loss function exists to evaluate ES forecasts on their own (\cite{Gneiting.2011}), \textcite{Fissler.2016} provide a class of loss functions that are consistent for joint VaR and ES forecasts:
\begin{align}\label{eq:joint_loss}
    L(x_{1}, x_{2}, y) = & ( \mathbbm{1}\{y \leq x_{1}\} - p) \times G_{1}(x_{1}) - \mathbbm{1}\{y<x_{1}\} \times G_{1}(y) \nonumber \\
    & + G_{2}(x_{2}) \times \left( x_{2} - x_{1} + \dfrac{1}{p} \mathbbm{1}\{y \leq x_{1}\} (x_{1}-y) \right) \\ 
    & - \mathcal{G}_{2}(x_{2}) + a(y), \nonumber
\end{align}
where \( G_{1}, G_{2}, \mathcal{G}_{2}, a:\mathbb{R} \rightarrow \mathbb{R}, \mathcal{G}_{2}^{'}=G_{2}\), \( G_{1} \) is increasing and \( \mathcal{G}_{2} \) is increasing and convex. \( L \) is strictly consistent if \( \mathcal{G}_{2} \) is strictly increasing and strictly convex.

We follow \textcite{Taylor.2020} and use three different parametrizations of Equation (\ref{eq:joint_loss}) that we present in the table below. First, a scoring function based on the asymmetric Laplace (AL) density that \textcite{Taylor.2019} proposes. Second, a parametrization that \textcite{Nolde.2017} suggest (NZ). Third, the loss function put forward by \textcite{Fissler.2016}, slightly adjusted as in \textcite{Taylor.2020} (FZG).\footnote{These parametrizations give different weights to the conditional quantile forecast and the forecast of the conditional mean of the truncated distribution. For a discussion of these parametrizations see e.g.\ \textcite{Taylor.2020}.}

\begin{table}[h!]
    \centering
    \scriptsize
    \caption{Parametrizations of the joint loss function}
    \label{tab:joint_loss}
    \begin{tabularx}{1.0\textwidth}{l X X X X}
                    & \( G_{1}(x)\)     & \( G_{2}(x) \)                & \( \mathcal{G}_{2}(x) \)    & \(a(y)\) \\ 
        \( AL \)    &  \( 0 \)          & \( -1/x \)                    & \( -ln(-x)  \)              & \(1-ln(1-p) \)  \\ 
        \( NZ \)    &  \( 0 \)          & \( 1/2 (-x)^{-1/2}  \)        & \( -(-x)^{1/2}  \)          & \( 0 \)\\ 
        \( FZG \)    &  \( x \)          & \( \exp(x) / (1+\exp(x)) \)   & \( ln(1+\exp(x)) \)         & \( 0 \) \\ 
    \end{tabularx}\\
\scriptsize{\textit{Notes:} This table summarizes the parametrizations of the loss function given by Equation \ref{eq:joint_loss}.}
\end{table}

\subsection{Testing Equal Predictive Ability}
To compare the global out-of-sample loss of competing models, we consider (i) the DM test by \textcite{Diebold.1995} in the case of two models, and (ii) the MCS procedure by \textcite{Hansen.2011} in the case of multiple models. Both procedures are based on the studentized loss differences between two or more models.

\subsubsection{Diebold-Mariano Test}\label{subsec:DM}
The DM test is a hugely popular test for the pairwise comparison of equal predictive ability. Let  \(d_{t, 12} \equiv l_{t, 1} - l_{t, 2} \) denote the loss difference between model 1 and 2 at time \( t\). The DM test uses the the relative performance variable
\begin{equation}
    \bar{d}_{t, 12} = \dfrac{1}{P} \displaystyle \sum^{P}_{t=1}  d_{t, 12} , t=1, \ldots, P,
\end{equation} 
and the statistic 
\begin{equation}
    t_{12} = \dfrac{\bar{d}_{12}}{\widehat{var}(\bar{d}_{12})} \overset{d}{\rightarrow} \mathcal{N}(0, 1)
\end{equation}
to test the null hypothesis \( H_{0} : \mathbb{E}[d_{t, 12}] = 0 \). Under comparatively mild mixing conditions on the \( d_{12} \) (\textcite{Giacomini.2006}), and if \( \widehat{var}(\bar{d}_{12}) \) is a consistent estimator of the variance of \( \bar{d}_{12} \), then the limiting distribution under the null is standard normal. We consider the two-tailed DM test, as this does not require a benchmark model, and as the true or optimal forecaster is unknown in practice. Moreover, this test statistic is used in the MCS testing procedure, which also does not specify a benchmark model. 

We examine the power of the DM test, defined as a rejection when \( sgn(\bar{d}_{12}) = sgn(\mathbb{E}[d_{t, 12}] ) \), and the type III error, which occurs if the test rejects but  \( sgn(\bar{d}_{12}) \neq sgn(\mathbb{E}[d_{t, 12}] ) \). Furthermore, we use the realizations \( t_{12} \) to examine the distribution of the underlying random variable.

\subsubsection{Model Confidence Set}\label{subsec:mcs}
The idea of the MCS is to reduce the set of candidate models to some smaller subset that contains the model(s) with the smallest expected loss, \( \mathcal{M}^{*} \), with a given level of confidence \( 1-\alpha \). Initially, the MCS testing procedure starts with a set of models \( \mathcal{M}^{0} \), on which it then performs a sequence of tests of equal predictive ability. Once the null hypothesis of equal predictive ability cannot be rejected anymore, the testing procedure halts. It delivers a subset \( \widehat{\mathcal{M}}^{*}_{1-\alpha} \), the so-called \textit{model confidence set}. This subset contains the models for which the testing procedure does not find statistically significant differences in the out-of-sample loss. 

Below, we briefly outline the framework of the MCS and introduce the necessary notation. The initial set \( \mathcal{M}^{0} \) consists of \( m_{0} \in \mathbb{N} \) competing models, \( \mathcal{M} \subset \mathcal{M}^{0} \) denotes the set that contains models \( i=1, \ldots, m, \; m \leq m_{0} \). We use a strictly consistent loss function \( L \) to evaluate the point forecasts. The loss at time \( t=1, \ldots, P \) corresponding to model \( i \) is denoted \( l_{t,i} \). We consider the relative performance variables 
\begin{equation}
    d_{t, ij} \equiv l_{t, i} - l_{t, j}  \text{  for } i, j \in  \mathcal{M}^{0},  t=1, \ldots, P,
\end{equation} 
and 
\begin{equation}\label{eq:d_idot}
    d_{t, i\cdot} \equiv l_{t, i} - \dfrac{1}{m} \displaystyle\sum_{j}^{m}  l_{t, j}  \text{  for } i, j \in  \mathcal{M}^{0},  t=1, \ldots , P.
\end{equation} 

\( \mu_{ij} \equiv \mathbb{E}[d_{t, ij}] \) and \( \mu_{i \cdot} \equiv \mathbb{E}[d_{t, i \cdot}] \) denote the expected loss differentials. The competing models are ranked by their expected losses: model \( i \) is \textit{superior} over model \( j \) if \( \mu_{ij} < 0 \). Under the null of equal predictive ability it holds that 
\begin{align}\label{eq:H_0_idot}
    H_{0, \mathcal{M} } : \mu_{ij} = 0 \text{ for all } i,j = 1, \ldots, m,
\end{align}
or equivalently\footnote{\textcite{Hansen.2011} show this equivalence in Section 3.1.2.},  
\begin{align}\label{eq:H_0_ij}
    H_{0, \mathcal{M}} : \mu_{i \cdot} = 0 \text{ for all } i = 1,\ldots, m .
\end{align}

To determine \( \mathcal{M}^{*} \), the MCS testing procedure uses an equivalence test \( \delta_{\mathcal{M}} \) and an elimination rule \( \epsilon_{\mathcal{M}} \). At any testing step, it performs the test \( \delta_{\mathcal{M}} \) to test \( H_{0, \mathcal{M}} \). If the test reject \( H_{0, \mathcal{M}} \), the elimination rule \( \epsilon_{\mathcal{M}} \) selects the model that is eliminated from the current set of candidate models \( \mathcal{M} \). 

Algorithm \ref{alg:MCS} below summarizes the MCS testing procedure. 
\begin{algorithm}\label{alg:MCS}
\begin{itemize}
    \item Step 0: Initially set \( \mathcal{M} = \mathcal{M}^{0}\).
    \item Step 1: Test \( H_{0, \mathcal{M}} \) using an equivalence test  \( \delta_{\mathcal{M}} \) at level \( \alpha \).
    \item Step 2: If \( H_{0, \mathcal{M}}  \) is not rejected, set \( \widehat{\mathcal{M}}^{*}_{1-\alpha} = \mathcal{M}  \).\\
    Otherwise, use an elimination rule \( \epsilon_{\mathcal{M}} \) to eliminate an object from \( \mathcal{M} \), repeat step 1 and step 2.
\end{itemize}
\end{algorithm}

We focus on the testing procedure that uses the \( T_{max, \mathcal{M}} \) statistic, which is based on the loss differential between model \( i \) and the average over all models. We defer results for the testing procedure that uses the \( T_{R, \mathcal{M}} \) statistic to the appendix. The \( T_{R, \mathcal{M}} \) statistic is based on the loss differentials between model \( i \) and model \( j \). In both cases, the MCS testing procedure eliminates the model that has the largest standardized excess loss as compared to its competitors.

\subsubsection*{Finite sample properties of the MCS}\label{subsubsec:mcs_properties}

In this paper we focus on the performance of the MCS procedure by means of \textit{potency} and \textit{power}. 

Potency is a concept that \textcite{Hendry.2014} employ in the context of model selection, and \textcite{Quaedvlieg.2021} uses it to describe the performance of a multi step MCS. While related to the usual notion of size, potency is defined as the frequency of \( \mathcal{M}^{*} \subset \widehat{\mathcal{M}}_{1-\alpha}^{*} \), i.e.\ as the frequency with which the MCS \( \widehat{\mathcal{M}}_{1-\alpha}^{*} \) includes the model(s) with the smallest expected loss.

In their abstract, \textcite{Hansen.2011} put forward the following interpretation of the MCS. "A MCS is a set of models that is constructed such that it will contain the best model with a given level of confidence. The MCS is in this sense analogous to a confidence interval for a parameter." Thus, we discuss the relationship between potency, the level of the test \( \alpha \) used in the MCS test and the level of confidence with which the MCS supposedly contains the best model.  

We first consider the relationship between potency and the strong control of the familywise error rate\footnote{`Strong' implies that the control of the familywise error rate holds for any \( \mathcal{M}^{*} \subset \mathcal{M}^{0} \), while `weak' control usually refers to the case that \( \mathcal{M}^{*} = \mathcal{M}^{0} \), i.e.\  \( \mu_{ij}= 0 \text{ for all } i,j \in \{1, \ldots, m \} \) (\textcite{Lehmann.2022}).}, i.e.\ the probability of eliminating one or more models with the smallest expected loss. The familywise error rate is bounded by the level \( \alpha \) used at every step of the MCS testing procedure. Yet, the level of the test \( \alpha \) has different implications if \( \mathcal{M}^{*} \) consists of a single best model as compared to if \( \mathcal{M}^{*} \) consists of several best models (\textcite{Hansen.2011}) - which, in reality, is not known. 

First, assume that \( \mathcal{M}^{*} \) consists of a single best model, i.e. exactly one model has the smallest expected loss. Then Corollary 1 of \textcite{Hansen.2011} states that potency approaches 1 for \( P \to \infty \), i.e.\ the best model is included with probability 1 with the out-of-sample size approaching infinity.   
Second, assume that \( \mathcal{M}^{*} \) consists of several best models, i.e.\ two or more models have the same, smallest expected loss. Then potency is asymptotically bounded from below by \( 1 - \alpha \), the level of confidence that the MCS contains the best models. Intuitively, if the number of observations is large enough, the MCS test eliminates all models except those with the smallest expected loss, and then performs a test at level \( \alpha \) under the null.

Power does not have a unique definition (e.g., \textcite{Romano.2005} discuss different notions of power). In general, we would like to know how many of the inferior models are eliminated from the \( \widehat{\mathcal{M}^{*}_{1-\alpha}} \), and which models are eliminated how frequently. We stick to the definition of power that \textcite{Hansen.2011} use in their paper, i.e., the average number of elements in \( \mathcal{ \widehat{M}}_{1-\alpha}^{*} \). A more detailed way of capturing the power property would be: which distance of the expected losses between the best model(s) and an inferior model \( j \) is associated with which rate of rejection of the null of equal predictive ability? Unfortunately, if the losses have different (co-)variances, there is no straightforward way to describe the distance from the null of EPA such that it relates directly to the testing procedure and the elimination frequencies.

Consequently, we report the frequency at which \( \mathcal{M}^{*} \subset \widehat{\mathcal{M}^{*}_{1-\alpha}} \), the average number of elements in \( \mathcal{ \widehat{M}}_{1-\alpha}^{*} \) - by which we follow \textcite{Hansen.2011} - and discuss individual rejection frequencies. The latter indicates how frequently model \( j \) is removed from  \( \mathcal{ \widehat{M}}_{1-\alpha}^{*} \). We argue that this is the most detailed and most informative measure. It helps understand the channels that determine if the testing procedure eliminates a model from the set of candidates. Those channels include the expected losses, variances and correlations of the losses.

\subsubsection{Hypotheses on the predictive ability}\label{framework:hypotheses}
The econometric literature considers two different ways of stating hypotheses about predictive ability. For simplicity, assume that the forecasts stem from parametric models with parameters \( \beta \). In the first case, the hypotheses make a statement about the expected losses at population values \( \beta^{*} = \plim \hat{\beta}_{T} = \lim_{T \to \infty } \hat{\beta} \), as e.g.\ in \textcite{Diebold.1995}, \textcite{West.1996} or \textcite{White.2000}. As \textcite{West.1996} notes, this implies that we use estimated parameters \( \hat{\beta} \) to infer about the population level predictive ability. We write \( \mu(\beta^{*}) \) to denote the expected losses at population values, and \( \mu(\hat{\beta}_{T}) \) to denote the expected loss at the estimated parameters \( \hat{\beta}_{T} \). Likely, it holds that \( \mu(\beta^{*}) \neq \mu(\hat{\beta}_{T}) \), and the ranking of two models in terms of their expected losses could be different for \( \mu(\beta^{*}) \) and \( \mu(\hat{\beta}_{T}) \).  

In the second case, the null is formulated in terms of \( \hat{\beta}_{T} \), i.e. it depends on the sample size and the estimation method, as e.g.\ in \textcite{Giacomini.2006}, \textcite{Clark.2015}. For a more detailed discussion of the two concepts, see e.g.\ \textcite{Clark.2013}.

Our approach differs from the two above in the sense that we do not estimate the parameters of the models that we use in our simulations, but use fixed parameters \( \bar{\beta} \) for the following reasons. Our focus is on evaluating how the losses impact the MCS test, and not how estimation error does. The goal of the simulation is to assess if the test for PA reveals differences in the expected losses, when the losses that we use for testing are sufficiently realistic for VaR and ES forecasts. If the losses that we use in the simulations are realistic is an empirical question which we discuss in Appendix \ref{app:data}. The goal is not to assess if one or another estimation method is superior.
By fixing the parameters of the competing models, we also ensure that the simulated samples correspond to our hypotheses: we use simulated losses with an expected value of \( \mu(\bar{\beta}) \) to test a hypothesis about \( \mu(\bar{\beta}) \).\footnote{Theoretically, we could try to formulate our hypothesis using \( \mu(\hat{\beta}_{T}) \). Yet, the computational cost would be extreme, the results could be unstable and they could depend on parameters such as the starting values of our models.}  

\subsubsection{Related simulation studies}  

\textcite{Nolde.2017} provide simulation results on the short sample properties of statistical loss functions applied to both VaR and joint VaR and Es forecasts in the supplementary material of their paper, and highlight the high sampling variability. They attribute the poor small sample properties mainly to cases when the data in the estimation window is not representative of the data in the evaluation period. Their simulation design is different from ours in the sense that they compare the optimal forecast to forecasting methods that use estimated parameters, which is more favorable to the true model.

\textcite{Deng.2021} examine the finite sample properties of the one-tailed DM test using different parametrizations of loss functions for joint VaR and ES forecasts, finding that the test seldom indicates an inferior model as superior, and that the finite sample properties improve when the quantile increases from \( p=0.025 \) to \( p=0.05 \), as well as when the evaluation period increases. 

\textcite{Aparicio.2018} employ the MCS procedure to compare different trading strategies and use excess returns as loss functions. They inquire if the MCS procedure reveals the best model within a year of trading as a function of Sharpe ratios. Yet, they find that the associated Sharpe ratios are unrealistically large. They also find that the models selected by the MCS show significantly worse out-of-sample performance than in-sample performance and thus suffer from backtest overfitting. Yet, as the authors note, they assume \textit{independent} strategies though correlations likely occur in practice, and although larger correlations lead to an easier distinction between models given the variances are held constant.

\section{Simulation}\label{sec:simulation}

In this section, we conduct a simulation study to infer if, and under which conditions, the DM test and MCS testing procedure reveal differences in the predictive ability between competing VaR and ES approaches based on statistical loss functions. 

We consider two scenarios that we examine in the subsections below, the first one is a static one, where the returns follow a  Student's t distribution, which implies iid loss differentials. The second scenario is dynamic; the returns display time varying volatility, which induces heterogeneity and temporal dependence of the loss differentials. 
In the static case, we perform a pairwise comparison using the DM test, while we examine both the DM test and the MCS testing procedure in the dynamic scenario. 
We evaluate one-day-ahead daily forecasts for VaR and ES at the quantile levels \( p \in \{0.01, 0.025, 0.05, 0.1 \} \) and focus on the out-of-samples sizes \( P \in \{ 251, 500, 1000, 2500\} \), which correspond to one year, two, four and ten years of daily observations, respectively. While the 1\% VaR and 2.5\% ES correspond to to the quantile levels that the Bank for International Settlements requires (see \textcites{Basel.2013}{Basel.2014}, respectively), and the 5\% level is popular in applied research (see, e.g., \textcites{Bernardi.2017}{Taylor.2019}), the evaluation at other quantile levels offers insights how the finite sample properties of the tests depend on \( p \). We perform the two-tailed DM test at the \(\alpha=0.05 \) level. 

\subsection{Static scenario}\label{sec:static_scenario}

First, we perform 50,000 simulations to examine the finite sample properties of the DM test when we simulate returns from a standardized student-t distribution with \( \nu=4 \) degrees of freedom, which is fat tailed and a popular choice to model financial returns in more complicated setting, see e.g., \textcite{Nolde.2017}. Understanding the static scenario with independent and identically distributed (iid) losses will help understand what we can expect in more complicated setting, e.g., with time varying volatility as captured by the next DGP. We compare the true forecasts of VaR and ES to (i) forecasts from the standard normal distribution, (ii) forecasts that over- and underreport risk, i.e., by submitting a forecast at a different quantile from the true distribution. We can think of the latter case as a bank that underreports risk in order to maximize its profits. From a regulatory perspective, it is then interesting to assess under which conditions such fraudulent forecasts can be detected using statistical losses. 
To put the distributional misspecification (i) into an economic perspective, we consider the 1\% VaR of a USD 1,000,000 portfolio when the log returns scaled by 100 are assumed to be either standardized Student's t or standard normally distributed. While the forecast based on the Student's t predicts a VaR of USD -26,147, the normal distribution predicts only USD -22,995.

The Tables \ref{tab:static_distributional_0_0} and \ref{tab:static_distributional_1_2} present the results for the comparison (i) for standalone VaR forecasts and joint VaR and ES forecasts, respectively 
Table \ref{tab:static_distributional_0_0} shows that at the most extreme quantile level \( p=0.01 \), the test has basically no power against this distributional misspecification for both 251 and 500 observations, for 1000 it is slightly above \( 0.05 \), which is only the nominal size of the test, while for 2500 observations it reaches 15\%. The power increases with \( p \) and \( P \), for instance, the power for \(p=0.1 \) is 42\% for an out-of-sample size \(P=1000\), while it reaches 77\% for 2500 observations.\footnote{We omit the VaR results for \(p=0.025 \) as the quantile of the normal and Student's t distribution are very close such that the calculation of \( t_{ij} \) is not reliable.}
For the joint evaluation of VaR and ES using the FZG score (Table \ref{tab:static_distributional_1_2}), the results are similarly poor for \( p=0.01 \) and \(P \in \{251, 500\} \). For \(P=1000\), the additional information about the shape of the tails improves the power slightly, while for \( P=2500\), the power is much larger at 36\% than the 15\% in the VaR case. For the larger quantile levels, testing the joint loss shows more power starting from \(P=251 \). We do not find substantial differences between the various parametrizations of the loss functions, for which we report results in Apendix \ref{app:static_case}.
For both the GPL loss and the joint loss functions, we find that the type III error is larger than the power for \( p \in \{0.01, 0.025 \} \) and the short out-of-sample sizes \( P \in \{251, 500 \} \), which we further discuss below.

\begin{table}[h!]
    \scriptsize
    \centering
    \caption{Static case, distributional misspecification: tick loss}    
    \label{tab:static_distributional_0_0}
    \begin{tabularx}{1\textwidth}{l X X X X X X X X }
        \toprule
                &\multicolumn{4}{c}{Power} & \multicolumn{4}{c}{Type III error}   \\
        \toprule
                \(p \backslash P \) & 251 & 500 & 1000 & 2500 & 251 & 500 & 1000 & 2500  \\
        \toprule
        0.01   &  0.009 &  0.023 &  0.054 &  0.152 &  0.038 &  0.038 &  0.014 &  0.003   \\
        0.05   &  0.113 &  0.118 &  0.147 &  0.257 &  0.003 &  0.003 &  0.002 &  0.000   \\
        0.1    &  0.182 &  0.260 &  0.415 &  0.768 &  0.001 &  0.000 &  0.000 &  0.000   \\
        \bottomrule
    \end{tabularx}
    \raggedright
    \scriptsize{ \textit{Notes:} This table displays the finite sample power and type III error of the DM test for one-step-ahead forecasts evaluated using the tick loss.}
\end{table}

\begin{table}[h!]
    \scriptsize
    \centering
    \caption{Static case, distributional misspecification: FZG score}
    \label{tab:static_distributional_1_2}
    \begin{tabularx}{1\textwidth}{l X X X X X X X X }
    \toprule
            &\multicolumn{4}{c}{Power} & \multicolumn{4}{c}{Type III error}   \\
    \toprule
    \(p \backslash P \) & 251 & 500 & 1000 & 2500 & 251 & 500 & 1000 & 2500  \\
    \toprule
    0.01    &  0.005 &  0.022 &  0.083 &  0.361 &  0.086 &  0.043 &  0.011 &  0.001   \\
    0.025   &  0.004 &  0.014 &  0.057 &  0.230 &  0.104 &  0.043 &  0.016 &  0.002   \\
    0.05    &  0.139 &  0.212 &  0.383 &  0.775 &  0.001 &  0.000 &  0.000 &  0.000   \\
    0.1     &  0.186 &  0.271 &  0.439 &  0.788 &  0.001 &  0.000 &  0.000 &  0.000   \\
    \bottomrule
    \end{tabularx}\\
    \raggedright
    \scriptsize{ \textit{Notes:} This table displays the finite sample power and type III error of the DM test for one-step-ahead forecasts evaluated using the FZG score.}
\end{table}

Additionally, we examine the distribution of the test statistics that are used in this DM test. Figure \ref{fig:static_distribution_of_t_ij} below presents the density plots of the t-statistics for different values of the quantile \( p \) and the out-of-sample window \( P \), with an fitted Student's t distribution probability density function (pdf), in the case of the tick loss. We do not find remarkable differences between the GPL loss and the joint loss functions, neither between the different parametrizations of the loss functions, and provide additional plots in Appendix \ref{app:static_case}. 
For small values of \( p \) and \(P \), their distributions are heavily skewed and the fitted Student's t distribution has very small degrees of freedom, i.e., test that rely upon the approximate normality of the test statistic may display a finite sample behaviour that is very different from their theoretical properties. 
\begin{figure}[!htb]
    \includegraphics[width=1\linewidth]{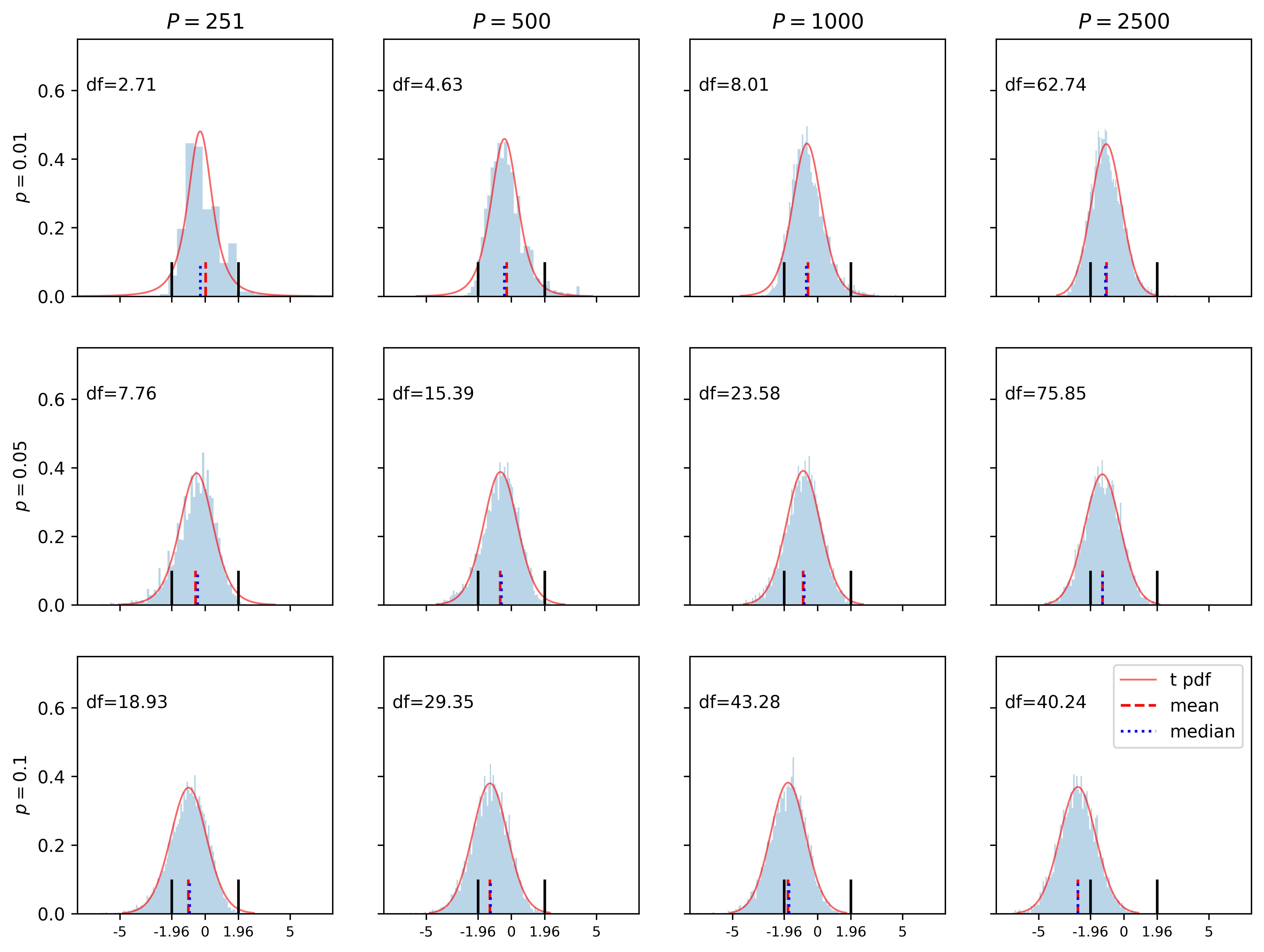}
    \caption{Density histogram of \(t_{12}\) for the tick loss}
    \label{fig:static_distribution_of_t_ij}
    \footnotesize{\textit{This figure displays density histograms of \(t_{ij} \) for one-step-ahead forecasts evaluated using the tick loss, where model \(i\) is the true Student's t model while model \(j\) is a misspecified model that assumes a normal distribution.}}
\end{figure}

Furthermore, Table \ref{tab:static_wrong_quantile_0_0} presents the results for comparison (ii) at the quantile levels \( p=0.01 \) and \( p=0.1\) using the tick loss. We find that too conservative forecast are much easier to detect, while there is very little power against fraudulent forecasters that submit quantiles for \(1-2.5\) \% higher levels, which holds across all levels that we consider, i.e., for \( p \in \{0.01, 0.025, 0.05, 0.1\} \). For instance, if \( p=0.01 \), the DM test has less than 5\% power against the fraudulent 2.5\% quantile forecast for the regulatory period of 251 observations, i.e., one year, which increases to 40\% in the case of 1000 out-of-sample observations. We provide results for other configurations in Appendix \ref{app:static_case}, which show the same pattern regarding the over-and underreporting of risk.

\begin{table}[h!]
    \scriptsize
    \centering
    \caption{Static case, wrong quantile: tick loss}
    \label{tab:static_wrong_quantile_0_0}
    \begin{tabularx}{0.75\textwidth}{l X X X X X }
        &\(p^{*} \backslash P \) & 251 & 500 & 1000 & 2500   \\ 
    \toprule 
    \multirow{ 7}{*}{\( p=0.01\)}   &  0.005 &  0.144 &  0.240 &  0.283 &  0.451   \\ 
            &  0.015 &  0.010 &  0.023 &  0.055 &  0.146   \\ 
            &  0.020 &  0.025 &  0.063 &  0.183 &  0.517   \\ 
            &  0.025 &  0.051 &  0.157 &  0.374 &  0.811   \\ 
            &  0.050 &  0.407 &  0.805 &  0.986 &  1.000   \\ 
            &  0.100 &  0.969 &  1.000 &  1.000 &  1.000   \\ 
            &  0.150 &  1.000 &  1.000 &  1.000 &  1.000   \\ 
    \bottomrule
    \multirow{ 7}{*}{\( p=0.1\)}   &  0.050 &  0.447 &  0.694 &  0.924 &  0.999   \\
        &  0.090 &  0.074 &  0.076 &  0.102 &  0.167   \\
        &  0.110 &  0.029 &  0.043 &  0.061 &  0.107   \\
        &  0.125 &  0.065 &  0.115 &  0.212 &  0.481   \\
        &  0.150 &  0.187 &  0.357 &  0.668 &  0.968   \\
        &  0.200 &  0.612 &  0.903 &  0.996 &  1.000   \\
        &  0.250 &  0.918 &  0.998 &  1.000 &  1.000   \\
    \bottomrule
    \end{tabularx}\\
    \raggedright
    \scriptsize{ \textit{Notes:} This table displays the finite sample power of the DM test against a forecast based on a wrong quantile for one-step-ahead forecasts evaluated using the tick loss. \(p^{*} \) denotes the wrong quantile of the same distribution as the true quantile \(p\).}
\end{table}

To illustrate how the tick loss impacts the test against distributional misspecification and against the underreporting of risk, Example \ref{example:tick_loss} demonstrates how the asymmetric loss functions affects finite sample estimates of the mean at the extreme quantile level \(p=0.01\).\footnote{The example applies analogously to the DGP in \ref{sec:dynamic_scenario} when comparing two models that have the same volatility specification but assume that the innovations follow a Student's t and normal distribution, respectively, such as the comparison between TGARCH-n and TGARCH-t.} 

\begin{example}\label{example:tick_loss} 
    Consider the tick loss defined in Equation \ref{eq:gpl_loss} at quantile level \( p=0.01 \). We compare the true, superior standardized Student's t (model \( i  \)) and the misspecified competitor (model \( j \)) that assumes standard normally distributed innovations \( z_{t} \). Let \( x_{i} \) and \( x_{j} \) denote the VaR forecasts that stem from models \( i \) and \( j \), respectively, while \( y \) denotes the daily financial return. We have \( x_{i} < x_{j} \) for \( \nu \in \{3, 4, 7, 12\} \) and thus \( d_{ij} | x_{j} < y = p \times \left(  (y - x_{i}) - (y - x_{j}) \right)= p \times (x_{j} - x_{i} ) > 0 \). Each time the financial return is above the VaR forecast from the normal distributions, the associated loss is smaller than the loss of the more conservative Student's t. Consequently, we rely on sufficiently many returns that are smaller than the VaR forecast from the normal distribution to correctly rank the two models. Yet, such observations are few at level \( p = 0.01 \) and for sample sizes \( P \leq 500 \). 
\end{example}

Moreover, we calculate expressions for the loss difference and its expectation in the case of the homogeneous parametric GPL loss (Equation \ref{eq:hom_gpl_loss}) in Appendix \ref{sec:loss_difference_iid} that characterize how the loss differences in our setting depend on \(p \). For small values of \( p \), usually only a tiny fraction of the loss differences correctly indicate the ranking of the two models, which explains the type III errors in small samples. 

Overall, even in the static scenario with iid loss differentials, we find that the estimates of the mean and the variance of \(\bar{d}_{12} \) are poor for small values of \(p \) and \(P \), and that the DM test has little power against too optimistic risk forecast in these settings.

\FloatBarrier

\subsection{Dynamic scenario}\label{sec:dynamic_scenario}

Next, we examine the finite sample properties of the DM test and the MCS in a time series setting with a more realistic DGP that captures the stylized facts of financial return data such as fat tails, time varying volatility, clustering and the leverage effect.

We focus on generalized autoregressive conditional heteroskedasticity (GARCH)-type models (\textcite{Bollerslev.1986}) due to their popularity to model financial returns (e.g., \textcites{Nolde.2017}{Li.2018}{Patton.2020}).  

As DGP, we choose a process with zero conditional mean but with time varying variance that follows a TGARCH(1,1) (or GJR-GARCH) with standardized Student's t distributed innovations:
\begin{align} 
    r_{t} &= z_{t} \sqrt{\sigma_t^2} \\ 
    \sigma_t^2 &= \omega + \alpha \varepsilon_{t-1}^2 + \gamma \mathbbm{1}_{\{\varepsilon_{t} < 0\}} \varepsilon^{2}_{t-1}
        +  \beta \sigma_{t-1}^2,\\ \label{eq:dgp}
        z_{t} & \sim sst_{\nu},
\end{align}
where \( \nu \) denotes the degrees of freedom, and \( sst_{\nu} \) denotes the scaled Student's t distribution with unit variance.\footnote{If \( T \sim t_{\nu}, \: \nu > 2 \) then we write \( Z \sim sst_{\nu} \) for the random variable \( Z = \sqrt{\dfrac{\nu - 2}{\nu}} T \) with unit variance.} 

\( \sigma^{2}=\dfrac{1}{1-\alpha - 0.5\gamma - \beta} \) denotes the unconditional variance of the returns.

We consider DGPs with different degrees of freedom \( \nu \in \{3, 4, 7, 12\} \). The motivation is that each DGP corresponds to a different state of the market, i.e.,  highly volatile, volatile and calm, respectively. We choose the model parameters close to parameters that we frequently observe on daily log-returns from closing prices adjusted for dividend and split. Specifically, we estimate the TGARCH on equally weighted indices composed of 30 randomly chosen stocks from the Dow Jones U.S. Small, Mid, and Large cap-indices.\footnote{We describe the data in more detail in Appendix \ref{app:data}.} The DGPs use the following parameters: 
\begin{table}[h!]
    \scriptsize
    \centering
    \caption{Parameters DGPs}
    \label{tab:dgp_parameters}
    \begin{tabularx}{0.75\textwidth}{X X X X X }
    \( \omega\) & \( \alpha\) & \( \gamma \) & \( \beta \) & \( \nu \)   \\ 
    \toprule 
    0.03 &  0.04 &  0.1 &  0.9 &  \( \{3, 4, 7, 12\} \)   \\ 
    \end{tabularx} \\
    \vspace{1ex}
    \raggedright
    \scriptsize{ \textit{Notes:} This table displays the parameters of the DGPs (TGARCH-t, Equation \ref{eq:dgp}) for the simulations that relate to the results in Section \ref{sec:dynamic_scenario}.}
\end{table}

\subsubsection{Pairwise comparison}\label{sec:pairwise_comparison}
We examine the finite sample properties of the DM test for a DGP with \( \nu=4 \) degrees of freedom, i.e., this simulation design is analogous to the one in the static case. We perform a pairwise comparison of the true TGARCH-t model vs (i) a TGARCH-n that assumes normally distributed innovations, and (ii) a GARCH-t that misspecifies the volatility. Compared to the static scenario in subsection \ref{sec:static_scenario}, the time varying volatility scales the loss differences and may introduce temporal dependence in setting (ii). We calculate expressions for the loss difference and its expectation conditional on  the filtration \( \mathcal{F}_{t-1}\) in the case of the homogeneous parametric GPL loss in Appendix \ref{sec:loss_difference_garch_case}. These calculations demonstrate the additional variability that the time varying volatility induces in finite samples. As in the MCS procedure, we bootstrap the variance of \( \bar{d}_{ij} \), and use \( B=2,500 \) bootstrap resamples, but perform \( 10,000 \) simulations. Below, we briefly discuss the results for the tick loss and FZG score, and defer result for the other parametrizations to Appendix \ref{app:dynamic_case}.

The Tables \ref{tab:dynamic_distributional_0_1} and \ref{tab:dynamic_distributional_1_2} present the results for comparison (i), while the Tables \ref{tab:dynamic_volatility_0_1} and \ref{tab:dynamic_volatility_1_2} provide the figures for (ii). The patterns are generally similar to the static scenario, though the power is usually slightly lower, while the type III errors are larger in the specifications with \(p \in \{0.01, 0.025\} \) and \( P \in \{251, 500\} \). The power against volatility is very low (<10\%) for 
\(p \in \{0.01, 0.025\} \) and \( P \in \{251, 500\} \) but much larger for the larger quantiles and out-of-sample sizes. Again, for out-of-sample sizes of \( P \geq 1,000 \), the joint forecast evaluation shows more power than evaluating standalone VaR forecasts. 

\begin{table}[h!]
    \scriptsize
    \centering
    \caption{Dynamic case, distributional misspecification: tick loss}
    \label{tab:dynamic_distributional_0_1}
    \begin{tabularx}{1\textwidth}{l X X X X X X X X }
        \toprule 
                &\multicolumn{4}{c}{Power} & \multicolumn{4}{c}{Type III error}   \\ 
        \toprule 
        \(p \backslash P \) & 251 & 500 & 1000 & 2500 & 251 & 500 & 1000 & 2500  \\
        \toprule 
        0.01   &  0.005 &  0.010 &  0.026 &  0.086 &  0.101 &  0.055 &  0.031 &  0.008   \\ 
        0.05   &  0.170 &  0.180 &  0.215 &  0.295 &  0.002 &  0.001 &  0.001 &  0.000   \\ 
        0.1    &  0.269 &  0.334 &  0.476 &  0.727 &  0.001 &  0.000 &  0.000 &  0.000   \\ 
        \bottomrule 
        \end{tabularx}
        \raggedright
        \scriptsize{ \textit{Notes:} This table displays the finite sample power and type III error of the DM test at the \(\alpha=0.05 \) level of the test for one-step-ahead forecasts evaluated using the tick loss}
        \end{table}

\begin{table}[h!]
    \scriptsize
    \centering
    \caption{Dynamic case, distributional misspecification: FZG score}
    \label{tab:dynamic_distributional_1_2}
    \begin{tabularx}{1\textwidth}{l X X X X X X X X }
        \toprule 
                &\multicolumn{4}{c}{Power} & \multicolumn{4}{c}{Type III error}   \\ 
        \toprule 
                \(p \backslash P \) & 251 & 500 & 1000 & 2500 & 251 & 500 & 1000 & 2500  \\
        
        \toprule 
        0.01    &  0.011 &  0.029 &  0.091 &  0.343 &  0.105 &  0.043 &  0.012 &  0.001   \\ 
        0.025   &  0.010 &  0.024 &  0.068 &  0.246 &  0.105 &  0.047 &  0.015 &  0.002   \\ 
        0.05    &  0.174 &  0.222 &  0.382 &  0.737 &  0.003 &  0.001 &  0.000 &  0.000   \\ 
        0.1     &  0.225 &  0.296 &  0.471 &  0.789 &  0.002 &  0.000 &  0.000 &  0.000   \\ 
        \bottomrule 
        \end{tabularx}
        \raggedright
        \scriptsize{ \textit{Notes:} This table displays the finite sample power and type III error of the DM test for one-step-ahead forecasts evaluated using the FZG score.}
\end{table}

\begin{table}[h!]
    \scriptsize
    \centering
    \caption{Dynamic case, volatility misspecification: tick loss}
    \label{tab:dynamic_volatility_0_1}
    \begin{tabularx}{1\textwidth}{l X X X X X X X X }
        \toprule 
                &\multicolumn{4}{c}{Power} & \multicolumn{4}{c}{Type III error}   \\ 
        \toprule 
        \(p \backslash P \) & 251 & 500 & 1000 & 2500 & 251 & 500 & 1000 & 2500  \\
        \toprule 
        0.01    &  0.007 &  0.013 &  0.028 &  0.083 &  0.099 &  0.058 &  0.029 &  0.007   \\ 
        0.025   &  0.004 &  0.006 &  0.006 &  0.008 &  0.116 &  0.096 &  0.077 &  0.058   \\ 
        0.05    &  0.178 &  0.180 &  0.205 &  0.306 &  0.002 &  0.001 &  0.001 &  0.000   \\ 
        0.1     &  0.265 &  0.351 &  0.472 &  0.736 &  0.001 &  0.000 &  0.000 &  0.000   \\ 
        \bottomrule 
        \end{tabularx}
    \raggedright\\
    \scriptsize{ \textit{Notes:} This table displays the finite sample power and type III error of the DM test at the \(\alpha=0.05 \) level of the test for one-step-ahead forecasts evaluated using the tick loss}
\end{table}

\begin{table}[h!]
    \scriptsize
    \centering
    \caption{Dynamic case, volatility misspecification: FZG score}
    \label{tab:dynamic_volatility_1_2}
    \begin{tabularx}{1\textwidth}{l X X X X X X X X }
    \toprule 
            &\multicolumn{4}{c}{Power} & \multicolumn{4}{c}{Type III error}   \\ 
    \toprule 
            \(p \backslash P \) & 251 & 500 & 1000 & 2500 & 251 & 500 & 1000 & 2500  \\

    \toprule 
    0.01    &  0.009 &  0.034 &  0.095 &  0.343 &  0.104 &  0.041 &  0.013 &  0.001   \\ 
    0.025   &  0.009 &  0.028 &  0.070 &  0.240 &  0.108 &  0.046 &  0.018 &  0.002   \\ 
    0.05    &  0.166 &  0.237 &  0.383 &  0.745 &  0.003 &  0.000 &  0.000 &  0.000   \\ 
    0.1     &  0.227 &  0.309 &  0.461 &  0.790 &  0.001 &  0.000 &  0.000 &  0.000   \\ 
    \bottomrule 
    \end{tabularx}
    \raggedright
    \scriptsize{ \textit{Notes:} This table displays the finite sample power and type III error of the DM test for one-step-ahead forecasts evaluated using the FZG score.}
\end{table}

Moreover, we examine the distributional properties of the \( t_{ij} \) for (i) and (ii), which are more extreme than in the static case, but show the same pattern. We thus provide the plots in Appendix \ref{app:dynamic_case}. In addition to using the bootstrapped variance, we plot the distribution of \(t_{ij} \) when we standardize \( \bar{d}_{ij} \) by the simulated variance of \( \bar{d}_{ij} \), which we consider correct. Comparing the density histograms of the differently standardized \(\bar{d}_{ij} \) confirms the insights from the calculations in Appendix \ref{sec:loss_difference_garch_case}: for small values of \(p \), short samples that estimate the wrong sign of \( \mathbb{E}[d_{t, ij}] \) often underestimate the \( var(\bar{d_{ij}})\), which leads to a skewed distribution of the test statistic and non-negligible type III errors. We take this as evidence that the bootstrap may not be reliable in these settings as many short samples are not representative of the DGP.

To summarize, the examination of the pairwise comparison problems in Sections \ref{sec:static_scenario} and \ref{sec:pairwise_comparison} illustrates how these comparison problems differ from standard ones for the extreme quantile \( p \) and short out-of-sample size \(P \) scenarios, where we find little power and heavily skewed test statistics, as well as non-negligible type III errors. 
These findings motivate us to assess if the finite sample properties of the MCS testing procedure are in line with its theoretical ones.

\FloatBarrier

\subsubsection{Model Confidence Set simulations}\label{sec:simulations_mcs}
This section provides Monte Carlo evidence on the finite sample properties of the MCS testing procedure. We discuss potency, power and individual rejection frequencies as a function of the quantile \( p \), the out-of-sample size \( P \), the number of models \( m \), and the market conditions captured by the degrees of freedom \( \nu \) of the DGP. In the main part, we restrict the discussion part to the level of the test \( \alpha=0.25 \). As the pairwise DM test shows little power at the conventional level of the test \( \alpha=0.05 \), we need to consider larger levels \( \alpha\) such that the multiple comparison procedure has power.

For the shorter out-of-sample sizes \(P \leq 500 \) and the smaller quantile levels \( p \in \{0.01, 0.025 \} \), we do not expect that the MCS test shows much power. Yet, we include sample sizes as short as \( P=61\) to assess if these are long enough to use the MCS for model averaging, as, e.g., done by \textcite{Bernardi.2017}. Thus, we are primarily interested in the potency and if the few rejections that occur correctly indicate models that have larger expected losses. 

We set the number of bootstrap resamples in the MCS to \( B=5,000 \) and perform \(2,500 \) Monte Carlo simulations for each DGP, where we consider \( \nu \in \{3, 7, 12\} \). 

We do not find strong differences between different parametrizations of the loss functions that we define in Equation \ref{eq:gpl_loss} and Equation \ref{eq:joint_loss}. In this section, we thus discuss results for the tick loss function (Equation \ref{eq:tick_loss}) when evaluating VaR forecasts, and for the FZG score (cf. Table \ref{tab:joint_loss}) when evaluating joint VaR and ES forecast. Appendix \ref{app:detailed_results} presents the results for alternative specifications such as the test statistic \( T_{R, \mathcal{M}}\), different levels of the test \( \alpha \) and different parametrizations of the loss functions.

We consider two sets of models, the first one contains \( m=5 \) competing models, while the second one consists of  \( m=10 \) models.
We characterize the first set of \( m=5 \) models as follows. The first model has a TGARCH conditional variance specification and \( sst_{\nu} \)-distributed innovations. We refer to this model as the `true' or `correctly specified' model. Additionally, we consider the following four misspecified models that do not correctly model the return process. First, a TGARCH(1,1) with \( z_{t} \sim \mathcal{N}(0,1)\) (TGARCH-n). Second, a GARCH(1,1) with \( z_{t} \sim sst_{\nu} \) (GARCH-t). Third, a GARCH(1,1) with \( z_{t} \sim \mathcal{N}(0,1) \). Fourth, a constant variance model with \( z_{t} \sim st_{\nu}(0,\sigma^{2}) \) (constant variance model).

The parameters of all models are set such that all models have the same unconditional variance \( \sigma^{2} \). For this set of models, the misspecification is interpretable: the misspecified competitors do not capture the fat tails, the leverage effect, or the time varying volatililty.

The second set of \( m=10\) competing forecasts is constructed by adding five models to the first set. These models are three Risk Metrics\( ^{TM} \) models with \( z \sim sst_{\nu^{\prime}} \), \( \nu^{\prime} \in \{3, 7, 12\} \), and two additional GARCH(1,1) with \( z \sim sst_{\nu^{\prime}}\), \( \nu^{\prime} \in \{3, 7, 12\} \setminus \{ \nu \}  \). Depending on the DGP, we add models that show both large and small expected losses relative to the initial models. 
We examine the larger set of models for the following reasons: 
(1) including additional models may lead to fewer eliminations of inferior models as follows. Let \( \mathcal{M}^{0}_{small} \) denote the initial set of competing models and \( \mathcal{M}^{0}_{large} \supset \mathcal{M}^{0}_{small} \) the larger set of models. The control of the familywise error rate is achieved through adjusted p-values that monotonically increase with each testing step. Consequently, eliminations become less likely with each testing step. If we add models that have comparatively large expected losses to \( \mathcal{M}^{0}_{small} \), we may thus observe that the MCS test eliminates inferior models from \( \mathcal{M}^{0}_{large} \) less frequently than it eliminates them from \( \mathcal{M}^{0}_{small} \).
(2) including models with comparatively small expected losses may facilitate identifying the worst models. Assume that model \( i=1 \) is inferior to models \( j \in \{ 2, \ldots, m \}  \). If we increase \( m \), we expect \( var (D_{i \cdot}) \) to decrease (cf. Equation \ref{eq:variance_formula} in Appendix \ref{app:variance_formula}) and \(\bar{d}_{i \cdot} \) to become smaller. As model \( i \) is inferior, including more alternatives with smaller expected losses than model \( i \) leads to a larger t-statistic \( t_{i \cdot } = \dfrac{\bar{d}_{i \cdot}}{\sqrt{\hat{var}(\bar{d}_{i \cdot})}} \) and we thus expect that the MCS test eliminates model \( i \) more frequently when we increase \( m \).

We find correlations between 0.95 and 0.99 for the GARCH-type models, and correlations between 0.8 and 0.95 between the constant variance model and the GARCH-type models, which are in line with the correlations among the GARCH-type models that we estimate on the Dow Jones Indices data (see Appendix \ref{app:data}).


\subsubsection*{Results for the first set of competing  models: \( m = 5 \).} \label{results_first_set}
Figure \ref{fig:true_model_var} below visualizes the VaR results. It shows potency in the upper row (Panels A, B, C) and power in the lower row (Panels D, E, F). The columns refer to the three choices of DGPs with different degrees of freedom \( \nu \in \{3, 7, 12 \} \).

\begin{figure}[!ht]
    \includegraphics[width=1\linewidth]{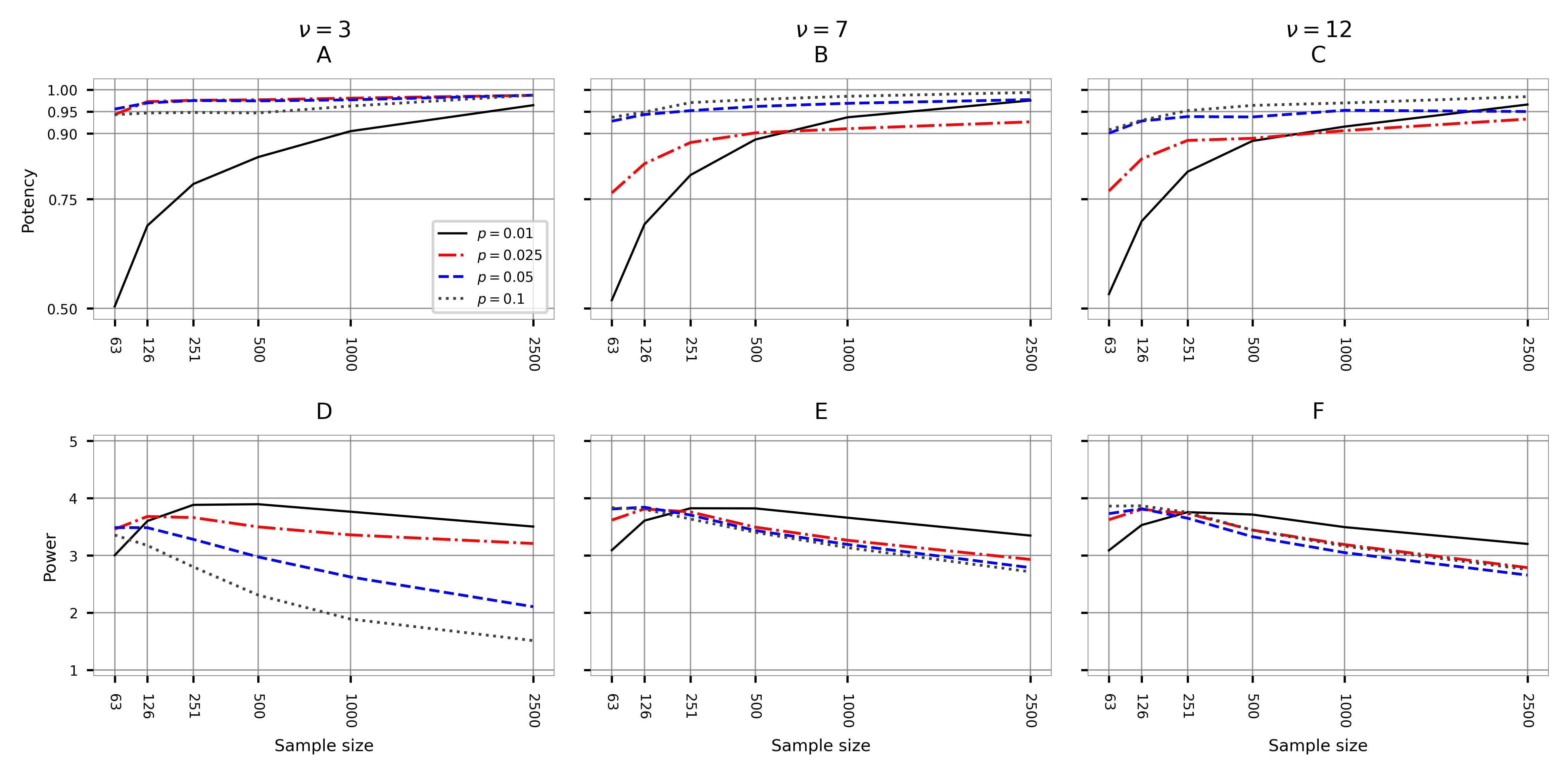}
    \caption{Finite sample properties for the tick loss}
    \label{fig:true_model_var}
    \footnotesize{This figure displays the finite sample properties of the MCS procedure in the following setting: one-day-ahead VaR forecasts evaluated using the tick loss, true model included, number of models \(m=5\), level of the test \( \alpha=0.25 \). The upper row displays the potency, i.e.\  the frequency of \( \mathcal{M}^{*} \subset \widehat{\mathcal{M}}_{1-\alpha}^{*} \). The lower row displays the power property, i.e.\  the average number of elements in \(\mathcal{ \widehat{M}}_{1-\alpha}^{*} \).}
\end{figure}

Panels A, B, C reveal that the MCS usually keeps the best model more often than the specified confidence level of 75\% (1-\( \alpha \)) even for the smallest out-of-sample size \( P=63 \). Potency then increases with the quantile level \( p \) and with the out-of-sample size \( P \). Its values are much lower for the extreme quantile level \( p=0.01 \) (solid dark grey line) and out-of-sample sizes \( P \leq 500 \) than for the other quantiles \(p \in \{0.025, 0.05, 0.1 \} \) (red dash-dotted line, blue dashed line and grey dotted line, respectively). 
At the most extreme quantile level \( p=0.01 \) (solid dark grey line), potency is below the specific confidence level \( 1-\alpha \) for shorter out-of-sample sizes \( P \leq 126 \). Starting from out of-sample sizes \( P \geq 251\), i.e.\ one year of daily data, the MCS test keeps the best model in around 80\% of our simulations. For out-of-sample size \( P=1000\), potency increases to 90\% for the DGPs with \( \nu  \in  \{3, 12\} \) degrees of freedom (Panels A, C), and to 95\% for the DGP with \( \nu=7 \) degrees of freedom (Panel B). 

At the quantile level \( p = 0.025  \), potency  but still above the confidence level \( 1-\alpha\): it increases from around 75\% to around 85\% from 63 observations to 500 observations for the DGPs with \( \nu \in \{7, 12\} \), while it is above 90\% for alle out-of-sample sizes for volatile markets (Panel A). 
At the larger quantile levels \( p \in \{0.05, 0.1 \} \), the MCS keeps the best model more often than 90\% for all out-of-sample sizes \( P \).  

To summarize, our findings suggest that - in comparable settings - potency of the MCS test at level \( \alpha=0.25 \) exceeds the confidence level of 75\% for half a year of daily data when we consider less extreme quantile levels \( p \geq 0.025 \), while the extreme quantile level \( p=0.01 \) requires one year or more daily data.

Panels D, E and F in Figure \ref{fig:true_model_var} display the power as the average number of models in the MCS. In general, the power increases with the quantile level \( p \) and out-of-sample size \( P \). 

At the most extreme quantile level \( p=0.01 \), we find the same patterns for all three DGPs, i.e.\ across the degrees of freedom \( \nu \): power decreases from 63 out-of sample observations to 251 out-of-sample observations and then increases with the out-of-sample size \( P \). Simultaneously, we observe a steep rise in potency (Panels A, B, C) for out-of-sample sizes \( P \leq 251 \), and thus suspect that the MCS test is not reliable for this quantile level and out-of-sample sizes \( P < 251 \), which is in line with the insights from subsection \ref{sec:pairwise_comparison} above.

For the DGP with \( \nu=3 \) degrees of freedom (Panel A), we observe the most disperse measures of power across different quantile levels. Starting at 126 observations, i.e.\  half a year of data, power increases monotonically with the sample size for the quantile levels \( p \in \{0.05, 0.1\} \). While for a sample size of 126, the MCS test keeps between 3.6 and 3.2 models for VaR forecasts at quantile level \( p=0.025\) and \(p=0.1\), respectively, it only keeps 3.5 and 2.2 models for a sample size of 500 observations, respectively. 

In the settings that relate to calmer market conditions, i.e.\  when \( \nu \in \{7, 12\} \), and for quantile levels \( p \in \{0.05, 0.1\} \), power displays a homogeneous pattern: the number of models that the MCS test eliminates increases slightly with the out-of-sample size \( P \). It eliminates 1 model for an out-of-sample size of \( P=126 \), and between 1 and 1.6 models for a sample size of \( P=500 \). 

To summarize, we conjecture that, at the level of the test \(\alpha=0.25\), one may use the MCS test to trim off some worst VaR models while ensuring good potency for small values of \(p\) for an out-of-sample period of about one year.


Figure \ref{fig:true_model_es} presents potency and power of the MCS when we use the FZG score to evaluate joint forecasts of VaR and ES - analogously to Figure \ref{fig:true_model_var} for VaR forecasts.

\begin{figure}[!ht]
    \includegraphics[width=1\linewidth]{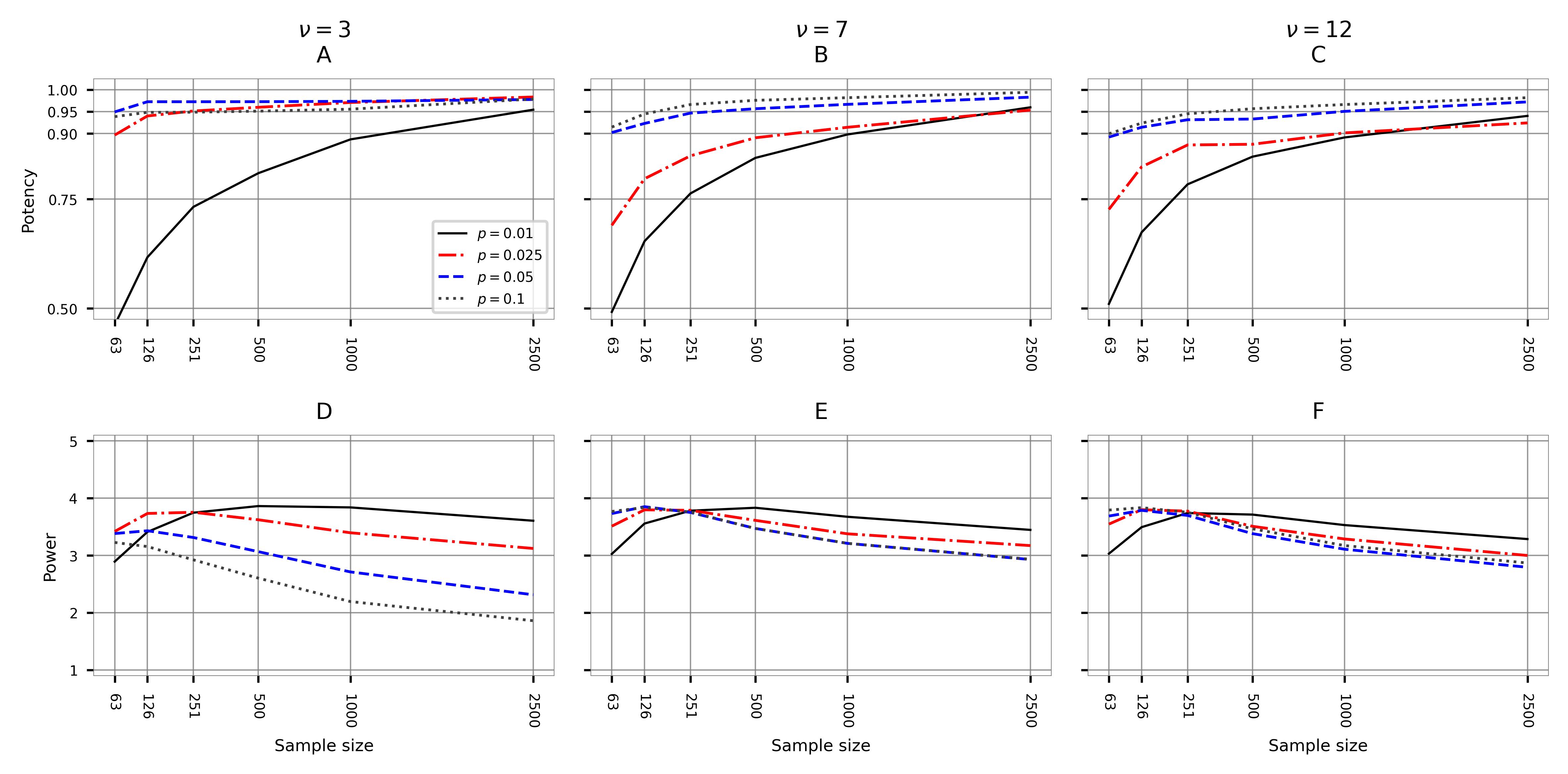}
    \caption{Finite sample properties for the FZG score}
    \label{fig:true_model_es}
    \footnotesize{This figure displays the finite sample properties of the MCS procedure in the following setting: one-day-ahead VaR and ES forecasts evaluated using the FZG score, true model included, number of models \(m=5\), level of the test \( \alpha=0.25 \). The upper panel displays the potency, i.e.\  the frequency of \( \mathcal{M}^{*} \subset \widehat{\mathcal{M}}_{1-\alpha}^{*} \). The lower panel displays the power property, i.e.\  the average number of elements in \(\mathcal{ \widehat{M}}_{1-\alpha}^{*} \).}
\end{figure}

The potency at quantile levels  \(p \in \{0.05, 0.1 \} \) is below the confidence level of 75\% for 63 out-of-sample observations, while at the quantile level \( p=0.01 \) it is even below 75\% for an out-of-sample size of 126 observations. For out-of-sample sizes \( P \leq 500 \), the MCS test consistently keeps the best model more frequently  at the quantile level \( p=0.025 \) than at the quantile level \(p=0.01\). This is analogous to the pattern which we observe for the tick loss in Figure \ref{fig:true_model_var} (Panels A, B, C).

For quantile levels \(p \in \{0.05, 0.1 \} \), potency is well above the confidence level of 75\%. Panel A, which corresponds to the DGP with \( \nu=3 \) degrees of freedom, e.g.\ shows that its values across all out-of-sample sizes exceed 95\%.  

The power is shown in Panels D, E, F of Figure \ref{fig:true_model_es}. Our findings are very similar to those shown in Figure \ref{fig:true_model_var} (Panels D, E, F) above for VaR forecasts. Yet, overall power is slightly lower.

In the settings that relate to calmer market conditions, i.e.\ for \( \nu \in \{7, 12 \} \) degrees of freedom, and joint VaR and ES forecast at quantile level \linebreak \( p \in \{0.025, 0.05, 0.1 \} \), we find that power increases slowly with the out-of-sample size \( P \) ( \(P \geq 126\) ) and the quantile level \( p \): the MCS eliminates 1 model for 126 out-of-sample observations and between 1.4 and 1.6 models for an out-of-sample size of two years. For forecasts at the quantile level \( p=0.01 \) (solid dark grey line), Panels D, E, F show that power increases with the out-of-sample size \( P \) only for \( P \geq 251 \), i.e.\ for more than one year of daily data.

To sum up, we find that the discriminatory power of the MCS increases with the quantile level \( p \), as in the DM test simulations. Yet, for the out-of-sample sizes of up to 10 years of daily observations, the MCS test does not reveal the single best model.

Table \ref{fig:rejection_frequencies_fin} below shows how often the MCS test eliminates each model, to which we refer as individual rejection frequencies. The left and right panel provides individual rejection frequencies for standalone VaR forecasts and joint VaR and ES forecast, respectively. The results are for the DGP with \( \nu=3 \) degrees of freedom and level of the test \( \alpha=0.25 \).

\begin{table}[H]
    \includegraphics[width=1\linewidth]{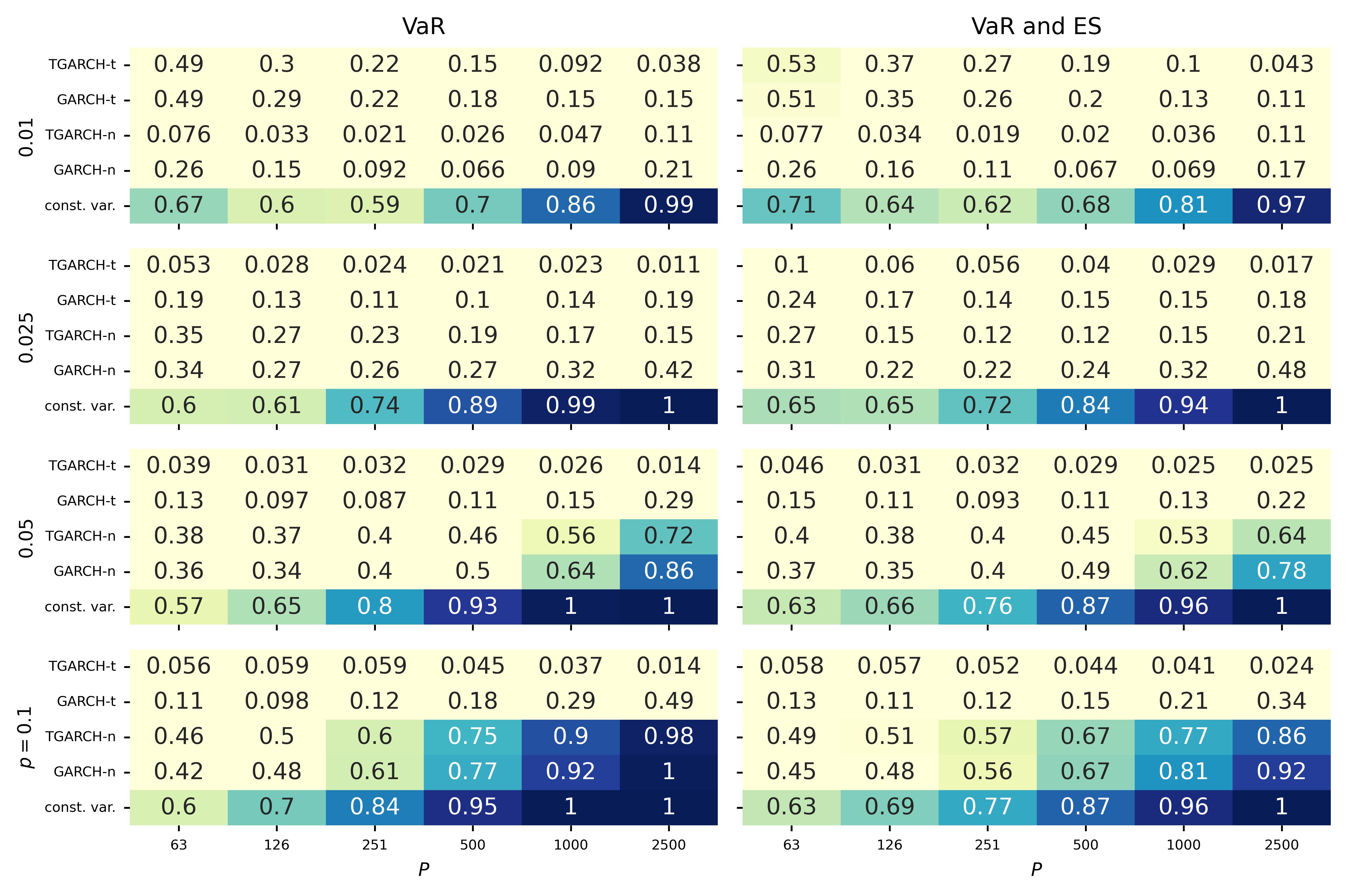}
    \caption{Individual rejection frequencies, \( \alpha=0.25 \)}
    \label{fig:rejection_frequencies_fin}
    \footnotesize{This table displays the individual rejection frequencies of the candidate models in the MCS, number of models \(m=5\), level of the test \( \alpha=0.25 \). DGP with \( \nu=3 \) degrees of freedom.}
\end{table}

At the most extreme quantile level \( p=0.01 \), the MCS test eliminates the constant variance model - which has by far the largest expected loss - most frequently. Starting from 251 observations, the rejection frequency increases monotonically with the number of observations until it reaches almost 1 for 2500 observations. We conclude that the MCS test reliably detects that this naive model has a larger expected loss than its competitors.  

Yet, at the quantile level \(p=0.01\) we also observe patterns in the individual rejection frequencies that are not consistent with the ranking of the models in terms of the expected losses. These findings are in line with the power and type III errors that we find in the pairwise comparison simulations in Section \ref{sec:pairwise_comparison}, and stem from imprecise estimates of the expected losses - see Figure \ref{fig:var_sign_d_ij} - and the finite sample behavior of the tick loss, which we discuss above.

First, all individual rejection frequencies decrease from 63 to 251 out-of-sample observations. While Figure \ref{fig:true_model_es} reveals that some rejection frequencies have to decrease, as the number of models in the MCS increases, it is remarkable that the MCS test eliminates \textit{all} models less frequently when the out-of-sample size increases. Second, for sample sizes of up to 4 years, i.e.\  1000 observations, the MCS test eliminates the true model and the GARCH-t more frequently than the TGARCH-n and GARCH-n, although the latter models have larger expected losses. This discrepancy between the expected losses and the individual rejection frequencies decreases with the out-of-sample size \( P \). We thus argue against using the MCS test for model averaging for very short out-of-sample periods.

Overall, we conjecture that statistical losses and the MCS test are not appropriate tools to detect VaR models that are too optimistic at the quantile level \( p=0.01\).
At the quantile levels \( p \in \{0.025, 0.05, 0.1\} \), however, the MCS test consistently eliminates models with larger expected losses more frequently than models with smaller expected losses. 

Consequently, the constant variance model is eliminated most frequently, followed by the GARCH-n, which has the second largest expected loss.  

Yet, the MCS test shows little power against the second and third best GARCH-t and TGARCH-n model, respectively, for sample sizes of up to \( P=500 \). On average, one of them becomes a candidate for elimination only once the MCS test eliminates the two other inferior models that have larger expected losses. Due to the control of the familywise error rate, such an elimination requires much stronger evidence than previous eliminations.


\subsubsection*{Second set of models - \( m = 10 \).}
Next, we discuss potency, power and individual rejection frequencies for the larger set of models, which we describe above. Figure \ref{fig:true_model_10_var} presents potency and power for VaR forecasts, while we provide results for joint forecasts in Appendix \ref{app:detail_results_m10}.

Potency - Panels A, B, C of Figure \ref{fig:true_model_10_var} -  increases with the quantile level \( p \) and the out-of-sample size \( P \). For the quantile levels \( p \in \{0.025, 0.05, 0.1 \} \), potency exceeds the confidence level of 75\% even for the smallest out of sample size \( P =63 \). For the most extreme quantile level \( p=0.01 \), potency exceeds 75\% starting from \(P=126 \) observations.

Potency takes on values between 95\% and 99\% for out-of-sample sizes \( P \geq 500 \), i.e.\ starting from two years of daily data. For this larger set of \( m=10 \) models, the MCS test keeps the model with the smallest expected loss more often than for the smaller set of \(m=5 \) models  (cf. Figure \ref{fig:true_model_var}).

\begin{figure}[!htb]
    \includegraphics[width=1\linewidth]{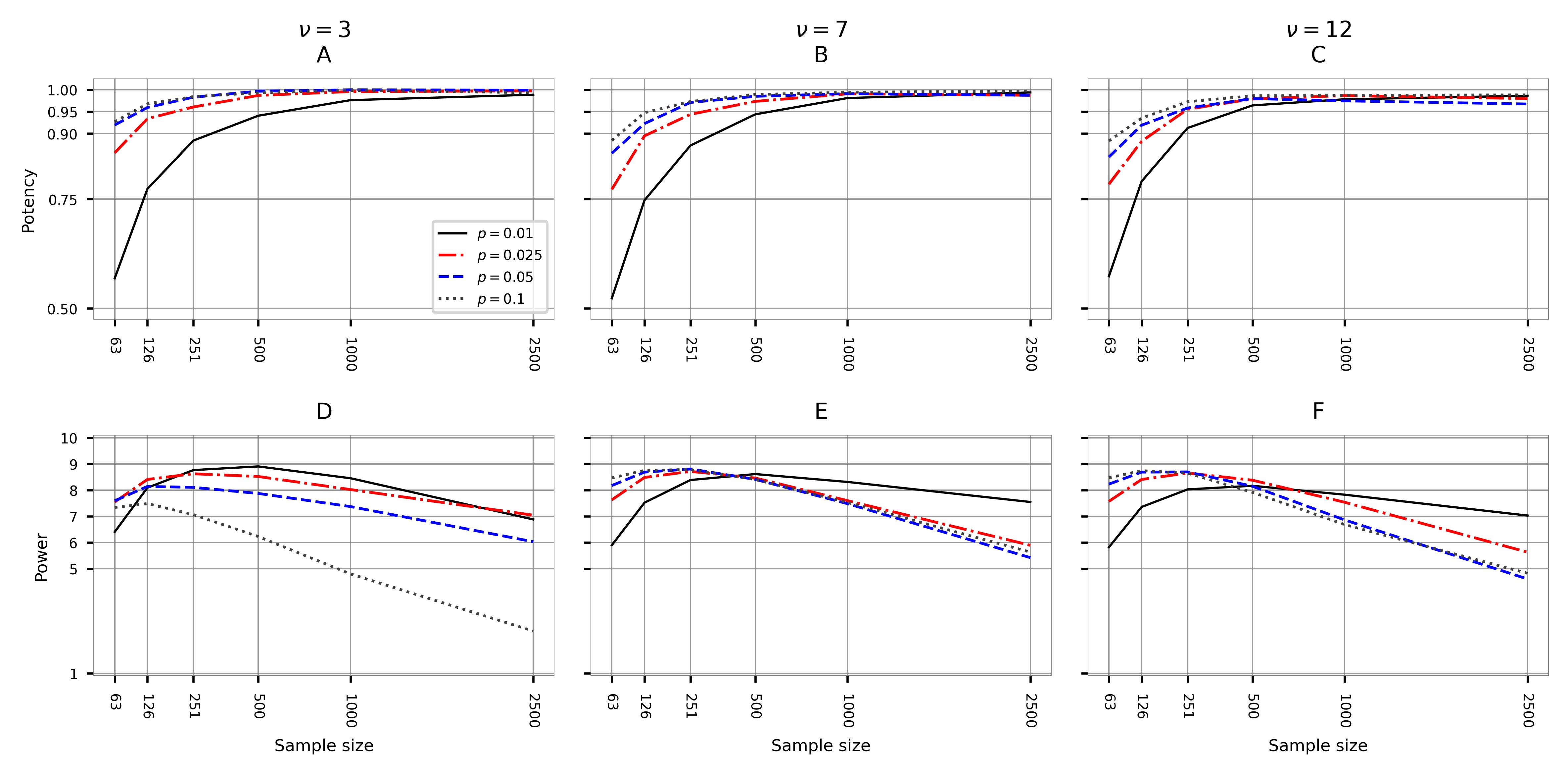}
    \caption{Finite sample properties for the tick loss}
    \label{fig:true_model_10_var}
    \footnotesize{This figure displays the finite sample properties of the MCS procedure in the following setting: one-day-ahead VaR forecasts evaluated using the tick loss, true model included, number of models \(m=10\), level of the test \( \alpha=0.25 \). The upper panel displays the potency, i.e.\  the frequency of \( \mathcal{M}^{*} \subset \widehat{\mathcal{M}}_{1-\alpha}^{*} \). The lower panel displays the power property, i.e.\  the average number of elements in \(\mathcal{ \widehat{M}}_{1-\alpha}^{*} \).}
\end{figure}

The power - Panels D, E, F of Figure \ref{fig:true_model_10_var} - behaves similarly as in the case with \( m=5 \) models depicted in Figure \ref{fig:true_model_var} above. We observe that the number of models in the MCS decreases starting from 126 out-of-sample observations for the quantile levels \( p \in \{0.025, 0.05, 0.1 \} \). For the most extreme quantile level \( p=0.01 \), the number of models in the MCS decreases for out-of-sample sizes \( P \geq 251 \). For the DGPs that relate to calmer markets conditions - panels E and F - the MCS test eliminates between 1.5 and 2 models for out-of-samples sizes of \( P=500 \) and across all quantile levels \( p\). 

We find more variation in panel A which corresponds to the DGP with \( \nu=3 \) degrees of freedom. For an out-of-sample size \( P=500 \) and quantile levels \( p \in \{0.01, 0.025, 0.05 \} \), the MCS test eliminates 1, 1.5 and 2 models, respectively. 

Overall, the MCS test still eliminates the constant variance model most often, though between 5-10\% less frequently than for the set of \(m=5\) candidate models. 
At the quantile levels \( p \in \{0.01, 0.025 \} \), no strong pattern emerges among the other inferior models, which all show elimination frequencies of less than 10\%. 

At the larger quantile levels \( p \in \{0.05, 0.1 \} \), the models that the MCS test eliminates most frequently after the constant variance model, are the GARCH-n, TGARCH-n and GARCH-\(t_{12}\). We note, however, that the elimination frequencies of the GARCH-n and the TGARCH-n are between 11\% and 20\% smaller for the larger set of \( m=10 \) models as compared to the smaller set of \(m=5 \) models (cf. Table \ref{fig:rejection_frequencies_fin} above). To summarize, the MCS test has less power against inferior alternatives when we enlarge the set of competing models.

\FloatBarrier

\section{Conclusion}\label{sec:conclusion}
This paper sheds light on the finite sample properties of two popular tests for equal predictive ability, the Diebold-Mariano (DM) test by \textcite{Diebold.1995} and the model confidence set (MCS) testing procedure by \textcite{Hansen.2011} applied to out-of-sample VaR and ES forecasts evaluated using asymmetric statistical loss functions.

Our findings may be summarized as follows. First, the tests show little power against models that underestimate the tail risk at the extreme quantile levels \( p \in \{0.01, 0.025 \} \), while the power generally increases with the quantile level \( p \) and the out-of-sample size \( P \). 
Second, for the small quantile levels \( p \in \{0.01, 0.025 \} \) and out-of-sample sizes of up to two years, we observe heavily skewed test statistics and non-negligible type III errors, which implies that researchers should be cautious about using standard normal or bootstrapped critical values. We demonstrate how these unfavorable finite sample results relate to the asymmetric loss functions and the time varying volatility inherent in financial return data.
Third, the power improves for two and more years of observations when one performs joint forecast evaluation of ES and VaR as rather than of standalone VaR forecasts. 
Overall, we conclude that the tests need long evaluation windows of several years to make reliable inference about the predictive ability of competing models for extreme quantiles, which highlights the usual tradeoff between long evaluation samples and structural breaks. 
Moreover, we find that the potency of the MCS is usually much higher than \( 1-\alpha\) for out of sample sizes of 4 years and more, which makes it seem safe to use large values of \( \alpha \) in applied work in order to observe power. 
The distinct parametrizations of the generalized piecewise linear (GPL) loss to evaluate VaR forecasts display no remarkable differences in potency and power. We neither find a uniform pattern between the different parametrizations of the loss functions to evaluate joint VaR and ES forecasts.

This paper leaves much room for future work, e.g., one may explore if conditional forecast evaluation along the lines of \textcite{Giacomini.2006} shows better discriminative ability, which may be particularly relevant under misspecification and from a regulatory perspective. Additionally, finite sample corrections of the variance estimates may mitigate the large type III errors for the shorter out-of-sample sizes of up to two years.

\subsection*{Acknowledgements}
The author is grateful to Timo Dimitriadis, Christian Gouri\'eroux, Roxana Halbleib, Ekaterina Kazak, Michael Massmann and Winfried Pohlmeier for helpful comments. All remaining errors are my own. Previous versions of this paper have been presented at the Young Scientists Workshop Statistische Woche Dortmund 2023, a Doctoral Workshop in Konstanz 2024, the QFFE in Marseille 2024, the ISF in Dijon 2024, the IAAE in Thessaloniki 2024, the COMPSTAT in Giessen 2024 and the CFE in London 2024. This work was performed on the computational resource bwUniCluster funded by the Ministry of Science, Research and the Arts Baden-Württemberg and the Universities of the State of Baden-Württemberg, Germany, within the framework program bwHPC.



\endgroup
\clearpage


\printbibliography
\clearpage

\appendix

\renewcommand{\thesection}{\Alph{section}}
\section{Data and DGP}\label{app:data_and_DGP}
\subsection{Data}\label{app:data}
The data on which we estimate the parameters of the DGPs in Section \ref{sec:dynamic_scenario} consist of three equally weighted indices composed of 30 randomly chosen stocks from the Dow Jones U.S. Small, Mid, and Large cap-indices. The source of the data is the Thomson Reuters Data Stream. We compute daily log-returns from closing prices adjusted for dividend and split.

The symbols of the stocks included are given below. The small cap index contains the stocks with the following symbols: AGL, AIR, AMR, ASH, BDN, BEZ, BIG, BIO, BRE, BXS, CBRL, CBT, COO, CTX, CW, DLX, ESL, GAS, HXL, ITG, LIZ, LPX, MDP, NEU, PBY, PCH, PPD, RLI, TXI, UNS. 

The mid cap index contains: ACV, ADSK, AMD, BCR, BDK, BMS, CBE, CCK, CEG, CSC, DBD, DOV, DTE, EK, DPL, GMGMQ, GR, GWW, HOT, MAS, MDC, MWV, NAV, NI, ROST, RSH, SWK, UNM, VFC, WEC.

The large cap index contains: ABT, ADBE, AMAT, APC, APD, AVP, BAC, BEN, BK, CA, CL, D, DD, EMR, FPL, ITW, JCI, JPM, LOW, MMM, MRK, OXY, PCAR, PEP, PFE, SO, SYY, TGT, WFT, XOM.

We estimate parameters from January 1987 to July 2009 using rolling windows of 4 years. Table \ref{tab:estimates_parameters} below present the results for the small, mid and large cap index in the first, second and third panel, respectively.

\begin{table}[h!]
  \scriptsize
  \centering
  \caption{Parameter estimates}
  \label{tab:estimates_parameters}
  \begin{tabularx}{0.75\textwidth}{l X X X X X }
              &\( \omega\) & \( \alpha\) & \( \gamma \) & \( \beta \) & \( \nu \)   \\ 
  \toprule 
  \multicolumn{6}{l}{Small } \\
  mean      &  0.069 &  0.014 &  0.106 &  0.793 &  15.869   \\ 
  5\% quantile    &  0.000 &  0.000 &  0.005 &  0.000 &  4.367   \\ 
  median    &  0.038 &  0.002 &  0.120 &  0.876 &  7.732   \\ 
  95\% quantile   &  0.324 &  0.057 &  0.198 &  0.992 &  94.372   \\ 
  \bottomrule 
  \multicolumn{6}{l}{Mid } \\
  mean      &  0.030 &  0.008 &  0.100 &  0.905 &  11.447   \\ 
  5\% quantile    &  0.000 &  0.000 &  0.016 &  0.829 &  5.044   \\ 
  median    &  0.028 &  0.000 &  0.112 &  0.909 &  8.044   \\ 
  95\% quantile   &  0.054 &  0.036 &  0.156 &  0.990 &  26.152   \\ 
  \bottomrule 
  \multicolumn{6}{l}{Large } \\
  mean      &  0.038 &  0.010 &  0.106 &  0.884 &  12.215   \\ 
  5\% quantile    &  0.003 &  0.000 &  0.029 &  0.836 &  4.440   \\ 
  median    &  0.020 &  0.000 &  0.104 &  0.918 &  8.011   \\ 
  95\% quantile   &  0.068 &  0.036 &  0.143 &  0.970 &  34.728   \\ 
  \bottomrule  \\
  \end{tabularx} \\
  \vspace{1ex}
  \raggedright
  \scriptsize{ \textit{Notes:} This table displays parameter estimates for the TGARCH-t (Equation \ref{eq:dgp}) that stem from a 4 year rolling window estimation from January 1987 to July 2009 on three equally weighted indices composed of 30 randomly chosen stocks from the Dow Jones U.S. Small, Mid, and Large cap-indices.}
\end{table}

\subsection{DGP}\label{app:dgp}
\subsubsection{Expected losses}\label{app:expected_losses}

In general, it is not straightforward to characterize the size of the expected losses and the expected loss differentials. First, the individual losses \( l_{t, i} \) associated with model \( i \) live on different scales as they have different variances in our simulations, which is in line with what we observe empirically. Second, scaling the losses is not informative as we test about the absolute difference in expectation, i.e. model \(i \) is superior to model \( j \) if \( \mu_{ij} < 0 \) regardless the variances of the two models. 

Third, scaling the expected loss differentials \( \mu_{ij} \) by the variance of \( D_{t, ij} \) may help understand the first elimination. Yet, the scaled loss which is most relevant in each subsequent testing step depends on the testing path, i.e.\ it depends on which model(s) the MCS test eliminates in the previous step.

Instead, we may compare the expected loss differentials to those that we observe empirically in absolute values or relative to the model with the smallest average or expected loss. In our simulations, the size of the expected losses of misspecified models varies between 101 \% and 110\% of the expected loss of the true model. These ratios correspond to the ratios of the average sample losses in \textcite{Bernardi.2017}, \textcite{Taylor.2020} and \textcite{Dimitriadis.2022}. We stress that `very similar' expected losses do not need to stem from `very similar' forecasts: consider e.g.\ example \ref{example:tick_loss}. In this example, we illustrate how the tick loss function evaluates VaR forecast of two models if one model consistently makes forecasts that are closer to zero than the forecasts of the other model. 

To illustrate how correlations affect the MCS test, assume that we compare only two models \( i \) and \( j \) and that the variances are fixed. It holds that \( var(D_{t, ij}) = var(L_{t, i}) + var(L_{t, j}) - 2 cov(L_{t, i}, L_{t, j}) \). Thus, the higher the correlations between the losses of model \( i \) and \( j \), the smaller the variances of the loss differential \( D_{t, ij} \). The MCS test uses the t-statistic \( t_{ij } = \dfrac{\bar{d}_{ij}}{\sqrt{\hat{var}(\bar{d}_{i j})}} \). Hence, high correlations lead to larger t-statistics, thus more rejections and imply that the MCS more easily detects differences in the expected losses.\footnote{\textcite{Hansen.2011} also discuss this example when all losses have unit variance.} 

In our simulations, we observe correlations between 0.95 and 0.99 for the GARCH-type models, and correlations between 0.8 and 0.95 between the constant variance model and the GARCH-type models. These correlations are in line with the correlations among the GARCH-type models that we estimate on the Dow Jones Indices data. 

\subsubsection{Variance Formula}\label{app:variance_formula}

For random variables \( X_{i}, i=1, ..., m \) and real numbers \( a_{i}, i=1, ..., m \) it holds that
\begin{align}
    var(\sum a_{i} X_{i}) &= \sum_{i=1}^{m} a_{i}^{2} var(X_{i}) + 2 \sum_{i,j=1, i \neq j }^{m} a_{i} a_{j} cov(X_{i}, X_{j}). 
\end{align}

For \( var(D_{i \cdot}) = var(L_{i} - \dfrac{1}{m} \sum_{j=1}^{m}L_{j}) \), we get

\begin{align}\label{eq:variance_formula}
    var(D_{i \cdot}) &= var(L_{i} - \dfrac{1}{m} \sum_{j=1}^{m}L_{j}) \\ \nonumber
    &=var \left( \dfrac{m-1}{m}L_{i} - \dfrac{1}{m}\sum_{j=1, j \neq i}^{m} L_{j}  \right)  \\ \nonumber
    &= \left( \dfrac{m-1}{m} \right) ^{2} var(L_{i}) + \dfrac{1}{m^{2}} var(\sum_{j=1, j \neq i}^{m} L_{j}) - \dfrac{2 (m-1)}{m^{2}} \sum_{j=1, j \neq i}^{m} cov(L_{i}, L_{j})  \\ \nonumber
    &= \left( \dfrac{m-1}{m} \right) ^{2} var(L_{i}) + \dfrac{1}{m^{2}} \sum_{j=1, j \neq i}^{m} var( L_{j}) \\ \nonumber
    &+ \dfrac{2}{m^{2}} \sum_{j=1, j \neq i, k=j+1, k\neq i}^{m}cov(L_{j}, L_{k}) - \dfrac{2(m-1)}{m^{2}} \sum_{j=1, j \neq i}^{m} cov(L_{i}, L_{j}). \nonumber
\end{align}

\clearpage

\subsection{Theoretical properties of the loss differential \( d_{ij}\)}\label{sec:loss_difference_iid}

Below, we calculate the loss differential and it theoretical value for some of the pairwise VaR comparison problems in Section \ref{sec:simulation}.
For ease of exposition, we focus on the tick loss, but note that the insight generally hold for members of the homogeneous parametric GPL familiy from Equation \ref{eq:hom_gpl_loss}, but that the scaling changes. 

For \( b = 1 \), Equation \ref{eq:hom_gpl_loss} yields the tick loss function.  
\begin{align}
    L(x, y) =  (\mathbbm{1}\{y \leq x\} - p) \times (\sgn (x) |x| - \sgn(y) |y|).
\end{align}

We now consider two competing forecasters that make  forecasts \( x_{1} \) and \( x_{2} \) of the p-quantile, writing down explicitly the loss difference  
\begin{align*}
    L(x_{1}, y) - L(x_{2}, y) &= (1-p)(x_{1}-y) \mathbbm{1}\{y<x_{1}\} \\
    & - p (x_{1}-y)  \mathbbm{1}\{y \geq x_{1}\} \\
    & - (1-p)(x_{2}-y) \mathbbm{1}\{y<x_{2}\} \\
    & + p (x_{2}-y)  \mathbbm{1}\{y \geq x_{2}\} 
\end{align*}

We first consider \( p \) so that \( q_{1, p} < q_{2, p} \), which is for instance the case when comparing the 0.01 quantile of the standardized Student's t distribution for \( \nu \) sufficiently small and a standard normal distribution. 
Let \( Y \) be a random variable with continuous cdf \( F \). 
In this case we have \( \{y<x_{1}\} \subset \{y<x_{2}\} \),  \( \{y \geq x_{2}\} \subset \{y  \geq x_{1}\} \) and thus 
\begin{align*}\label{eq:simplemost_loss_difference}
    L(x_{1}, y) - L(x_{2}, y) &= (1-p) [(x_{1}-y) - (x_{2}-y)] \mathbbm{1}\{y<x_{1}\} \\
    & + p [(x_{2}-y) -(x_{1}-y)]   \mathbbm{1}\{y \geq x_{2}\} \\
    & - [(1-p)(x_{2}-y) - p (x_{1}-y)] \mathbbm{1}\{ x_{1} \leq  y<x_{2}\} \\
    &= (1-p) [x_{1}- x_{2}] \mathbbm{1}\{y<x_{1}\} \\
    & + p [x_{2} - x_{1}]   \mathbbm{1}\{y \geq x_{2}\} \\
    & +[ p(x_{2} - x_{1}) -(x_{2}-y) ] \mathbbm{1}\{ x_{1} \leq  y<x_{2}\} \; * 
\end{align*}
where the  term \( * \) ist equal to zero when \( p(x_{2} - x_{1}) = x_{2}-y \), and thus the loss difference is smaller than zero for \( y <x_{2} - p (x_{2} - x_{1}) \), which is observed with probability \( F(x_{2} - p (x_{2} - x_{1})) \), which is the share of observations that correctly indicate that model 1 is superior to model 2.

Taking the expectation, we obtain
\begin{align*}
    \mathbb{E}[L(x_{1}, y) - L(x_{2}, y) ]&= (1-p) [q_{1}- q_{2}] F(q_{1})  \\
        &+ p [q_{2} - q_{1}]  (1 - F(q_{2})) \\
        &+[ p(q_{2} - q_{1})  (F(q_{2}) - F(q_{1})) \\
        &     - \mathbb{E}[(q_{2}-z) ] \mathbbm{1}\{ q_{1} \leq  z<q_{2}\} ] \\
        &= (1-p) [q_{1}- q_{2}] F(q_{1})  \\
        &+ p [q_{2} - q_{1}]  (1 - F(q_{2})) \\
        &+ p(q_{2} - q_{1})  (F(q_{2}) - F(q_{1})) \\
        &   - (q_{2}- \mathbb{E}[z| q_{1} \leq  z<q_{2} ]) (F(q_{2}) - F(q_{1})) \\
        & \overset{*}{=} (1-p) \Delta q p - p \Delta q  (1 - F(q_{2})) -p \Delta q (F(q_{2}) - p) \\
        &-(q_{2}- \mathbb{E}[z| q_{1} \leq  z<q_{2} ]) (F(q_{2}) - p) \\
        &=  \Delta q p [ 1-p  - 1 + F(q_{2})  - F(q_{2}) + p ] \\
        &   - (q_{2}- \mathbb{E}[z| q_{1} \leq  z<q_{2} ]) (F(q_{2}) - p) \\
        &= - (q_{2}- \mathbb{E}[z| q_{1} \leq  z<q_{2} ]) (F(q_{2}) - p) \\
\end{align*}

where at \(*\) we use that model 1 is the true model and denote \( \Delta q = q_{1} - q_{2}\).

Moreover, we see from the expression above and the expression for the loss difference that an observations of a loss difference \( d_{t, ij} >0 \) contributes much less to the estimated variance even if the expected value of \( d_{t, ij} \) was known than those with \( d_{t, ij} < 0 \), i.e., samples that lack sufficiently many of the latter observations will tend to additionally underestimate the variance. Given the high variability in shorter samples at the extreme quantiles \( p \in \{0.01, 0.025\} \), this happens frequently and explains the large number of type III errors that we observe in Section \ref{sec:static_scenario}.

If we consider \( p \) so that \( q_{2, p} < q_{1, p} \), which is for instance the case when comparing the 0.1 quantile of the standardized Student's t distribution for \( \nu \) sufficiently small and a standard normal distribution, the properties of the loss difference reverse, and it becomes much easier to correctly identify the true model.

\subsubsection{Dynamic Scenario}\label{sec:loss_difference_garch_case}
\subsubsection*{Distributional misspecification}

Next assume a GARCH setting, where we compare the true model with one that assumes the wrong distribution of the innovations, i.e., we have that 
\begin{align} 
    x_{t, 1} &= \sigma_{t} q_{1, p} \\ 
    x_{t, 2} &= \sigma_{t} q_{2, p}, 
\end{align}

where \(z_{t} \) follows some distribution with unit variance, zero mean and continuous CDF \( F \), \(q_{1, p} = F^{-1}(p) \), \( q_{2, p} = F^{-1}_{2}(p)\) with \( F_{2} \neq F \) being some misspecified distribution. The time varying volatility \( \sigma_{t} \) is measurable with respect to the sigma algebra \( \mathcal{F}_{t-1} \), which is the information set available at time \( t-1 \).

If we plug in to to expression for the loss difference according to the equations above, we obtain
\begin{align*}
L(x_{1}, y) - L(x_{2}, y) &= \sigma_{t} [ (1-p) [q_{1}- q_{2}] \mathbbm{1}\{z<q_{1}\}  \\
    &+ p [q_{2} - q_{1}]   \mathbbm{1}\{z \geq q_{2}\} \\
    &+[ p(q_{2} - q_{1}) -(q_{2}-z) ] \mathbbm{1}\{ q_{1} \leq  z<q_{2}\} ]
\end{align*}
which shows that the losses live on on a time varying scale, implying they are all weighted differently when using the out-of-sample mean to make inference about the mean of the loss difference.

Taking the expectation conditional on \( \mathcal{F}_{t-1} \), we obtain
\begin{align*}
    \mathbb{E}[L(x_{1}, y) - L(x_{2}, y)|\mathcal{F}_{t-1} ] &= - \sigma_{t} (q_{2}- \mathbb{E}[z| q_{1} \leq  z<q_{2} ]) (F(q_{2}) - p), \\
    \end{align*}
which gives us the same insights as in the iid scenario above but shows how the time varying volatility induces additional variability.

\subsubsection*{Volatility misspecification}
\begin{align}  
    y &= \sigma_{t} z_{t} \\
    x_{t, 1} &= \sigma_{t} q_{p} \\ 
    x_{t, 2} &= \sigma_{t, 2} q_{p}, 
\end{align}
where \(z_{t} \) follows some distribution with unit variance, zero mean and continuous CDF \( F \), \(q_{p} = F^{-1}(p) \), \( \sigma_{t, 2} \neq \sigma_{t} \) with \(\sigma_{t, 2} \) being some misspecified conditional volatiltity.

Assuming that \( \sigma_{t} > \sigma_{t, 2}\) and plugging in to the final expression above  we obtain

\begin{align*}
    L(x_{1}, y) - L(x_{2}, y) &= (1-p) [\sigma_{t} q_{p}- \sigma_{t, 2} q_{p}] \mathbbm{1}\{\sigma_{t} z <\sigma_{t} q_{p}\} \\
    & + p [\sigma_{t, 2} q_{p} - \sigma_{t} q_{p}]   \mathbbm{1}\{\sigma_{t} z \geq \sigma_{t, 2} q_{p}\} \\
    & +[ p(\sigma_{t, 2} q_{p} - \sigma_{t} q_{p}) -(\sigma_{t, 2} q_{p}-\sigma_{t} z) ] \mathbbm{1}\{ \sigma_{t} q_{p} \leq  \sigma_{t} z <\sigma_{t, 2} q_{p}\} \\
    &= \sigma_{t} [ (1-p) [ q_{p}- \dfrac{\sigma_{t, 2}}{\sigma_{t}} q_{p}] \mathbbm{1}\{ z < q_{p}\} \\
    & + p [\dfrac{\sigma_{t, 2}}{\sigma_{t}} q_{p} - q_{p}]   \mathbbm{1}\{z \geq \dfrac{\sigma_{t, 2}}{\sigma_{t}} q_{p}\} \\
    & +[ p( \dfrac{\sigma_{t, 2}}{\sigma_{t}} q_{p} - q_{p}) -(\dfrac{\sigma_{t, 2}}{\sigma_{t}} q_{p}- z) ] \mathbbm{1}\{ q_{p} \leq  z < \dfrac{\sigma_{t, 2}}{\sigma_{t}} q_{p}\} ] \\
    &= \sigma_{t} [ (1-p) q_{p} (1- \dfrac{\sigma_{t, 2}}{\sigma_{t}}) \mathbbm{1}\{ z < q_{p}\} \\
    & + p q_{p} (\dfrac{\sigma_{t, 2}}{\sigma_{t}} - 1)    \mathbbm{1}\{z \geq \dfrac{\sigma_{t, 2}}{\sigma_{t}} q_{p}\} \\
    & +[ p q_{p} (\dfrac{\sigma_{t, 2}}{\sigma_{t}} - 1) -(\dfrac{\sigma_{t, 2}}{\sigma_{t}} q_{p}- z) ] \mathbbm{1}\{ q_{p} \leq  z < \dfrac{\sigma_{t, 2}}{\sigma_{t}} q_{p}\} ]  \; * 
\end{align*}
where the  term \( * \) ist equal to zero when \( p q_{p} (\dfrac{\sigma_{t, 2}}{\sigma_{t}} - 1)  = \dfrac{\sigma_{t, 2}}{\sigma_{t}} q_{p} - z \), and thus the loss difference is smaller than zero for \( z <  q_{p} \dfrac{\sigma_{t, 2}}{\sigma_{t}} +  q_{p} p (1- \dfrac{\sigma_{t, 2}}{\sigma_{t}})  \), which is observed with probability \( F(q_{p} \dfrac{\sigma_{t, 2}}{\sigma_{t}} +  q_{p} p (1- \dfrac{\sigma_{t, 2}}{\sigma_{t}})) \), which is the share of observations that correctly indicate that model 1 is superior to model 2.

The expected loss difference conditional on  \( \mathcal{F}_{t-1} \) is similar to the smaller quantile setting above, we denote \( c = \dfrac{\sigma_{t, 2}}{\sigma_{t}} \) - suppressing \( t \) - to obtain

\begin{align*}
    \mathbb{E}[L(x_{1}, y) - L(x_{2}, y)|\mathcal{F}_{t-1} ] &= \sigma_{t} E[ (1-p) q (1- c) \mathbbm{1}\{ z < q\} \\
    & + p q (c - 1)    \mathbbm{1}\{z \geq c q\} \\
    & +[ p q (c - 1) -(c q- z) ] \mathbbm{1}\{ q \leq  z < c q\} ]  \\
    &= \sigma_{t} [ (1-p) q (1- c) F(q) \\
    & + p q (c - 1) (1-F(c q)) \\
    & +[ p q (c - 1) - E [(c q- z)| q \leq  z < c q ]] (F(cq) - F(q))]  \; | \; F(q)  = p \\
    &= \sigma_{t} [ - c q- E [z| q \leq  z < c q ] (F(cq) - p)  ].    \\
\end{align*}

\section{Additional results}\label{app:detailed_results}

\subsection{Additional results: static case}\label{app:static_case}

\begin{table}[h!]
        \scriptsize
        \centering
        \caption{Static case, distributional misspecification: GPL with $b=0.5$}
        \begin{tabularx}{1\textwidth}{l X X X X X X X X }
        \toprule
                &\multicolumn{4}{c}{Power} & \multicolumn{4}{c}{Type III error}   \\
        \toprule
        \(p \backslash P \) & 251 & 500 & 1000 & 2500 & 251 & 500 & 1000 & 2500  \\
        \toprule
        0.01   &  0.010 &  0.023 &  0.055 &  0.153 &  0.037 &  0.038 &  0.014 &  0.003   \\
        0.05   &  0.113 &  0.120 &  0.150 &  0.254 &  0.004 &  0.002 &  0.001 &  0.000   \\
        0.1    &  0.176 &  0.257 &  0.410 &  0.755 &  0.001 &  0.000 &  0.000 &  0.000   \\
        \bottomrule
        \end{tabularx}
        \raggedright
        \scriptsize{ \textit{Notes:} This table displays the finite sample power and type III error of the DM test for one-step-ahead forecasts evaluated using the GPL with $b=0.5$.}
        \end{table}


        \begin{table}[h!]
        \scriptsize
        \centering
        \caption{Static case, distributional misspecification: GPL with $b=2$}
        \begin{tabularx}{1\textwidth}{l X X X X X X X X }
        \toprule
                &\multicolumn{4}{c}{Power} & \multicolumn{4}{c}{Type III error}   \\
        \toprule
        \(p \backslash P \) & 251 & 500 & 1000 & 2500 & 251 & 500 & 1000 & 2500  \\
        \toprule
        0.01   &  0.010 &  0.022 &  0.053 &  0.143 &  0.038 &  0.039 &  0.015 &  0.003   \\
        0.05   &  0.115 &  0.122 &  0.153 &  0.263 &  0.003 &  0.002 &  0.001 &  0.000   \\
        0.1    &  0.185 &  0.273 &  0.436 &  0.787 &  0.001 &  0.000 &  0.000 &  0.000   \\
        \bottomrule
        \end{tabularx}
        \raggedright
        \scriptsize{ \textit{Notes:} This table displays the finite sample power and type III error of the DM test for one-step-ahead forecasts evaluated using the GPL with $b=2$.}
        \end{table}

\begin{table}[h!]
        \scriptsize
        \centering
        \caption{Static case, distributional misspecification: AL score}
        \begin{tabularx}{1\textwidth}{l X X X X X X X X }
        \toprule 
                &\multicolumn{4}{c}{Power} & \multicolumn{4}{c}{Type III error}   \\ 
        \toprule 
                \(p \backslash P \) & 251 & 500 & 1000 & 2500 & 251 & 500 & 1000 & 2500  \\

        \toprule 
        0.01    &  0.006 &  0.024 &  0.081 &  0.315 &  0.070 &  0.038 &  0.011 &  0.000   \\ 
        0.025   &  0.004 &  0.016 &  0.055 &  0.233 &  0.105 &  0.048 &  0.016 &  0.001   \\ 
        0.05    &  0.167 &  0.241 &  0.399 &  0.758 &  0.001 &  0.000 &  0.000 &  0.000   \\ 
        0.1     &  0.185 &  0.275 &  0.439 &  0.788 &  0.001 &  0.000 &  0.000 &  0.000   \\ 
        \bottomrule 
        \end{tabularx}
        \raggedright
        \scriptsize{ \textit{Notes:} This table displays the finite sample power and type III error of the DM test for one-step-ahead forecasts evaluated using the AL score.}
        \end{table}

\begin{table}[h!]
\scriptsize
\centering
\caption{Static case, distributional misspecification: NZ score}
\begin{tabularx}{1\textwidth}{l X X X X X X X X }
\toprule 
        &\multicolumn{4}{c}{Power} & \multicolumn{4}{c}{Type III error}   \\ 
\toprule 
\(p \backslash P \) & 251 & 500 & 1000 & 2500 & 251 & 500 & 1000 & 2500  \\

\toprule 
0.01    &  0.008 &  0.026 &  0.078 &  0.269 &  0.053 &  0.034 &  0.011 &  0.001   \\ 
0.025   &  0.004 &  0.015 &  0.053 &  0.219 &  0.105 &  0.049 &  0.017 &  0.001   \\ 
0.05    &  0.157 &  0.202 &  0.297 &  0.558 &  0.001 &  0.001 &  0.000 &  0.000   \\ 
0.1     &  0.181 &  0.270 &  0.429 &  0.775 &  0.001 &  0.000 &  0.000 &  0.000   \\ 
\bottomrule 
\end{tabularx}
\raggedright
\scriptsize{ \textit{Notes:} This table displays the finite sample power and type III error of the DM test for one-step-ahead forecasts evaluated using the NZ score.}
\end{table}



\begin{table}[h!]
\scriptsize
\centering
\caption{Static case, wrong quantile: tick loss}
\begin{tabularx}{0.67\textwidth}{l X X X X X }
                        &\(p^{*} \backslash P \) & 251 & 500 & 1000 & 2500   \\ 
\toprule 
\multirow{ 7}{*}{\( p=0.01\)}   &  0.005 &  0.144 &  0.240 &  0.283 &  0.451   \\ 
                        &  0.015 &  0.010 &  0.023 &  0.055 &  0.146   \\ 
                        &  0.020 &  0.025 &  0.063 &  0.183 &  0.517   \\ 
                        &  0.025 &  0.051 &  0.157 &  0.374 &  0.811   \\ 
                        &  0.050 &  0.407 &  0.805 &  0.986 &  1.000   \\ 
                        &  0.100 &  0.969 &  1.000 &  1.000 &  1.000   \\ 
                        &  0.150 &  1.000 &  1.000 &  1.000 &  1.000   \\ 
\bottomrule 
\multirow{ 7}{*}{\( p=0.025\)}   &  0.010 &  0.312 &  0.415 &  0.621 &  0.924   \\ 
                                &  0.020 &  0.100 &  0.104 &  0.126 &  0.193   \\ 
                                &  0.030 &  0.013 &  0.026 &  0.046 &  0.093   \\ 
                                &  0.040 &  0.044 &  0.092 &  0.215 &  0.537   \\ 
                                &  0.050 &  0.105 &  0.240 &  0.540 &  0.907   \\
                                &  0.100 &  0.729 &  0.969 &  1.000 &  1.000   \\
                                &  0.150 &  0.984 &  1.000 &  1.000 &  1.000   \\
\bottomrule
\multirow{ 7}{*}{\( p=0.05\)}   &  0.010 &  0.749 &  0.944 &  0.998 &  1.000   \\
                        &  0.040 &  0.114 &  0.115 &  0.159 &  0.283   \\
                        &  0.060 &  0.026 &  0.045 &  0.081 &  0.169   \\
                        &  0.075 &  0.080 &  0.160 &  0.331 &  0.716   \\
                        &  0.100 &  0.278 &  0.544 &  0.852 &  1.000   \\
                        &  0.150 &  0.792 &  0.980 &  1.000 &  1.000   \\
                        &  0.200 &  0.982 &  1.000 &  1.000 &  1.000   \\
\bottomrule
\multirow{ 7}{*}{\( p=0.1\)}   &  0.050 &  0.447 &  0.694 &  0.924 &  0.999   \\
                        &  0.090 &  0.074 &  0.076 &  0.102 &  0.167   \\
                        &  0.110 &  0.029 &  0.043 &  0.061 &  0.107   \\
                        &  0.125 &  0.065 &  0.115 &  0.212 &  0.481   \\
                        &  0.150 &  0.187 &  0.357 &  0.668 &  0.968   \\
                        &  0.200 &  0.612 &  0.903 &  0.996 &  1.000   \\
                        &  0.250 &  0.918 &  0.998 &  1.000 &  1.000   \\
\bottomrule
\end{tabularx}\\
\raggedright
\scriptsize{ \textit{Notes:} This table displays the finite sample power of the DM test against a forecast based on a wrong quantile for one-step-ahead forecasts evaluated using the tick loss.\(p^{*} \) denotes the wrong quantile of the same distribution as the true quantile \(p\).}
\end{table}

\begin{table}[h!]
        \scriptsize
        \centering
        \caption{Static case, wrong quantile: FZG score}
        \begin{tabularx}{0.67\textwidth}{l X X X X X }
                                        &\(p^{*} \backslash P \) & 251 & 500 & 1000 & 2500   \\ 
        \toprule 
        \multirow{ 7}{*}{\( p=0.01\)}   &  0.005 &  0.166 &  0.234 &  0.269 &  0.405   \\ 
                                        &  0.015 &  0.008 &  0.022 &  0.049 &  0.146   \\ 
                                        &  0.020 &  0.022 &  0.062 &  0.178 &  0.511   \\ 
                                        &  0.025 &  0.046 &  0.149 &  0.379 &  0.814   \\ 
                                        &  0.050 &  0.412 &  0.820 &  0.990 &  1.000   \\ 
                                        &  0.100 &  0.978 &  1.000 &  1.000 &  1.000   \\ 
                                        &  0.150 &  1.000 &  1.000 &  1.000 &  1.000   \\ 
        \bottomrule 
        \multirow{ 7}{*}{\( p=0.025\)}   &  0.010 &  0.301 &  0.381 &  0.563 &  0.880   \\
                                                &  0.020 &  0.125 &  0.108 &  0.124 &  0.189   \\
                                                &  0.030 &  0.012 &  0.024 &  0.043 &  0.092   \\
                                                &  0.040 &  0.040 &  0.095 &  0.210 &  0.539   \\
                                                &  0.050 &  0.099 &  0.237 &  0.534 &  0.909   \\
                                                &  0.100 &  0.741 &  0.973 &  1.000 &  1.000   \\
                                                &  0.150 &  0.988 &  1.000 &  1.000 &  1.000   \\ 
        \bottomrule
        \multirow{ 7}{*}{\( p=0.05\)}   &  0.010 &  0.668 &  0.891 &  0.992 &  1.000   \\
                                        &  0.040 &  0.114 &  0.115 &  0.154 &  0.272   \\
                                        &  0.060 &  0.025 &  0.043 &  0.079 &  0.163   \\
                                        &  0.075 &  0.077 &  0.153 &  0.324 &  0.713   \\
                                        &  0.100 &  0.271 &  0.539 &  0.857 &  1.000   \\
                                        &  0.150 &  0.799 &  0.983 &  1.000 &  1.000   \\
                                        &  0.200 &  0.985 &  1.000 &  1.000 &  1.000   \\
        \bottomrule
        \multirow{ 7}{*}{\( p=0.1\)}   &  0.050 &  0.415 &  0.651 &  0.890 &  0.997   \\
                                        &  0.090 &  0.075 &  0.081 &  0.105 &  0.165   \\
                                        &  0.110 &  0.028 &  0.042 &  0.063 &  0.106   \\
                                        &  0.125 &  0.064 &  0.111 &  0.212 &  0.464   \\
                                        &  0.150 &  0.183 &  0.352 &  0.663 &  0.966   \\
                                        &  0.200 &  0.612 &  0.906 &  0.996 &  1.000   \\
                                        &  0.250 &  0.922 &  0.999 &  1.000 &  1.000   \\
        \bottomrule
        \end{tabularx}\\
        \raggedright
        \scriptsize{ \textit{Notes:} This table displays the finite sample power of the DM test against a forecast based on a wrong quantile for one-step-ahead forecasts evaluated using the FZG score.\(p^{*} \) denotes the wrong quantile of the same distribution as the true quantiel \(p\).}
        \end{table}

\FloatBarrier

\subsection{Additional results: dynamic case, pairwise comparison}\label{app:dynamic_case}

\begin{table}[h!]
        \scriptsize
        \centering
        \caption{Dynamic case, distributional misspecification: GPL with $b=0.5$}
        \begin{tabularx}{1\textwidth}{l X X X X X X X X }
        \toprule 
                &\multicolumn{4}{c}{Power} & \multicolumn{4}{c}{Type III error}   \\ 
        \toprule 
                \(p \backslash P \) & 251 & 500 & 1000 & 2500 & 251 & 500 & 1000 & 2500  \\

        \toprule 
        0.01   &  0.012 &  0.024 &  0.049 &  0.141 &  0.061 &  0.045 &  0.018 &  0.004   \\ 
        0.05   &  0.145 &  0.156 &  0.184 &  0.277 &  0.004 &  0.002 &  0.002 &  0.000   \\ 
        0.1    &  0.236 &  0.305 &  0.448 &  0.760 &  0.002 &  0.000 &  0.000 &  0.000   \\ 
        \bottomrule 
        \end{tabularx}
        \raggedright
        \scriptsize{ \textit{Notes:} This table displays the finite sample power and type III error of the DM test for one-step-ahead forecasts evaluated using the GPL with $b=0.5$.}
\end{table}



\begin{table}[h!]
        \scriptsize
        \centering
        \caption{Dynamic case, distributional misspecification: GPL with $b=2$}
        \begin{tabularx}{1\textwidth}{l X X X X X X X X }
        \toprule 
                &\multicolumn{4}{c}{Power} & \multicolumn{4}{c}{Type III error}   \\ 
        \toprule 
                \(p \backslash P \) & 251 & 500 & 1000 & 2500 & 251 & 500 & 1000 & 2500  \\

        \toprule 
        0.01   &  0.001 &  0.002 &  0.003 &  0.008 &  0.148 &  0.093 &  0.056 &  0.026   \\ 
        0.05   &  0.226 &  0.242 &  0.278 &  0.329 &  0.000 &  0.000 &  0.000 &  0.000   \\ 
        0.1    &  0.326 &  0.387 &  0.480 &  0.619 &  0.000 &  0.000 &  0.000 &  0.000   \\ 
        \bottomrule 
        \end{tabularx}
        \raggedright
        \scriptsize{ \textit{Notes:} This table displays the finite sample power and type III error of the DM test for one-step-ahead forecasts evaluated using the GPL with $b=2$.}
\end{table}

\begin{table}[h!]
\scriptsize
\centering
\caption{Dynamic case, distributional misspecification: AL score}
\begin{tabularx}{1\textwidth}{l X X X X X X X X }
\toprule 
        &\multicolumn{4}{c}{Power} & \multicolumn{4}{c}{Type III error}   \\ 
\toprule 
        \(p \backslash P \) & 251 & 500 & 1000 & 2500 & 251 & 500 & 1000 & 2500  \\

\toprule 
0.01    &  0.014 &  0.038 &  0.096 &  0.326 &  0.071 &  0.040 &  0.010 &  0.001   \\ 
0.025   &  0.009 &  0.027 &  0.066 &  0.242 &  0.106 &  0.053 &  0.016 &  0.002   \\ 
0.05    &  0.196 &  0.254 &  0.408 &  0.764 &  0.003 &  0.001 &  0.000 &  0.000   \\ 
0.1     &  0.218 &  0.294 &  0.458 &  0.798 &  0.002 &  0.001 &  0.000 &  0.000   \\ 
\bottomrule 
\end{tabularx}
\raggedright
\scriptsize{ \textit{Notes:} This table displays the finite sample power and type III error of the DM test for one-step-ahead forecasts evaluated using the AL score.}
\end{table}

\begin{table}[h!]
\scriptsize
\centering
\caption{Dynamic case, distributional misspecification: NZ score}
\begin{tabularx}{1\textwidth}{l X X X X X X X X }
\toprule 
        &\multicolumn{4}{c}{Power} & \multicolumn{4}{c}{Type III error}   \\ 
\toprule 
        \(p \backslash P \) & 251 & 500 & 1000 & 2500 & 251 & 500 & 1000 & 2500  \\

\toprule 
0.01    &  0.012 &  0.031 &  0.071 &  0.249 &  0.079 &  0.042 &  0.012 &  0.002   \\ 
0.025   &  0.008 &  0.020 &  0.051 &  0.202 &  0.116 &  0.060 &  0.017 &  0.002   \\ 
0.05    &  0.200 &  0.237 &  0.328 &  0.567 &  0.002 &  0.001 &  0.000 &  0.000   \\ 
0.1     &  0.242 &  0.315 &  0.473 &  0.782 &  0.002 &  0.000 &  0.000 &  0.000   \\ 
\bottomrule 
\end{tabularx}
\raggedright
\scriptsize{ \textit{Notes:} This table displays the finite sample power and type III error of the DM test for one-step-ahead forecasts evaluated using the NZ score.}
\end{table}




\begin{table}[h!]
        \scriptsize
        \centering
        \caption{Dynamic case, volatility misspecification: GPL with $b=0.5$}
        \begin{tabularx}{1\textwidth}{l X X X X X X X X }
        \toprule 
                &\multicolumn{4}{c}{Power} & \multicolumn{4}{c}{Type III error}   \\ 
        \toprule 
                \(p \backslash P \) & 251 & 500 & 1000 & 2500 & 251 & 500 & 1000 & 2500  \\

        \toprule 
        0.01    &  0.012 &  0.022 &  0.048 &  0.133 &  0.061 &  0.046 &  0.022 &  0.005   \\ 
        0.025   &  0.007 &  0.007 &  0.011 &  0.013 &  0.103 &  0.074 &  0.066 &  0.048   \\ 
        0.05    &  0.151 &  0.149 &  0.188 &  0.286 &  0.004 &  0.002 &  0.002 &  0.000   \\ 
        0.1     &  0.235 &  0.304 &  0.460 &  0.759 &  0.001 &  0.000 &  0.000 &  0.000   \\ 
        \bottomrule 
        \end{tabularx}
        \raggedright
        \scriptsize{ \textit{Notes:} This table displays the finite sample power and type III error of the DM test for one-step-ahead forecasts evaluated using the GPL with $b=0.5$.}
\end{table}


\begin{table}[h!]
        \scriptsize
        \centering
        \caption{Dynamic case, volatility misspecification: GPL with $b=2$}
        \begin{tabularx}{1\textwidth}{l X X X X X X X X }
        \toprule 
                &\multicolumn{4}{c}{Power} & \multicolumn{4}{c}{Type III error}   \\ 
        \toprule 
                \(p \backslash P \) & 251 & 500 & 1000 & 2500 & 251 & 500 & 1000 & 2500  \\

        \toprule 
        0.01    &  0.001 &  0.001 &  0.002 &  0.009 &  0.152 &  0.104 &  0.066 &  0.027   \\ 
        0.025   &  0.001 &  0.001 &  0.001 &  0.002 &  0.173 &  0.149 &  0.140 &  0.122   \\ 
        0.05    &  0.236 &  0.246 &  0.286 &  0.343 &  0.000 &  0.000 &  0.000 &  0.000   \\ 
        0.1     &  0.322 &  0.390 &  0.493 &  0.620 &  0.000 &  0.000 &  0.000 &  0.000   \\ 
        \bottomrule 
        \end{tabularx}
        \raggedright
        \scriptsize{ \textit{Notes:} This table displays the finite sample power and type III error of the DM test for one-step-ahead forecasts evaluated using the GPL with $b=2$.}
\end{table}

\begin{table}[h!]
        \scriptsize
        \centering
        \caption{Dynamic case, volatility misspecification: AL score}
        \begin{tabularx}{1\textwidth}{l X X X X X X X X }
        \toprule 
                &\multicolumn{4}{c}{Power} & \multicolumn{4}{c}{Type III error}   \\ 
        \toprule 
                \(p \backslash P \) & 251 & 500 & 1000 & 2500 & 251 & 500 & 1000 & 2500  \\

        \toprule 
        0.01    &  0.013 &  0.038 &  0.102 &  0.330 &  0.075 &  0.039 &  0.013 &  0.001   \\ 
        0.025   &  0.009 &  0.026 &  0.072 &  0.247 &  0.110 &  0.051 &  0.019 &  0.002   \\ 
        0.05    &  0.188 &  0.257 &  0.414 &  0.764 &  0.002 &  0.001 &  0.000 &  0.000   \\ 
        0.1     &  0.220 &  0.287 &  0.453 &  0.790 &  0.001 &  0.000 &  0.000 &  0.000   \\ 
        \bottomrule 
        \end{tabularx}
        \raggedright
        \scriptsize{ \textit{Notes:} This table displays the finite sample power and type III error of the DM test for one-step-ahead forecasts evaluated using the AL score.}
\end{table}

\begin{table}[h!]
        \scriptsize
        \centering
        \caption{Dynamic case, volatility misspecification: NZ score}
        \begin{tabularx}{1\textwidth}{l X X X X X X X X }
        \toprule 
                &\multicolumn{4}{c}{Power} & \multicolumn{4}{c}{Type III error}   \\ 
        \toprule 
                \(p \backslash P \) & 251 & 500 & 1000 & 2500 & 251 & 500 & 1000 & 2500  \\

        \toprule 
        0.01    &  0.013 &  0.029 &  0.076 &  0.248 &  0.082 &  0.041 &  0.014 &  0.002   \\ 
        0.025   &  0.007 &  0.019 &  0.054 &  0.201 &  0.123 &  0.058 &  0.022 &  0.004   \\ 
        0.05    &  0.200 &  0.241 &  0.326 &  0.565 &  0.001 &  0.000 &  0.000 &  0.000   \\ 
        0.1     &  0.241 &  0.309 &  0.465 &  0.776 &  0.001 &  0.000 &  0.000 &  0.000   \\ 
        \bottomrule 
        \end{tabularx}
        \raggedright
        \scriptsize{ \textit{Notes:} This table displays the finite sample power and type III error of the DM test for one-step-ahead forecasts evaluated using the NZ score.}
\end{table}



\FloatBarrier
\clearpage
\begin{figure}[!htb]
        \includegraphics[width=\linewidth]{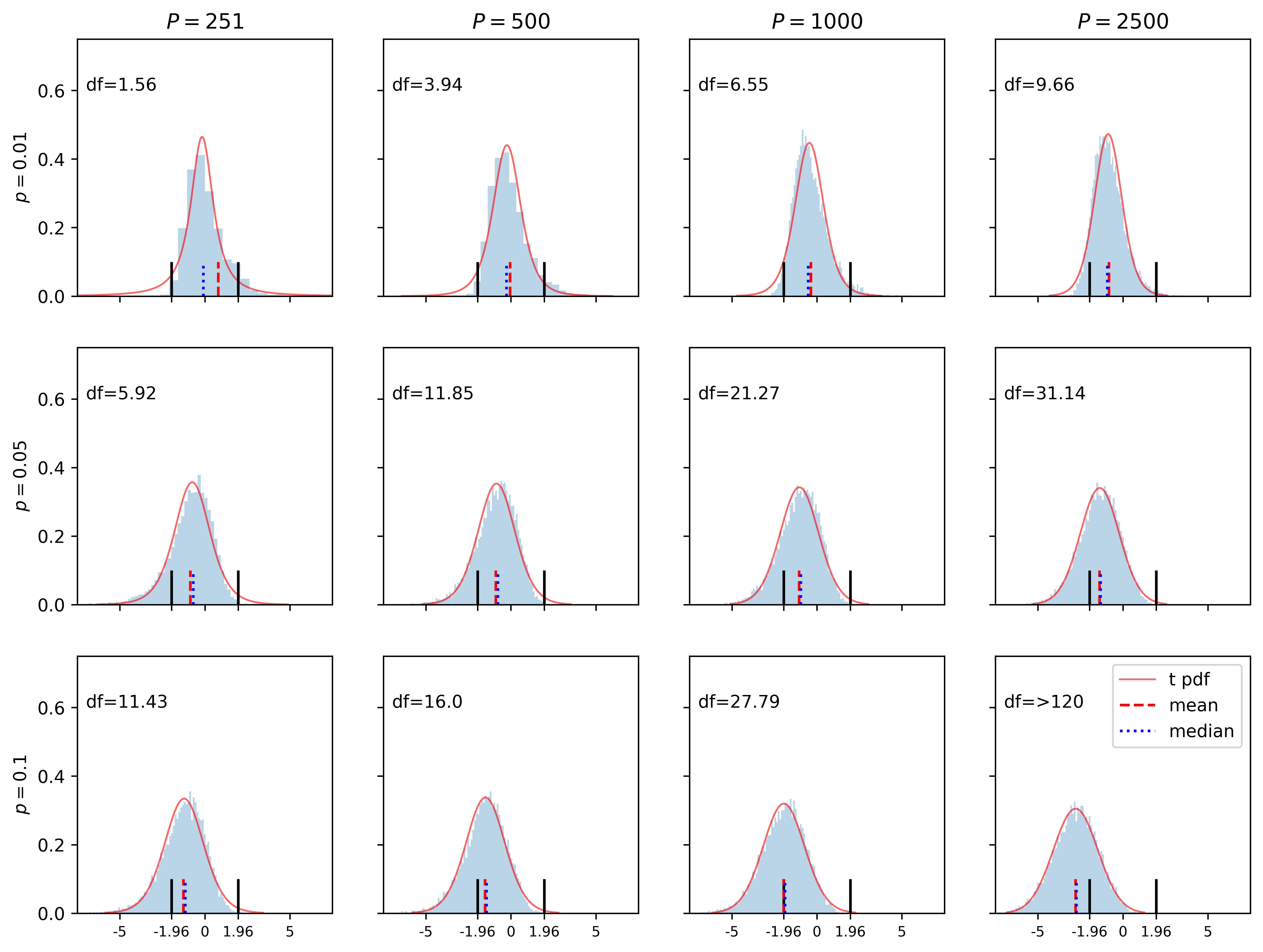}
        \caption{Density histogram of \(t_{ij}\): tick loss, bootstrapped variance}
        \footnotesize{\textit{This figure displays density histograms of \(t_{ij} \) for one-step-ahead forecasts evaluated using the tick loss, where model \(i\) is the true model as in Equation \ref{eq:dgp} while model \(j\) assumes normally distributed innovations. We standardize by the bootstrapped variance of \( \bar{d}_{ij}\).}}
\end{figure}

\begin{figure}[!htb]
        \includegraphics[width=\linewidth]{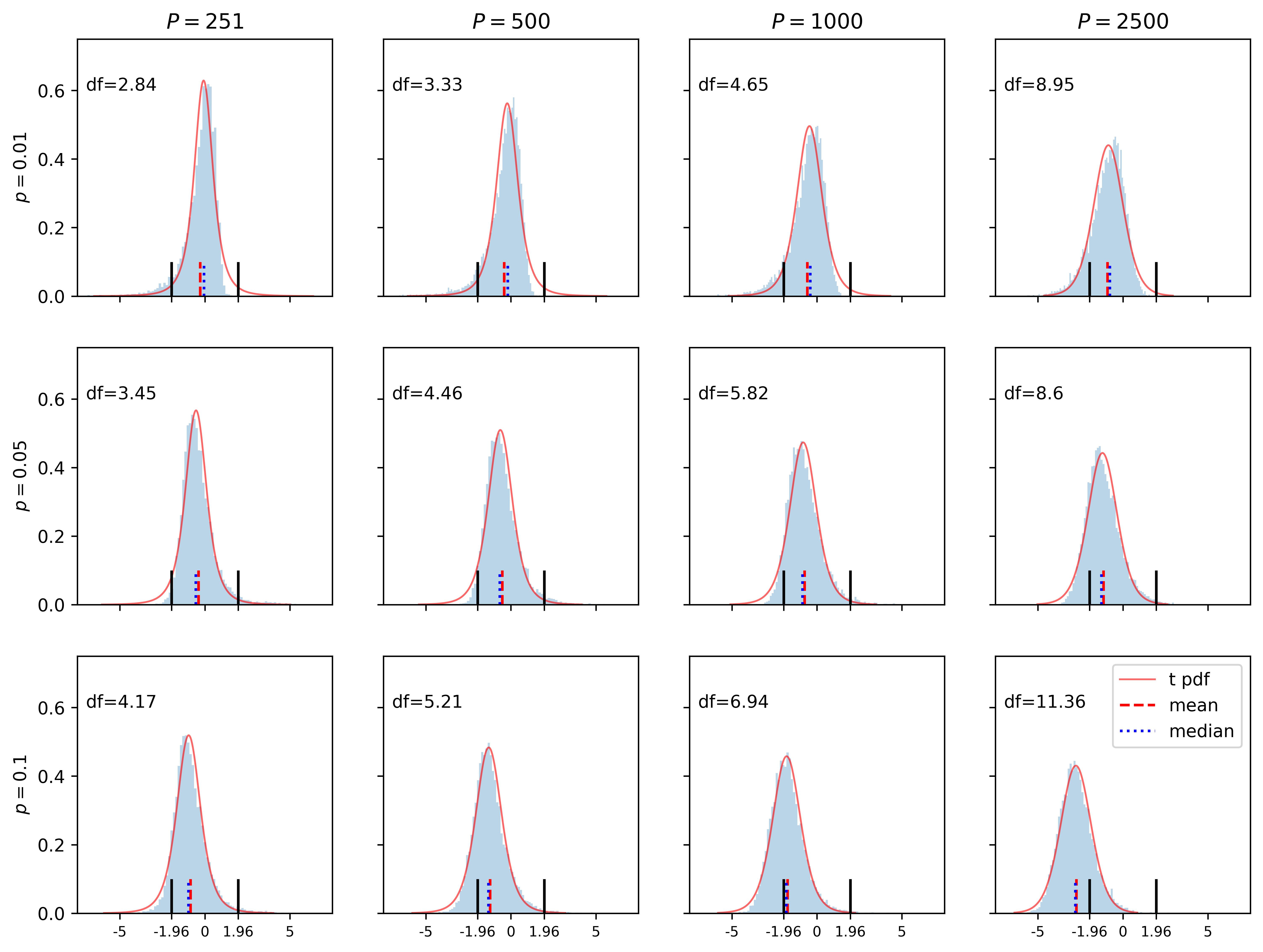}
        \caption{Density histogram of \(t_{ij}\): tick loss, simulated variance}
        \footnotesize{\textit{This figure displays density histograms of \(t_{ij} \) for one-step-ahead forecasts evaluated using the tick loss, where model \(i\) is the true model as in Equation \ref{eq:dgp} while model \(j\) assumes normally distributed innovations. Crucially, we standardize by the simulated variance of \( \bar{d}_{ij}\), which we consider correct.}}
\end{figure}

\begin{figure}[!htb]
        \includegraphics[width=\linewidth]{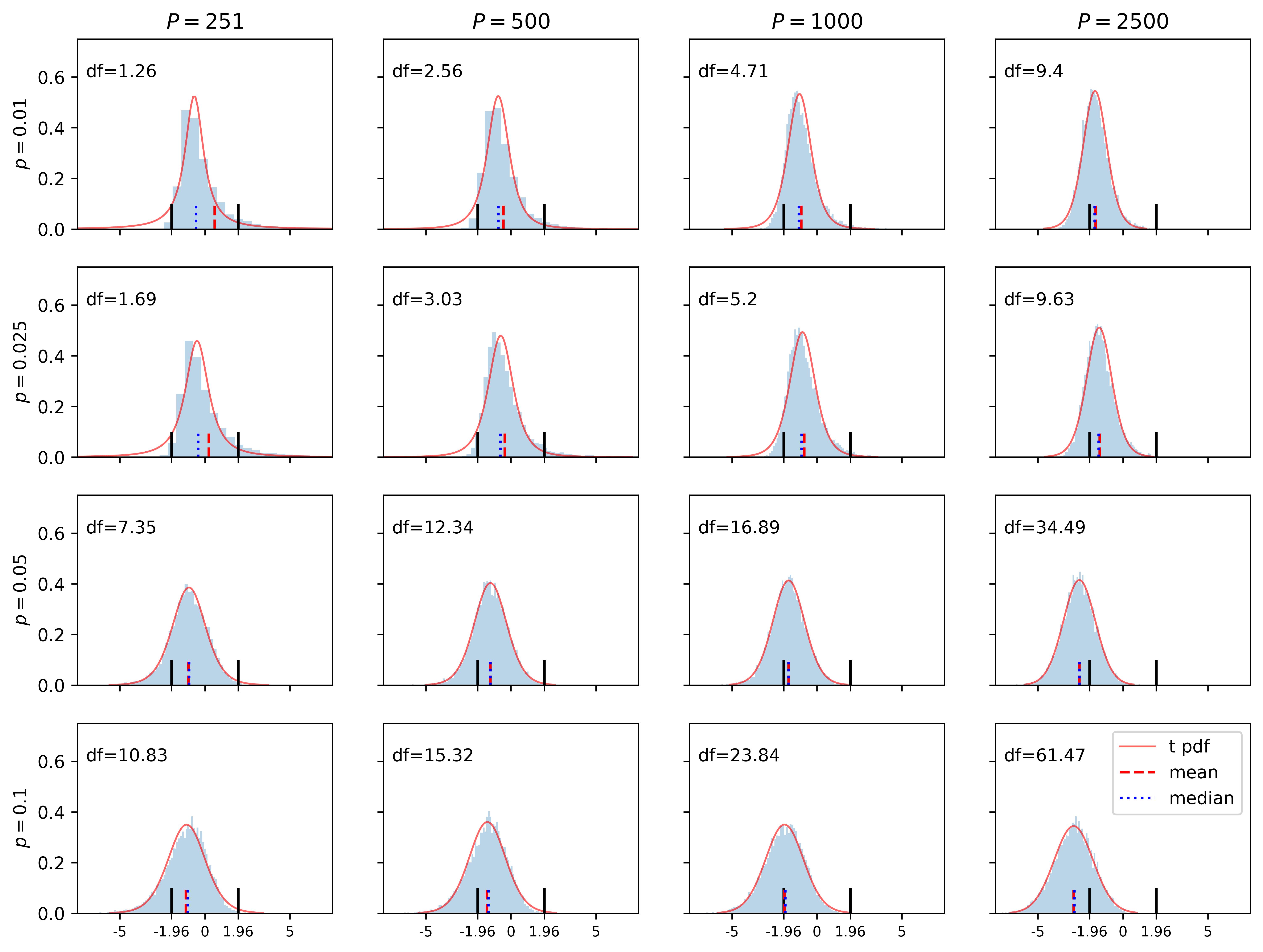}
        \caption{Density histogram of \(t_{ij}\): FZG score, bootstrapped variance}
        \footnotesize{\textit{This figure displays density histograms of \(t_{ij} \) for one-step-ahead forecasts evaluated using the FZG score, where model \(i\) is the true model as in Equation \ref{eq:dgp} while model \(j\) is a competitor with a wrong volatility misspecification.}}
\end{figure}

\begin{figure}[!htb]
        \includegraphics[width=\linewidth]{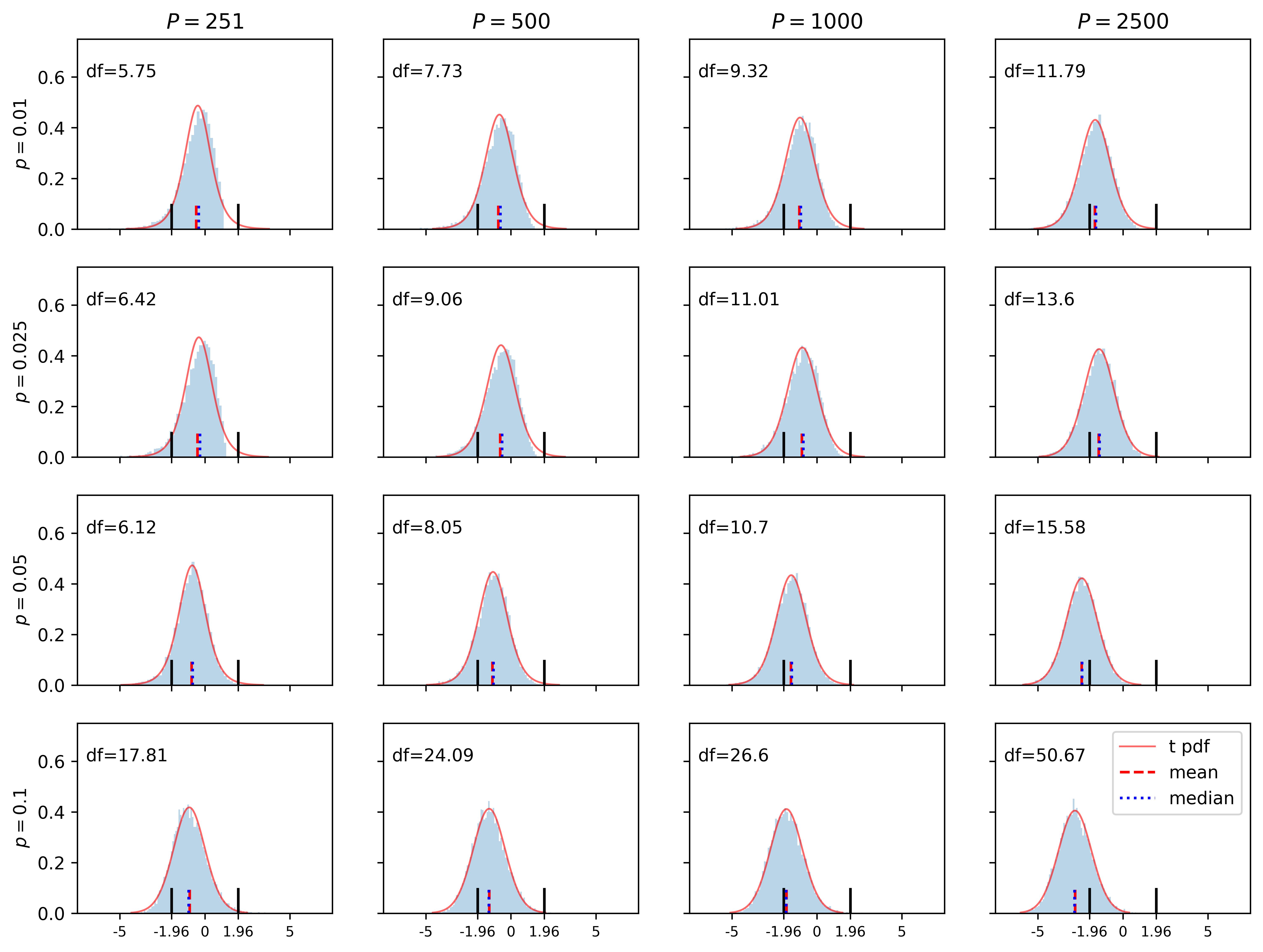}
        \caption{Density histogram of \(t_{ij}\): FZG score, simulated variance}
        \footnotesize{\textit{This figure displays density histograms of \(t_{ij} \) for one-step-ahead forecasts evaluated using the FZG score, where model \(i\) is the true model as in Equation \ref{eq:dgp} while model \(j\) is a competitor with a wrong volatility misspecification.}}
\end{figure}

\clearpage
\FloatBarrier

\subsection{Additional results - set of \( m=5\) models} \label{app:detail_results_m5}
This section provides power and potency of the MCS test for the first set of models that we describe in Section \ref{sec:simulations_mcs}. 

\subsubsection{Results for \( T_{R, \mathcal{M}}\)} \label{app:t_r_m}

\begin{figure}[!ht]
        \includegraphics[width=1\linewidth]{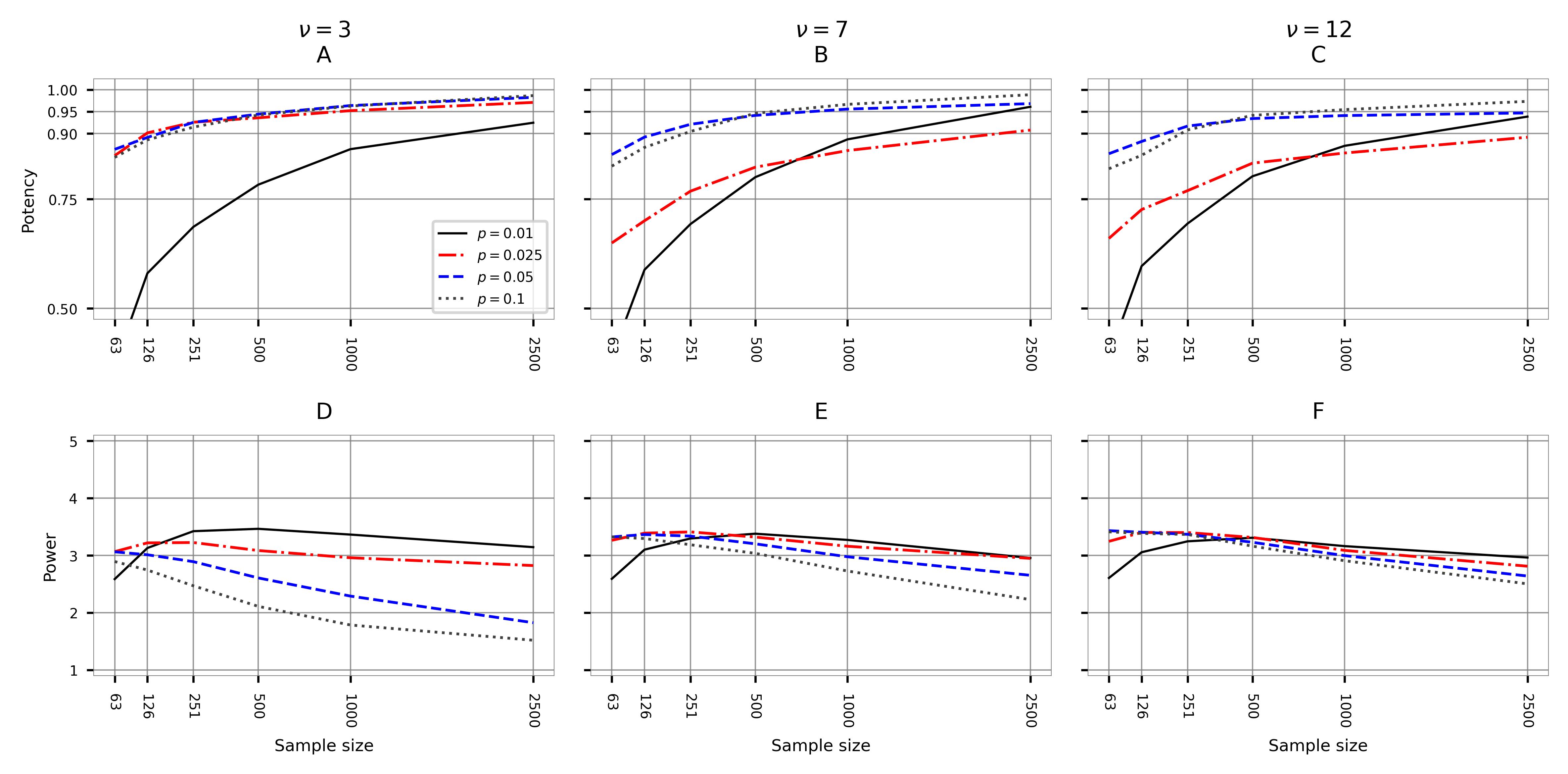}
        \caption{Finite sample properties for the tick loss}
        \label{fig:true_model_var_t_ij}
        \footnotesize{This figure displays the finite sample properties of the MCS procedure using \( T_{R, \mathcal{M}}\) in the following setting: one-day-ahead VaR forecasts evaluated using the tick loss, true model included, number of models \(m=5\), level of the test \( \alpha=0.25 \). The upper row displays the potency, i.e.\ the frequency of \( \mathcal{M}^{*} \subset \widehat{\mathcal{M}}_{1-\alpha}^{*} \). The lower row displays the power property, i.e.\ the average number of elements in \(\mathcal{ \widehat{M}}_{1-\alpha}^{*} \).}
\end{figure}

\begin{figure}[!ht]
        \includegraphics[width=1\linewidth]{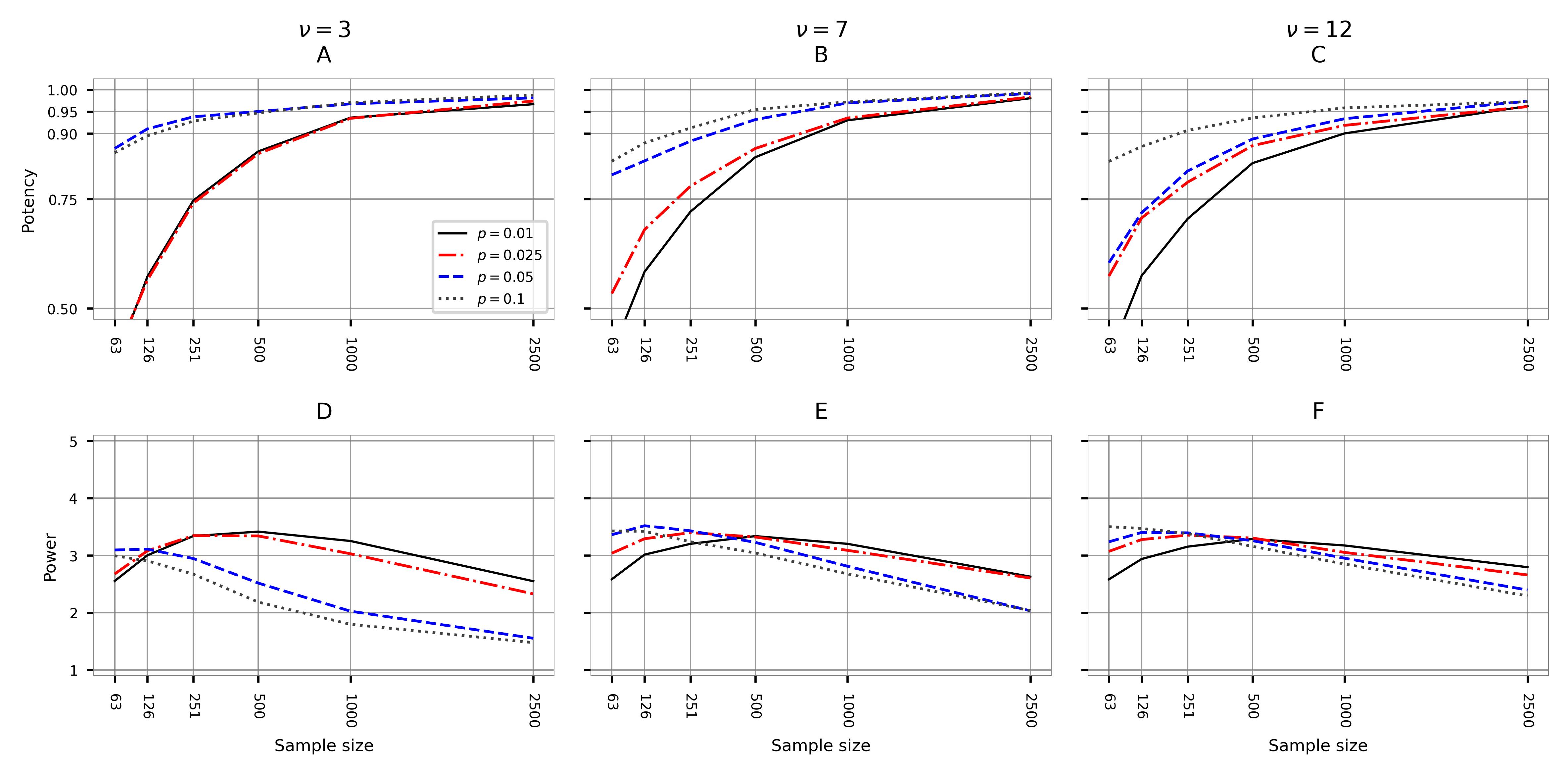}
        \caption{Finite sample properties for the FZG score}
        \label{fig:true_model_es_t_ij}
        \footnotesize{This figure displays the finite sample properties of the MCS procedure using \( T_{R, \mathcal{M}}\) in the following setting: one-day-ahead VaR and ES forecasts evaluated using the FZG score, true model included, number of models \(m=5\), level of the test \( \alpha=0.25 \). The upper panel displays the potency, i.e.\  the frequency of \( \mathcal{M}^{*} \subset \widehat{\mathcal{M}}_{1-\alpha}^{*} \). The lower panel displays the power property, i.e.\  the average number of elements in \(\mathcal{ \widehat{M}}_{1-\alpha}^{*} \).}
\end{figure}
\FloatBarrier

\subsection{Results for varying level of the test \( \alpha \) }
\begin{landscape}
\begin{figure}[!ht]
        \includegraphics[width=20cm, keepaspectratio]{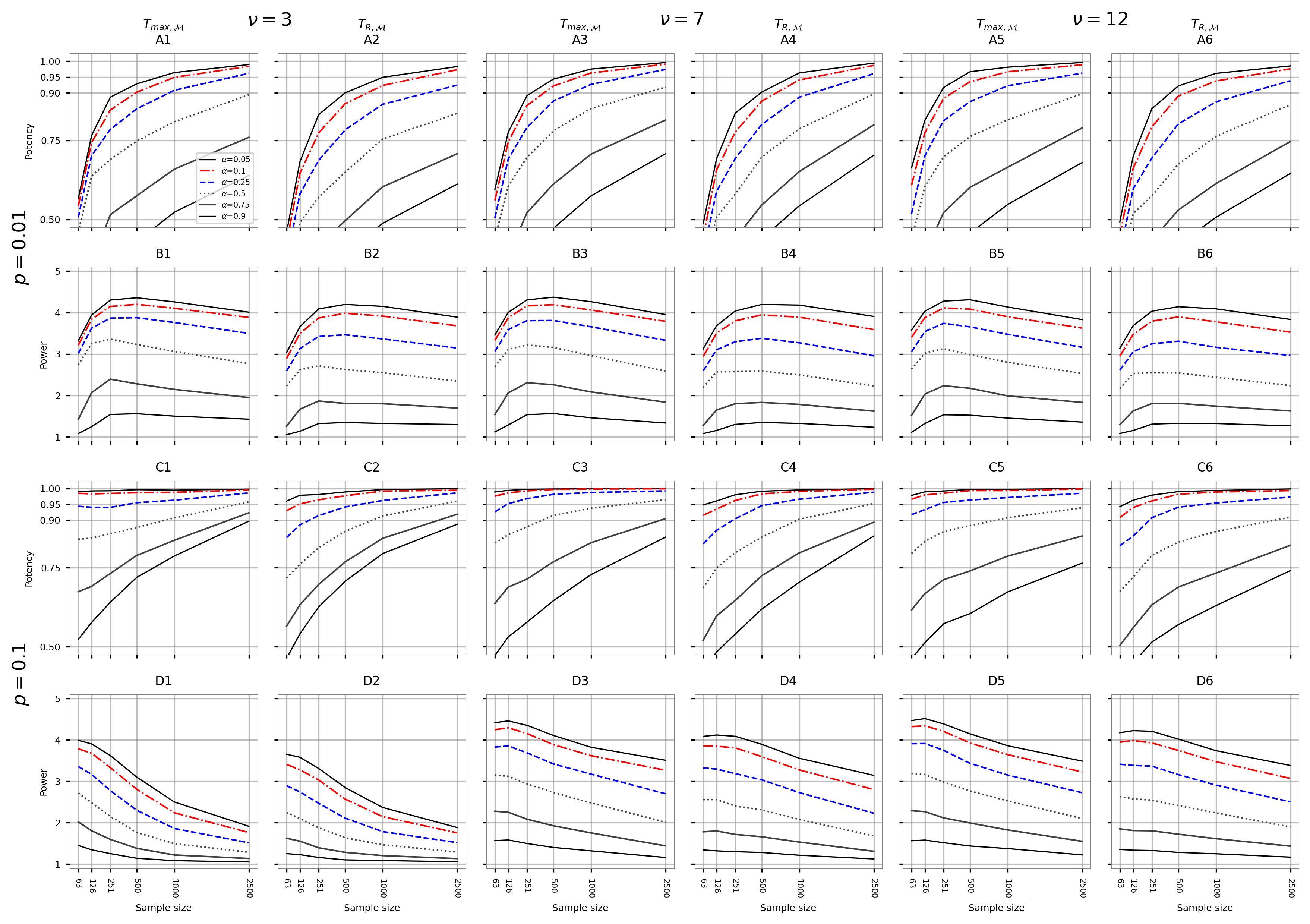}
        \caption{Varying the level of the test \(\alpha\) for the tick loss}
        \label{fig:true_model_var_varying_alpha}
        \footnotesize{This figure displays the finite sample properties of the MCS procedure for different levels of the test \( \alpha\) in the following setting: one-day-ahead VaR forecasts evaluated using the tick loss, true model included, number of models \(m=5\). Columns 1 and 2 present the results for the DGP with \( \nu=3 \), columns 3 and 4 for the DGP with \( \nu=7 \) and columns 5 and 6 for the DGP with \( \nu=12 \). Within each DGP, the two columns show the results for  \( T_{max, \mathcal{M}} \) and \( T_{R, \mathcal{M}}\), respectively. The two upper rows display the results for the quantile level \(p=0.01\) while the two lower rows show the results for the quantile level \(p=0.1\).}
\end{figure}
\begin{figure}[!ht]
        \includegraphics[width=20cm, keepaspectratio]{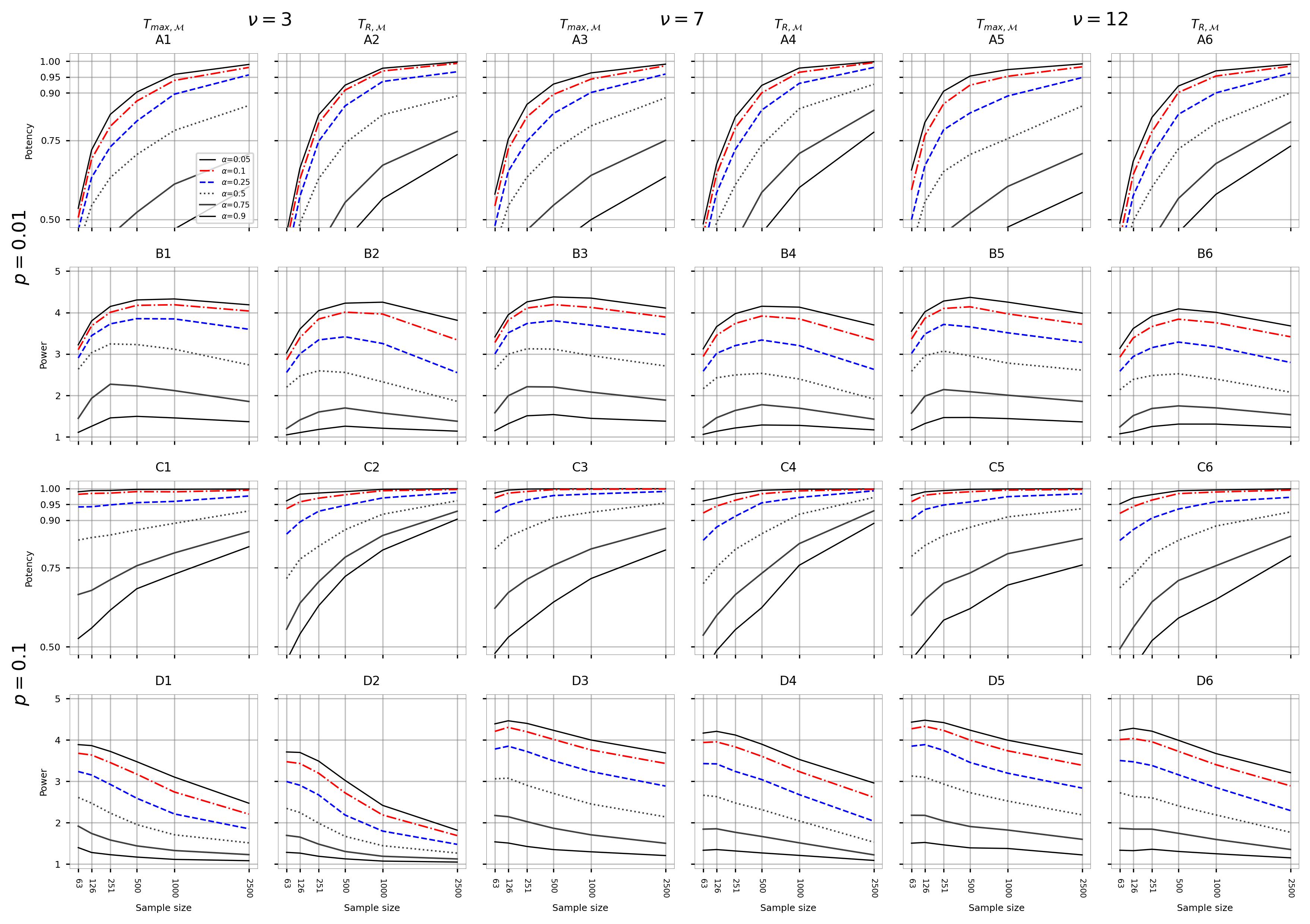}
        \caption{Varying the level of the test \(\alpha\) for the FZG score}
        \label{fig:true_model_es_varying_alpha}
        \footnotesize{This figure displays the finite sample properties of the MCS procedure for different levels of the test \( \alpha\) in the following setting: one-day-ahead VaR and ES forecasts evaluated using the FZG score, true model included, number of models \(m=5\). Columns 1 and 2 present the results for the DGP with \( \nu=3 \), columns 3 and 4 for the DGP with \( \nu=7 \) and columns 5 and 6 for the DGP with \( \nu=12 \). Within each DGP, the two columns show the results for  \( T_{max, \mathcal{M}} \) and \( T_{R, \mathcal{M}}\), respectively. The two upper rows display the results for the quantile level \(p=0.01\) while the two lower rows show the results for the quantile level \(p=0.1\).}
\end{figure}

\end{landscape}

\FloatBarrier


\subsubsection{Results for \( T_{max, \mathcal{M}} \) and differently parametrized loss functions}
\begin{table}[h!]
\scriptsize
\centering
\caption{Simulation results for the set of \(m=5\) models: VaR forecasts evaluated using the GPL loss with \(b=0.5\).}
\label{tab:first}
\begin{tabularx}{1\textwidth}{l X X X X X X X X X X X X X }
\toprule 
        &  & \multicolumn{6}{c}{Potency} & \multicolumn{6}{c}{Power}   \\ 
                               &\(p \backslash P \) & 63 & 126 & 251 & 500 & 1000 & 2500 & 63 & 126 & 251 & 500 & 1000 & 2500   \\ 
\toprule 
\multirow{ 4}{*}{\( \nu=3\)}   &  0.010 &  0.503 &  0.699 &  0.798 &  0.879 &  0.926 &  0.968 &  3.018 &  3.636 &  3.911 &  3.974 &  3.813 &  3.482   \\ 
                               &  0.025 &  0.947 &  0.972 &  0.978 &  0.976 &  0.983 &  0.990 &  3.485 &  3.716 &  3.692 &  3.553 &  3.360 &  3.182   \\ 
                               &  0.050 &  0.963 &  0.970 &  0.975 &  0.976 &  0.980 &  0.988 &  3.580 &  3.543 &  3.340 &  3.018 &  2.626 &  1.976   \\ 
                               &  0.100 &  0.947 &  0.941 &  0.942 &  0.952 &  0.963 &  0.990 &  3.416 &  3.201 &  2.766 &  2.260 &  1.806 &  1.461   \\ 
\bottomrule 
\multirow{ 4}{*}{\( \nu=7\)}   &  0.010 &  0.525 &  0.702 &  0.823 &  0.882 &  0.933 &  0.982 &  3.127 &  3.649 &  3.896 &  3.876 &  3.650 &  3.322   \\ 
                               &  0.025 &  0.766 &  0.827 &  0.893 &  0.906 &  0.919 &  0.938 &  3.673 &  3.824 &  3.797 &  3.502 &  3.248 &  2.924   \\ 
                               &  0.050 &  0.928 &  0.944 &  0.962 &  0.965 &  0.972 &  0.982 &  3.870 &  3.895 &  3.734 &  3.425 &  3.187 &  2.736   \\ 
                               &  0.100 &  0.928 &  0.949 &  0.967 &  0.974 &  0.987 &  0.996 &  3.856 &  3.846 &  3.680 &  3.391 &  3.146 &  2.628   \\ 
\bottomrule 
\multirow{ 4}{*}{\( \nu=12\)}   &  0.010 &  0.514 &  0.691 &  0.816 &  0.881 &  0.934 &  0.971 &  3.063 &  3.537 &  3.752 &  3.670 &  3.493 &  3.200   \\ 
                                &  0.025 &  0.781 &  0.838 &  0.890 &  0.904 &  0.921 &  0.930 &  3.652 &  3.788 &  3.718 &  3.447 &  3.169 &  2.801   \\ 
                                &  0.050 &  0.908 &  0.928 &  0.949 &  0.952 &  0.956 &  0.955 &  3.812 &  3.835 &  3.662 &  3.365 &  3.084 &  2.610   \\ 
                                &  0.100 &  0.913 &  0.935 &  0.955 &  0.966 &  0.976 &  0.985 &  3.947 &  3.918 &  3.760 &  3.483 &  3.173 &  2.663   \\ 
\bottomrule 
\end{tabularx}
\raggedright
\scriptsize{ \textit{Notes:} This table displays the finite sample properties of the MCS for VaR forecasts evaluated using the GPL loss with \(b=0.5\) and the set of \(m=5\) models. The level of the test is \( \alpha \)=0.25, the number of simulations is 2,500. Potency is the frequency at which \( \mathcal{M}^{*} \subset \widehat{\mathcal{M}}^{*}_{1-\alpha} \), power is defined as the average of \(|\widehat{\mathcal{M}}^{*}_{1-\alpha}|\).}
\end{table}

\begin{table}[h!]
\scriptsize
\centering
\caption{Simulation results for the set of \(m=5\) models: VaR forecasts evaluated using the tick loss.}
\label{tab:first}
\begin{tabularx}{1\textwidth}{l X X X X X X X X X X X X X }
\toprule 
        &  & \multicolumn{6}{c}{Potency} & \multicolumn{6}{c}{Power}   \\ 
                               &\(p \backslash P \) & 63 & 126 & 251 & 500 & 1000 & 2500 & 63 & 126 & 251 & 500 & 1000 & 2500   \\ 
\toprule 
\multirow{ 4}{*}{\( \nu=3\)}   &  0.010 &  0.504 &  0.689 &  0.784 &  0.846 &  0.905 &  0.964 &  3.004 &  3.601 &  3.883 &  3.894 &  3.764 &  3.504   \\ 
                               &  0.025 &  0.943 &  0.973 &  0.976 &  0.977 &  0.981 &  0.987 &  3.458 &  3.680 &  3.661 &  3.499 &  3.359 &  3.209   \\ 
                               &  0.050 &  0.956 &  0.970 &  0.975 &  0.974 &  0.977 &  0.987 &  3.486 &  3.485 &  3.281 &  2.974 &  2.625 &  2.105   \\ 
                               &  0.100 &  0.943 &  0.947 &  0.948 &  0.947 &  0.962 &  0.987 &  3.356 &  3.174 &  2.800 &  2.308 &  1.890 &  1.514   \\ 
\bottomrule 
\multirow{ 4}{*}{\( \nu=7\)}   &  0.010 &  0.519 &  0.692 &  0.805 &  0.886 &  0.937 &  0.975 &  3.091 &  3.608 &  3.823 &  3.821 &  3.659 &  3.347   \\ 
                               &  0.025 &  0.764 &  0.831 &  0.879 &  0.901 &  0.911 &  0.926 &  3.618 &  3.812 &  3.760 &  3.494 &  3.265 &  2.930   \\ 
                               &  0.050 &  0.928 &  0.943 &  0.952 &  0.962 &  0.969 &  0.977 &  3.811 &  3.841 &  3.706 &  3.435 &  3.192 &  2.787   \\ 
                               &  0.100 &  0.937 &  0.949 &  0.970 &  0.978 &  0.985 &  0.994 &  3.834 &  3.805 &  3.636 &  3.403 &  3.136 &  2.715   \\ 
\bottomrule 
\multirow{ 4}{*}{\( \nu=12\)}   &  0.010 &  0.533 &  0.699 &  0.812 &  0.883 &  0.916 &  0.966 &  3.088 &  3.530 &  3.756 &  3.714 &  3.494 &  3.201   \\ 
                                &  0.025 &  0.768 &  0.842 &  0.884 &  0.889 &  0.906 &  0.933 &  3.625 &  3.805 &  3.733 &  3.443 &  3.190 &  2.788   \\ 
                                &  0.050 &  0.900 &  0.928 &  0.938 &  0.938 &  0.953 &  0.950 &  3.730 &  3.815 &  3.650 &  3.328 &  3.049 &  2.657   \\ 
                                &  0.100 &  0.908 &  0.930 &  0.952 &  0.964 &  0.970 &  0.984 &  3.861 &  3.866 &  3.753 &  3.442 &  3.162 &  2.754   \\ 
\bottomrule 
\end{tabularx}
\raggedright
\scriptsize{ \textit{Notes:} This table displays the finite sample properties of the MCS for VaR forecasts evaluated using the tick loss and the set of \(m=5\) models. The level of the test is \( \alpha \)=0.25, the number of simulations is 2,500. Potency is the frequency at which \( \mathcal{M}^{*} \subset \widehat{\mathcal{M}}^{*}_{1-\alpha} \), power is defined as the average of \(|\widehat{\mathcal{M}}^{*}_{1-\alpha}|\).}
\end{table}

\clearpage
\begin{table}[h!]
\scriptsize
\centering
\caption{Simulation results for the set of \(m=5\) models: VaR forecasts evaluated using the GPL loss with \(b=2\).}
\label{tab:first}
\begin{tabularx}{1\textwidth}{l X X X X X X X X X X X X X }
\toprule 
        &  & \multicolumn{6}{c}{Potency} & \multicolumn{6}{c}{Power}   \\ 
                               &\(p \backslash P \) & 63 & 126 & 251 & 500 & 1000 & 2500 & 63 & 126 & 251 & 500 & 1000 & 2500   \\ 
\toprule 
\multirow{ 4}{*}{\( \nu=3\)}   &  0.010 &  0.478 &  0.655 &  0.746 &  0.815 &  0.844 &  0.910 &  2.921 &  3.503 &  3.761 &  3.818 &  3.655 &  3.555   \\ 
                               &  0.025 &  0.933 &  0.966 &  0.972 &  0.971 &  0.978 &  0.982 &  3.364 &  3.623 &  3.629 &  3.513 &  3.322 &  3.194   \\ 
                               &  0.050 &  0.952 &  0.961 &  0.958 &  0.964 &  0.966 &  0.980 &  3.411 &  3.387 &  3.234 &  3.006 &  2.692 &  2.380   \\ 
                               &  0.100 &  0.935 &  0.928 &  0.933 &  0.944 &  0.948 &  0.973 &  3.238 &  3.087 &  2.814 &  2.520 &  2.134 &  1.797   \\ 
\bottomrule 
\multirow{ 4}{*}{\( \nu=7\)}   &  0.010 &  0.500 &  0.667 &  0.775 &  0.825 &  0.876 &  0.943 &  3.053 &  3.573 &  3.804 &  3.766 &  3.617 &  3.429   \\ 
                               &  0.025 &  0.740 &  0.796 &  0.866 &  0.868 &  0.881 &  0.903 &  3.588 &  3.756 &  3.790 &  3.520 &  3.298 &  3.094   \\ 
                               &  0.050 &  0.912 &  0.934 &  0.954 &  0.954 &  0.965 &  0.983 &  3.779 &  3.844 &  3.754 &  3.469 &  3.254 &  3.003   \\ 
                               &  0.100 &  0.922 &  0.944 &  0.970 &  0.973 &  0.982 &  0.990 &  3.778 &  3.804 &  3.733 &  3.432 &  3.221 &  2.929   \\ 
\bottomrule 
\multirow{ 4}{*}{\( \nu=12\)}   &  0.010 &  0.504 &  0.675 &  0.786 &  0.838 &  0.887 &  0.942 &  3.020 &  3.510 &  3.733 &  3.656 &  3.480 &  3.292   \\ 
                                &  0.025 &  0.749 &  0.823 &  0.870 &  0.882 &  0.894 &  0.908 &  3.578 &  3.768 &  3.735 &  3.508 &  3.258 &  2.987   \\ 
                                &  0.050 &  0.898 &  0.919 &  0.938 &  0.951 &  0.964 &  0.973 &  3.754 &  3.799 &  3.686 &  3.448 &  3.219 &  2.901   \\ 
                                &  0.100 &  0.906 &  0.930 &  0.953 &  0.964 &  0.976 &  0.985 &  3.873 &  3.891 &  3.806 &  3.560 &  3.316 &  2.982   \\ 
\bottomrule 
\end{tabularx}
\raggedright
\scriptsize{ \textit{Notes:} This table displays the finite sample properties of the MCS for VaR forecasts evaluated using the GPL loss with \(b=2\) and the set of \(m=5\) models. The level of the test is \( \alpha \)=0.25, the number of simulations is 2,500. Potency is the frequency at which \( \mathcal{M}^{*} \subset \widehat{\mathcal{M}}^{*}_{1-\alpha} \), power is defined as the average of \(|\widehat{\mathcal{M}}^{*}_{1-\alpha}|\).}
\end{table}

\begin{table}[h!]
\scriptsize
\centering
\caption{Simulation results for the set of \(m=5\) models: Var and ES forecasts evaluated using the AL score.}
\label{tab:first}
\begin{tabularx}{1\textwidth}{l X X X X X X X X X X X X X }
\toprule 
        &  & \multicolumn{6}{c}{Potency} & \multicolumn{6}{c}{Power}   \\ 
                               &\(p \backslash P \) & 63 & 126 & 251 & 500 & 1000 & 2500 & 63 & 126 & 251 & 500 & 1000 & 2500   \\ 
\toprule 
\multirow{ 4}{*}{\( \nu=3\)}   &  0.010 &  0.499 &  0.674 &  0.797 &  0.898 &  0.962 &  0.980 &  3.017 &  3.584 &  3.956 &  4.166 &  4.101 &  3.106   \\ 
                               &  0.025 &  0.660 &  0.790 &  0.870 &  0.938 &  0.975 &  0.982 &  3.398 &  3.747 &  3.876 &  3.869 &  3.540 &  2.633   \\ 
                               &  0.050 &  0.958 &  0.979 &  0.984 &  0.984 &  0.978 &  0.983 &  3.607 &  3.655 &  3.484 &  3.028 &  2.433 &  1.705   \\ 
                               &  0.100 &  0.955 &  0.968 &  0.973 &  0.966 &  0.970 &  0.987 &  3.472 &  3.390 &  3.051 &  2.528 &  1.983 &  1.501   \\ 
\bottomrule 
\multirow{ 4}{*}{\( \nu=7\)}   &  0.010 &  0.522 &  0.692 &  0.810 &  0.912 &  0.958 &  0.989 &  3.114 &  3.641 &  3.915 &  3.988 &  3.744 &  3.116   \\ 
                               &  0.025 &  0.682 &  0.798 &  0.867 &  0.930 &  0.960 &  0.984 &  3.542 &  3.836 &  3.853 &  3.705 &  3.437 &  3.047   \\ 
                               &  0.050 &  0.901 &  0.926 &  0.941 &  0.964 &  0.976 &  0.994 &  3.861 &  3.930 &  3.754 &  3.427 &  3.076 &  2.516   \\ 
                               &  0.100 &  0.916 &  0.945 &  0.967 &  0.979 &  0.985 &  0.997 &  3.901 &  3.920 &  3.733 &  3.407 &  3.059 &  2.496   \\ 
\bottomrule 
\multirow{ 4}{*}{\( \nu=12\)}   &  0.010 &  0.527 &  0.682 &  0.805 &  0.893 &  0.940 &  0.978 &  3.088 &  3.523 &  3.767 &  3.768 &  3.592 &  3.193   \\ 
                                &  0.025 &  0.729 &  0.834 &  0.891 &  0.925 &  0.942 &  0.968 &  3.596 &  3.817 &  3.779 &  3.563 &  3.278 &  2.823   \\ 
                                &  0.050 &  0.884 &  0.879 &  0.908 &  0.930 &  0.953 &  0.980 &  3.790 &  3.820 &  3.639 &  3.322 &  2.980 &  2.461   \\ 
                                &  0.100 &  0.910 &  0.924 &  0.948 &  0.957 &  0.973 &  0.984 &  3.876 &  3.829 &  3.679 &  3.345 &  2.999 &  2.441   \\ 
\bottomrule 
\end{tabularx}
\raggedright
\scriptsize{ \textit{Notes:} This table displays the finite sample properties of the MCS for Var and ES forecasts evaluated using the AL score and the set of \(m=5\) models. The level of the test is \( \alpha \)=0.25, the number of simulations is 2,500. Potency is the frequency at which \( \mathcal{M}^{*} \subset \widehat{\mathcal{M}}^{*}_{1-\alpha} \), power is defined as the average of \(|\widehat{\mathcal{M}}^{*}_{1-\alpha}|\).}
\end{table}

\begin{table}[h!]
\scriptsize
\centering
\caption{Simulation results for the set of \(m=5\) models: Var and ES forecasts evaluated using the NZ score.}
\label{tab:first}
\begin{tabularx}{1\textwidth}{l X X X X X X X X X X X X X }
\toprule 
        &  & \multicolumn{6}{c}{Potency} & \multicolumn{6}{c}{Power}   \\ 
                               &\(p \backslash P \) & 63 & 126 & 251 & 500 & 1000 & 2500 & 63 & 126 & 251 & 500 & 1000 & 2500   \\ 
\toprule 
\multirow{ 4}{*}{\( \nu=3\)}   &  0.010 &  0.500 &  0.679 &  0.790 &  0.890 &  0.955 &  0.980 &  3.022 &  3.598 &  3.926 &  4.091 &  3.969 &  3.087   \\ 
                               &  0.025 &  0.922 &  0.954 &  0.965 &  0.974 &  0.985 &  0.990 &  3.525 &  3.830 &  3.829 &  3.635 &  3.286 &  2.666   \\ 
                               &  0.050 &  0.956 &  0.976 &  0.980 &  0.981 &  0.977 &  0.985 &  3.549 &  3.572 &  3.365 &  2.991 &  2.489 &  1.843   \\ 
                               &  0.100 &  0.952 &  0.958 &  0.963 &  0.957 &  0.967 &  0.990 &  3.427 &  3.286 &  2.911 &  2.395 &  1.896 &  1.483   \\ 
\bottomrule 
\multirow{ 4}{*}{\( \nu=7\)}   &  0.010 &  0.525 &  0.693 &  0.808 &  0.904 &  0.953 &  0.985 &  3.102 &  3.616 &  3.887 &  3.934 &  3.714 &  3.200   \\ 
                               &  0.025 &  0.709 &  0.817 &  0.872 &  0.924 &  0.949 &  0.979 &  3.571 &  3.833 &  3.811 &  3.614 &  3.353 &  2.999   \\ 
                               &  0.050 &  0.910 &  0.930 &  0.952 &  0.968 &  0.973 &  0.992 &  3.820 &  3.890 &  3.716 &  3.387 &  3.064 &  2.576   \\ 
                               &  0.100 &  0.918 &  0.948 &  0.970 &  0.976 &  0.985 &  0.995 &  3.872 &  3.871 &  3.724 &  3.390 &  3.081 &  2.568   \\ 
\bottomrule 
\multirow{ 4}{*}{\( \nu=12\)}   &  0.010 &  0.529 &  0.687 &  0.805 &  0.886 &  0.931 &  0.976 &  3.084 &  3.514 &  3.746 &  3.731 &  3.548 &  3.191   \\ 
                                &  0.025 &  0.748 &  0.841 &  0.890 &  0.912 &  0.928 &  0.960 &  3.611 &  3.815 &  3.758 &  3.490 &  3.226 &  2.786   \\ 
                                &  0.050 &  0.902 &  0.923 &  0.940 &  0.944 &  0.963 &  0.982 &  3.772 &  3.820 &  3.626 &  3.299 &  2.961 &  2.479   \\ 
                                &  0.100 &  0.911 &  0.928 &  0.952 &  0.962 &  0.970 &  0.981 &  3.880 &  3.848 &  3.704 &  3.366 &  3.041 &  2.540   \\ 
\bottomrule 
\end{tabularx}
\raggedright
\scriptsize{ \textit{Notes:} This table displays the finite sample properties of the MCS for Var and ES forecasts evaluated using the NZ score and the set of \(m=5\) models. The level of the test is \( \alpha \)=0.25, the number of simulations is 2,500. Potency is the frequency at which \( \mathcal{M}^{*} \subset \widehat{\mathcal{M}}^{*}_{1-\alpha} \), power is defined as the average of \(|\widehat{\mathcal{M}}^{*}_{1-\alpha}|\).}
\end{table}

\begin{table}[h!]
\scriptsize
\centering
\caption{Simulation results for the set of \(m=5\) models: Var and ES forecasts evaluated using the FZG score.}
\label{tab:first}
\begin{tabularx}{1\textwidth}{l X X X X X X X X X X X X X }
\toprule 
        &  & \multicolumn{6}{c}{Potency} & \multicolumn{6}{c}{Power}   \\ 
                               &\(p \backslash P \) & 63 & 126 & 251 & 500 & 1000 & 2500 & 63 & 126 & 251 & 500 & 1000 & 2500   \\ 
\toprule 
\multirow{ 4}{*}{\( \nu=3\)}   &  0.010 &  0.462 &  0.617 &  0.732 &  0.809 &  0.886 &  0.954 &  2.892 &  3.411 &  3.746 &  3.862 &  3.838 &  3.606   \\ 
                               &  0.025 &  0.896 &  0.940 &  0.951 &  0.960 &  0.971 &  0.983 &  3.419 &  3.733 &  3.753 &  3.623 &  3.396 &  3.122   \\ 
                               &  0.050 &  0.949 &  0.972 &  0.972 &  0.972 &  0.973 &  0.977 &  3.382 &  3.431 &  3.314 &  3.067 &  2.712 &  2.314   \\ 
                               &  0.100 &  0.938 &  0.948 &  0.949 &  0.951 &  0.955 &  0.979 &  3.227 &  3.155 &  2.923 &  2.604 &  2.195 &  1.861   \\ 
\bottomrule 
\multirow{ 4}{*}{\( \nu=7\)}   &  0.010 &  0.492 &  0.654 &  0.763 &  0.844 &  0.898 &  0.960 &  3.026 &  3.557 &  3.781 &  3.832 &  3.675 &  3.446   \\ 
                               &  0.025 &  0.690 &  0.796 &  0.849 &  0.890 &  0.914 &  0.953 &  3.511 &  3.796 &  3.788 &  3.612 &  3.378 &  3.172   \\ 
                               &  0.050 &  0.902 &  0.923 &  0.946 &  0.956 &  0.966 &  0.983 &  3.731 &  3.851 &  3.750 &  3.471 &  3.211 &  2.929   \\ 
                               &  0.100 &  0.914 &  0.944 &  0.966 &  0.976 &  0.982 &  0.994 &  3.768 &  3.837 &  3.758 &  3.471 &  3.212 &  2.933   \\ 
\bottomrule 
\multirow{ 4}{*}{\( \nu=12\)}   &  0.010 &  0.510 &  0.674 &  0.784 &  0.847 &  0.891 &  0.940 &  3.031 &  3.494 &  3.741 &  3.713 &  3.530 &  3.286   \\ 
                                &  0.025 &  0.726 &  0.824 &  0.874 &  0.875 &  0.901 &  0.924 &  3.547 &  3.798 &  3.774 &  3.509 &  3.287 &  3.000   \\ 
                                &  0.050 &  0.892 &  0.914 &  0.931 &  0.933 &  0.950 &  0.972 &  3.688 &  3.787 &  3.698 &  3.381 &  3.108 &  2.794   \\ 
                                &  0.100 &  0.899 &  0.924 &  0.945 &  0.956 &  0.966 &  0.982 &  3.792 &  3.834 &  3.753 &  3.465 &  3.172 &  2.868   \\ 
\bottomrule 
\end{tabularx}
\raggedright
\scriptsize{ \textit{Notes:} This table displays the finite sample properties of the MCS for Var and ES forecasts evaluated using the FZG score and the set of \(m=5\) models. The level of the test is \( \alpha \)=0.25, the number of simulations is 2,500. Potency is the frequency at which \( \mathcal{M}^{*} \subset \widehat{\mathcal{M}}^{*}_{1-\alpha} \), power is defined as the average of \(|\widehat{\mathcal{M}}^{*}_{1-\alpha}|\).}
\end{table}

\clearpage

\FloatBarrier
\subsection{Additional results - set of \( m=10\) models} \label{app:detail_results_m10}

\begin{table}[h!]
\scriptsize
\centering
\caption{Simulation results for the set of \(m=10\) models: VaR forecasts evaluated using the tick loss.}
\label{tab:first}
\begin{tabularx}{1\textwidth}{l X X X X X X X X X X X X X }
\toprule 
        &  & \multicolumn{6}{c}{Potency} & \multicolumn{6}{c}{Power}   \\ 
                               &\(p \backslash P \) & 63 & 126 & 251 & 500 & 1000 & 2500 & 63 & 126 & 251 & 500 & 1000 & 2500   \\ 
\toprule 
\multirow{ 4}{*}{\( \nu=3\)}   &  0.010 &  0.569 &  0.773 &  0.884 &  0.940 &  0.976 &  0.988 &  6.399 &  8.090 &  8.763 &  8.903 &  8.452 &  6.877   \\ 
                               &  0.025 &  0.856 &  0.934 &  0.960 &  0.987 &  0.996 &  0.997 &  7.544 &  8.405 &  8.623 &  8.518 &  8.020 &  7.040   \\ 
                               &  0.050 &  0.919 &  0.959 &  0.983 &  0.996 &  1.000 &  0.999 &  7.587 &  8.138 &  8.103 &  7.871 &  7.370 &  6.024   \\ 
                               &  0.100 &  0.927 &  0.968 &  0.984 &  0.993 &  0.998 &  0.993 &  7.335 &  7.482 &  7.062 &  6.223 &  4.794 &  2.611   \\ 
\bottomrule 
\multirow{ 4}{*}{\( \nu=7\)}   &  0.010 &  0.523 &  0.747 &  0.872 &  0.944 &  0.981 &  0.994 &  5.893 &  7.518 &  8.384 &  8.611 &  8.310 &  7.544   \\ 
                               &  0.025 &  0.772 &  0.894 &  0.944 &  0.973 &  0.990 &  0.987 &  7.625 &  8.482 &  8.713 &  8.459 &  7.595 &  5.889   \\ 
                               &  0.050 &  0.855 &  0.922 &  0.971 &  0.985 &  0.991 &  0.987 &  8.172 &  8.686 &  8.804 &  8.408 &  7.480 &  5.416   \\ 
                               &  0.100 &  0.884 &  0.947 &  0.973 &  0.989 &  0.994 &  0.997 &  8.464 &  8.751 &  8.785 &  8.401 &  7.527 &  5.628   \\ 
\bottomrule 
\multirow{ 4}{*}{\( \nu=12\)}   &  0.010 &  0.574 &  0.790 &  0.912 &  0.964 &  0.978 &  0.986 &  5.814 &  7.354 &  8.026 &  8.163 &  7.823 &  7.026   \\ 
                                &  0.025 &  0.784 &  0.882 &  0.955 &  0.979 &  0.986 &  0.980 &  7.568 &  8.404 &  8.651 &  8.377 &  7.534 &  5.625   \\ 
                                &  0.050 &  0.846 &  0.919 &  0.958 &  0.980 &  0.974 &  0.967 &  8.227 &  8.686 &  8.693 &  8.145 &  6.858 &  4.600   \\ 
                                &  0.100 &  0.883 &  0.935 &  0.972 &  0.986 &  0.987 &  0.988 &  8.469 &  8.741 &  8.646 &  7.914 &  6.680 &  4.819   \\ 
\bottomrule 
\end{tabularx}
\raggedright
\scriptsize{ \textit{Notes:} This table displays the finite sample properties of the MCS for VaR forecasts evaluated using the tick loss and the set of \(m=10\) models. The level of the test is \( \alpha \)=0.25, the number of simulations is 2,500. Potency is the frequency at which \( \mathcal{M}^{*} \subset \widehat{\mathcal{M}}^{*}_{1-\alpha} \), power is defined as the average of \(|\widehat{\mathcal{M}}^{*}_{1-\alpha}|\).}
\end{table}

\begin{table}[h!]
\scriptsize
\centering
\caption{Simulation results for the first set of models: ES evaluated using the FZG score.}
\label{tab:first}
\begin{tabularx}{1\textwidth}{l X X X X X X X X X X X X X }
\toprule 
        &  & \multicolumn{4}{c}{Potency} & \multicolumn{4}{c}{Power}   \\ 
                               &\(p \backslash P \) & 63 & 126 & 251 & 500 & 1000 & 2500 & 63 & 126 & 251 & 500 & 1000 & 2500   \\ 
\toprule 
\multirow{ 4}{*}{\( \nu=3\)}   &  0.010 &  0.521 &  0.716 &  0.876 &  0.950 &  0.988 &  0.998 &  6.233 &  7.883 &  8.915 &  9.106 &  8.722 &  7.161   \\ 
                               &  0.025 &  0.694 &  0.824 &  0.925 &  0.966 &  0.993 &  0.998 &  7.490 &  8.447 &  9.050 &  9.139 &  8.752 &  6.274   \\ 
                               &  0.050 &  0.871 &  0.936 &  0.980 &  0.992 &  0.998 &  1.000 &  7.896 &  8.568 &  8.857 &  8.748 &  8.172 &  6.260   \\ 
                               &  0.100 &  0.911 &  0.958 &  0.982 &  0.996 &  1.000 &  0.998 &  7.618 &  7.944 &  7.937 &  7.468 &  6.608 &  3.514   \\ 
\bottomrule 
\multirow{ 4}{*}{\( \nu=7\)}   &  0.010 &  0.531 &  0.724 &  0.851 &  0.918 &  0.956 &  0.986 &  5.910 &  7.211 &  8.016 &  8.295 &  8.137 &  7.597   \\ 
                               &  0.025 &  0.766 &  0.881 &  0.933 &  0.961 &  0.980 &  0.981 &  7.600 &  8.418 &  8.685 &  8.479 &  7.875 &  6.434   \\ 
                               &  0.050 &  0.843 &  0.912 &  0.958 &  0.980 &  0.992 &  0.995 &  8.070 &  8.644 &  8.774 &  8.567 &  7.997 &  6.495   \\ 
                               &  0.100 &  0.854 &  0.914 &  0.965 &  0.988 &  0.997 &  0.996 &  8.208 &  8.589 &  8.771 &  8.625 &  8.230 &  7.004   \\ 
\bottomrule 
\multirow{ 4}{*}{\( \nu=12\)}   &  0.010 &  0.583 &  0.788 &  0.893 &  0.944 &  0.965 &  0.977 &  5.679 &  7.066 &  7.629 &  7.661 &  7.277 &  6.290   \\ 
                                &  0.025 &  0.800 &  0.912 &  0.952 &  0.970 &  0.969 &  0.954 &  7.548 &  8.412 &  8.577 &  8.262 &  7.309 &  5.488   \\ 
                                &  0.050 &  0.843 &  0.914 &  0.957 &  0.980 &  0.984 &  0.987 &  8.168 &  8.630 &  8.767 &  8.441 &  7.522 &  5.405   \\ 
                                &  0.100 &  0.862 &  0.920 &  0.963 &  0.985 &  0.988 &  0.993 &  8.335 &  8.698 &  8.764 &  8.424 &  7.605 &  5.613   \\ 
\bottomrule 
\end{tabularx}
\raggedright
\scriptsize{ \textit{Notes:} This table displays the finite sample properties of the MCS for ES forecasts evaluated using the FZG score and the first set of models (crossref?). The level of the test is \( \alpha \)=0.25, the number of simulations is 1,000. Potency is the frequency at which \( \mathcal{M}^{*} \subset \widehat{\mathcal{M}}^{*}_{1-\alpha} \), power is defined as the average of \(|\widehat{\mathcal{M}}^{*}_{1-\alpha}|\).}
\end{table}


\FloatBarrier

\subsection{Plots of \( t_{i \cdot} \) }\label{app:variance_estimates}

\begin{minipage}{.95\linewidth}
        \begin{figure}[H]
                \includegraphics*[width=1\linewidth]{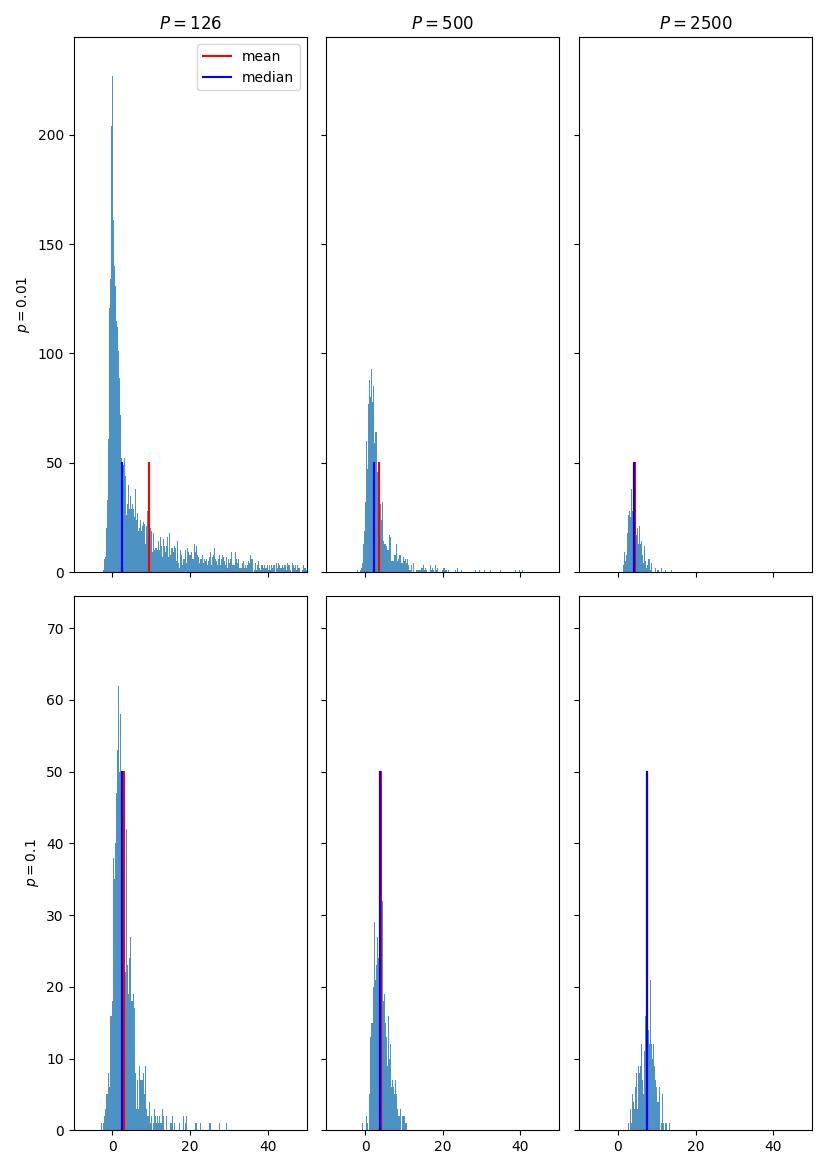}
                \caption{Density plot of \( t_{i\cdot} \)}
                \label{fig:density_t_idot}
                \footnotesize{This figure shows density plots of \( t_{i\cdot} \) based on the bootstrapped estimate \( \widehat{var}(\bar{d}_{i \cdot}) \) that the MCS test uses. Model \( i \) is the constant variance model. Scenario where \( \nu=3 \), VaR forecasts evaluated using the tick loss function. The out-of-sample size \(P\) is displayed on top, the quantile level \( p \) on the left. Mind the different scales on the y-axes.}
        \end{figure}
\end{minipage}

\subsection{Sign of \(d_{ij}\)}\label{app:sign_d_ij} 

\begin{table}[H]
        \includegraphics*[width=.8\linewidth]{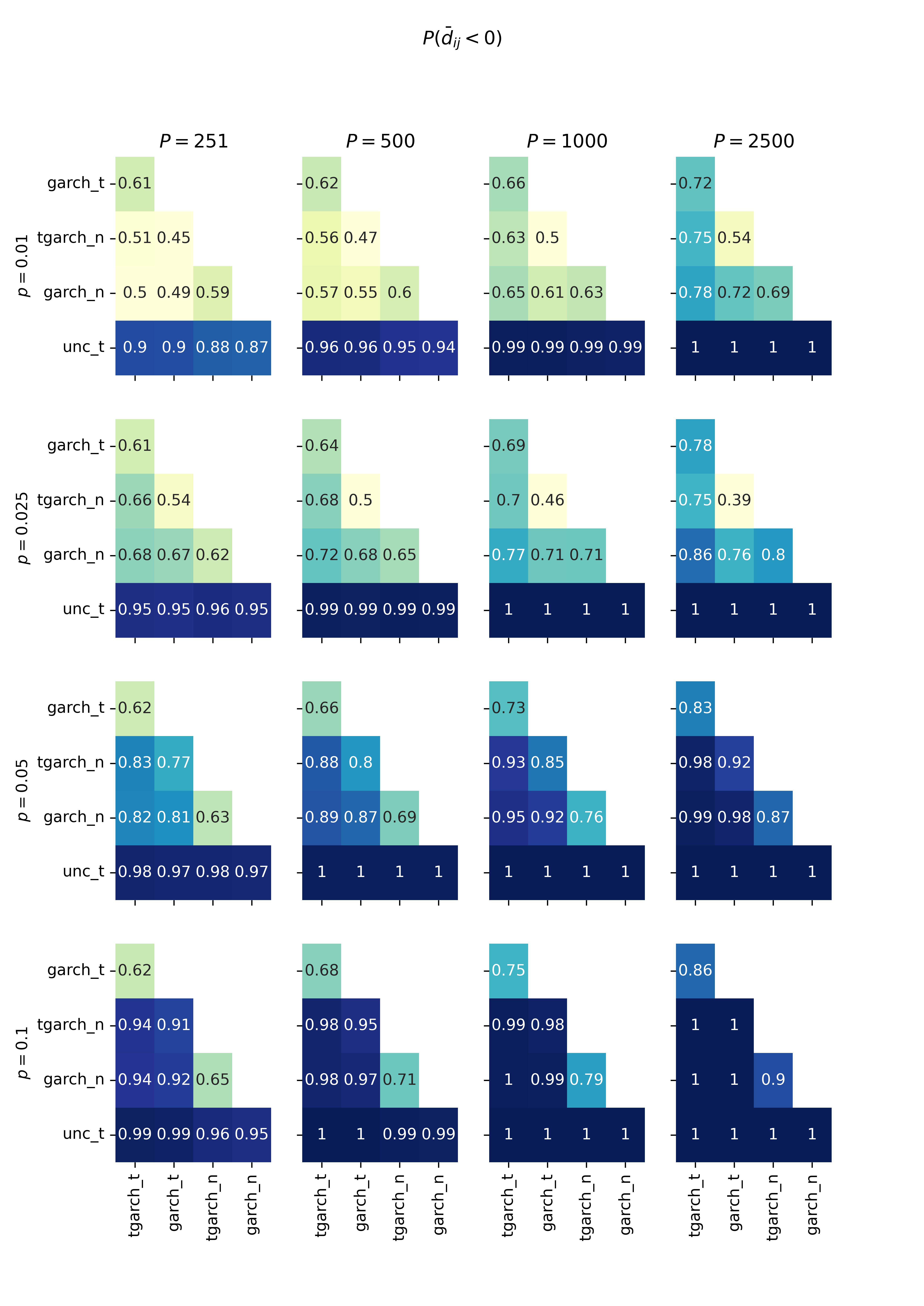}
        \caption{Sign of the \( \bar{d}_{ij} \)}
        \label{fig:var_sign_d_ij}
        \footnotesize{This table displays how often we observe samples for which the average sample loss differential \( \bar{d}_{ij} < 0. \), i.e.\  how often we find evidence that model \( i \) is superior to model \( j \). Model \( i \) is on the x-axis, model \( j \) on the y-axis. Scenario where \( \nu=3 \), VaR forecasts evaluated using the tick loss. The out-of-sample size \(P\) is displayed on top, the VaR level \( p \) on the left.}
\end{table}
\begin{table}[H]
        \includegraphics*[width=.8\linewidth]{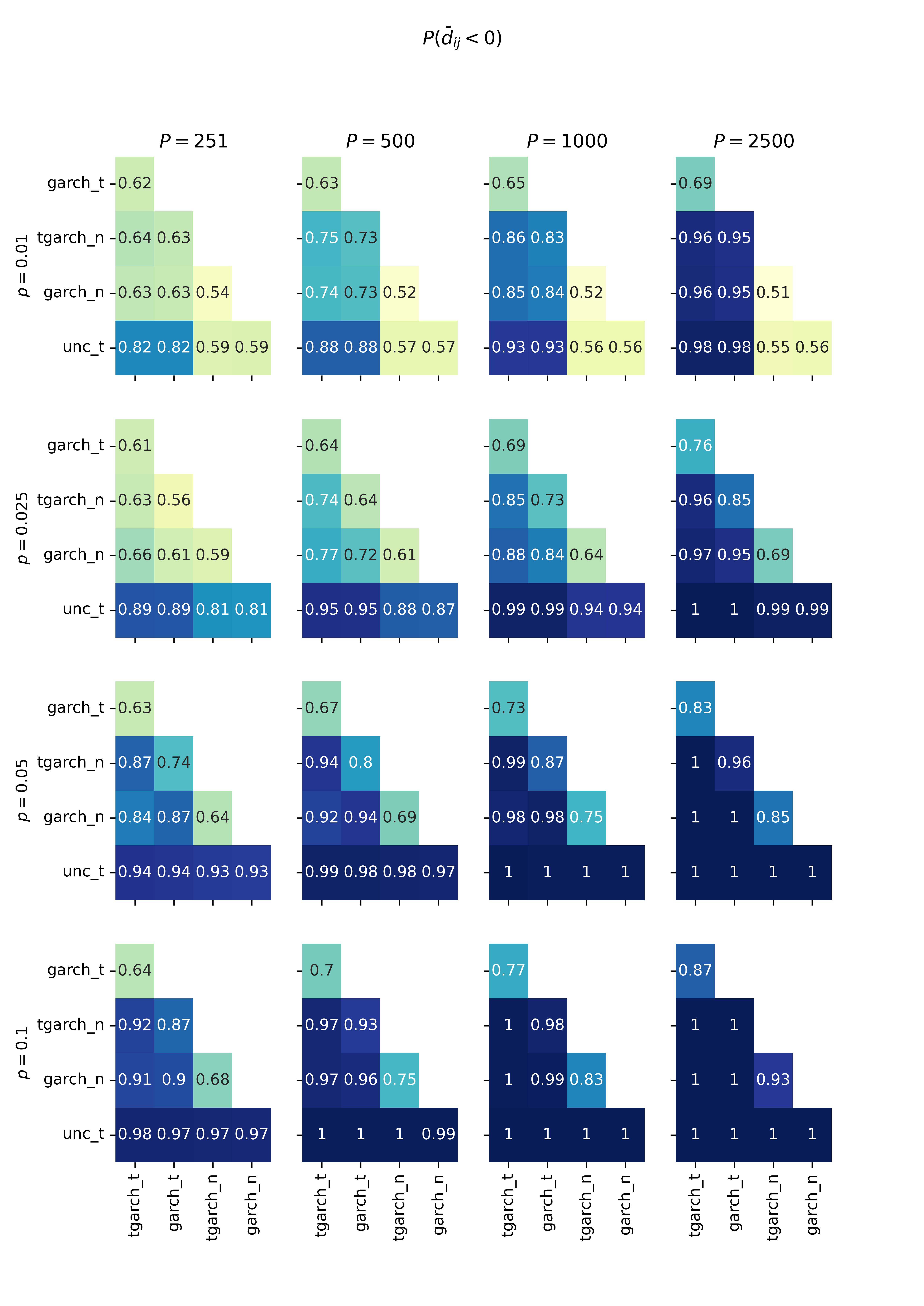}
        \caption{Sign of the \( \bar{d}_{ij} \)}
        \label{fig:es_sign_d_ij}
        \footnotesize{This table displays how often we observe samples for which the average sample loss differential \( \bar{d}_{ij} < 0. \), i.e.\  how often we find evidence that model \( i \) is superior to model \( j \). Model \( i \) is on the x-axis, model \( j \) on the y-axis. Scenario where \( \nu=3 \), VaR and ES forecasts evaluated using the \textit{AL} loss. The out-of-sample size \(P\) is displayed on top, the VaR level \( p \) on the left.}
\end{table}


\end{document}